\documentclass[11pt,a4paper]{article}

\usepackage[utf8]{inputenc}
\usepackage[T1]{fontenc}
\usepackage{amsmath,amssymb,amsthm}
\usepackage{bm}
\usepackage{booktabs}
\usepackage{array}
\usepackage{geometry}
\usepackage{enumitem}
\usepackage{algorithm}
\usepackage{algorithmic}
\usepackage{graphicx}
\usepackage{natbib}
\usepackage{float}
\usepackage{placeins}
\usepackage{xcolor}
\usepackage{hyperref}

\newtheorem{theorem}{Theorem}
\newtheorem{proposition}{Proposition}[section]
\newtheorem{corollary}[proposition]{Corollary}
\newtheorem{lemma}[proposition]{Lemma}
\theoremstyle{remark}
\newtheorem{remark}[proposition]{Remark}
\theoremstyle{definition}
\newtheorem{definition}[proposition]{Definition}

\DeclareMathOperator{\E}{\mathbb{E}}
\DeclareMathOperator{\Var}{Var}
\DeclareMathOperator{\Cov}{Cov}
\newcommand{\lvl}[1]{\textsf{\footnotesize[#1]}}

\title{What Paid Triangles Identify about Reserve Uncertainty:\\
       Information Boundaries in Loss Reserving}

\author{Robin Van Oirbeek\\[3pt] \normalsize University of Antwerp, Department of Mathematics, Antwerp, Belgium}

\date{\today}

\begin{document}
\maketitle

\begin{abstract}
Paid triangles identify some components of reserve uncertainty while
others enter through modelling assumptions. We distinguish population
identification, recovery by free accident-year-level and two-stage
second-moment fits, and precision at practical triangle sizes. In a
compound frequency--severity hierarchy, the population law identifies
the row dispersion governing non-diversifying reserve variance. With one
free level per accident year, its profile likelihood attains its supremum
only at homogeneity: the levels absorb realised row heterogeneity on
every triangle, in amounts and in counts. Anchoring reopens this channel,
but supplies only one observation per accident year. On a
$10 \times 10$ triangle the anchored estimate remains uncertain by a
factor of several. Within-row allocation dispersion survives conditioning
on row totals and hence survives free levels. In the predictive
experiments, the treatment of mean-structure estimation error separated the procedures by
several times more than the choice among plausible variance laws. A law
of the correct form improved coverage only when its dispersion came from
a prior instead of the free-level fit. On an outcome-free Schedule~P
panel, no amount-only procedure approaches nominal coverage. Strongly
one-sided misses reveal a displacement of realised reserves relative to
the fitted predictive centres that changing the amount-only variance law
cannot repair.
\end{abstract}


\section{Introduction}\label{sec:intro}

A stochastic reserving method turns a paid loss triangle into a
distribution for the outstanding reserve. It is rarely clear how much
of that distribution is learned from the triangle and how much is
assumed. We ask which parameters the law of the paid increments
determines, which survive a fit with free accident-year levels, and how
precisely a triangle of ordinary size estimates those that survive.

The question has a sharp answer for the parameter that matters most. The
reserve variance of a paid triangle separates into a severity term, a
diversifying cell term and a non-diversifying term generated by
heterogeneity shared across an accident year. The last one grows with the
square of the number of future cells in a row and dominates at high
heterogeneity. Its dispersion is identified by
the population law of the triangle, through the covariance of cells in
the same row. The free-level fit loses this signal regardless of sample size: one
free level per accident year absorbs that year's realised frailty, and
the profile likelihood of its dispersion prefers homogeneity on every
triangle.
The mechanism is the incidental-parameter problem of
\citet{neymanscott1948} \citep[see][]{Lancaster2000}, long familiar in the
econometrics of count panels \citep{HausmanHallGriliches1984,
AllisonWaterman2002}; what is specific to reserving is that the absorbed
parameter is the one that governs the reserve.

The argument of the paper is four sentences long. The population law
identifies the row dispersion, through the dependence between cells of
the same accident year. The standard reserving fit does not recover it,
because a free accident-year level absorbs the realised row effect.
Anchoring the levels reopens that channel, but a $10 \times 10$ triangle
then supplies only ten row totals. A conditional predictive must
therefore either import the row dispersion or price the realised row
levels diffusely. The remaining sections establish this argument and
examine its consequences.

We label results according to four distinct questions:
\begin{description}[leftmargin=1.2cm, labelwidth=0.8cm, itemsep=1pt]
\item[\lvl{P}] \emph{population identification}: the joint law of the
  paid increments, as if it were observed without limit rather than
  through one realised triangle, determines the parameter (for the
  variance parameters, with the cell means taken as known);
\item[\lvl{F}] \emph{free-level maximum likelihood}: the likelihood of the
  triangle with the accident-year levels as free parameters recovers it;
\item[\lvl{S}] \emph{the two-stage second-moment fit}: the
  quasi-likelihood mean fit followed by a regression of squared residuals
  (Section~\ref{sec:free:fits}) recovers it;
\item[\lvl{N}] \emph{precision at practical size}: a $10 \times 10$
  triangle determines it usefully.
\end{description}
A parameter can pass \lvl{P} and fail \lvl{F}, or pass \lvl{F} and fail
\lvl{N}. Predictive coverage is a separate question: weak \lvl{N}
limits the attainable width of a calibrated interval, while coverage
itself is measured in Section~\ref{sec:cost}. The results concern the
compound hierarchy of Section~\ref{sec:model} and the Dirichlet
allocation model of Section~\ref{sec:dirichlet}. ``Paid triangle'' below
means a triangle generated by one of these models. At \lvl{P}, the
population law identifies the mean structure; a single triangle still
has to separate row levels from development effects. On a
$10 \times 10$ triangle that estimation error is of the order of the
process variance (Section~\ref{sec:cost}), a result at level \lvl{N}.

\subsection{Contributions}\label{sec:intro:contrib}

\begin{enumerate}[label=(\arabic*), leftmargin=*, itemsep=4pt]
\item \emph{Identification and the free-level boundary.} The population
law identifies the severity composite $A = (1 + 1/\alpha)\bar Y$ and the
heterogeneity composite $B$ through the variance function, the row/cell
split through the within-row covariance and the severity split through
the third cumulant (Proposition~\ref{prop:identification},
Corollary~\ref{cor:cumulant}), and the reserve variance is dominated by
the row term (Proposition~\ref{prop:reserve_variance}). With one free
level per accident year the profile likelihood of the row dispersion
$\kappa_r$ attains its supremum only at $\kappa_r = \infty$, on every
triangle with an informative row, in amounts and in counts
(Theorem~\ref{thm:boundary}, with the attainment step in
Lemma~\ref{lem:attain} and the jointly profiled case in
Corollary~\ref{cor:joint_profile}); the two-stage second-moment
estimator, whose first stage reproduces the row totals, has a target at
the cap $B = 0$ and a severity composite deflated by the mean cell share
(Proposition~\ref{prop:two_stage_target}), and estimates
the cell dispersion instead (Proposition~\ref{prop:two_kappa_id}).
Conditionally on the triangle, $\kappa_r$ enters a predictive only
through the posterior of each realised row intensity
(Proposition~\ref{prop:conditional}): a diffuse treatment needs no
$\kappa_r$ and over-states the level uncertainty of well-observed rows, a
proper prior needs it.
\item \emph{A dispersion that survives free levels.} In the Dirichlet
allocation model the concentration of the within-row allocation is
carried by the likelihood conditional on the row totals, is not absorbed
by free levels and is consistently estimated
(Proposition~\ref{prop:dirichlet_c}). Where a dispersion lives in the
hierarchy decides whether free levels absorb it.
\item \emph{Precision at practical triangle size.} The channels that
survive are weak at $I = 10$: the Godambe information of the curvature
channel, with the cell means known, implies an asymptotic coefficient of
variation of the dispersion estimate of at least $0.8$
(Proposition~\ref{prop:godambe}; the precision of the cell dispersion
after free-level fitting is not derived), and with the
levels anchored the central $80\%$ of the maximum-likelihood estimates of
$\kappa_r$ spans a factor of at least $4.7$ (Table~\ref{tab:boundary}).
The information map of Table~\ref{tab:idmap} collects the levels reached
by every parameter.
\item \emph{Predictive and empirical consequences.} Placing every
procedure on two organising axes, its variance law and its treatment of
the estimation error of the mean structure, and replacing the inputs of
one benchmark construction by the truth (Section~\ref{sec:cost}) shows
that, in the constructions compared and at the calibration studied, the
second axis separated the procedures most. These comparisons rank
complete implementations: propagating the estimation error raises
coverage by $23$ to $30$ points, the variance law moves it by at most
$8$, and a variance law of the correct form improved coverage, in the
implementations compared, only where its dispersion was supplied by a
prior rather than estimated through the free-level fit. On an outcome-free
Schedule~P panel no amount-only procedure approaches nominal coverage,
and the realised reserves fall below the intervals far more often than
above them, a displacement of the outcomes relative to the fitted
predictive centres that no amount-only variance law repairs.
\end{enumerate}

In the population, a paid triangle therefore supplies a mean structure and
a two-moment variance law, both of which a triangle of practical size
estimates with error; the row dispersion that dominates the unconditional
reserve variance, and sharpens the level term of a conditional predictive
on its thin latest years (Proposition~\ref{prop:conditional}), comes from
anchored levels, from a richer data scope or from a prior, or is priced
diffusely, as the ODP bootstrap prices it. Counts do not remove the
boundary. They separate frequency from
severity and sharpen the cell dispersion, and anchoring addresses the
row level. Section~\ref{sec:model} sets out the hierarchy,
Section~\ref{sec:population} population identification,
Section~\ref{sec:free} the free-level fit and the boundary,
Section~\ref{sec:dirichlet} the allocation concentration,
Section~\ref{sec:precision} precision and the information map,
Section~\ref{sec:illus} the numerical and empirical illustrations,
Section~\ref{sec:cost} what the imports cost a predictive and
Section~\ref{sec:implications} the implications for reserving. Proofs,
the higher moments, the Dirichlet material in full, further evidence on
the free-level fit, the predictive constructions in full and the
simulation and Schedule~P details are in the Supplement.

\section{A Structural Representation of Paid Triangles}\label{sec:model}

\subsection{The hierarchy}

Claims of accident year $i$ that are paid in development year $j$ arrive
as a Poisson process whose intensity carries two sources of
heterogeneity: a frailty shared by every cell of the accident year and a
shock specific to the cell. Each claim pays an independent Gamma severity.

\begin{definition}[CNBG hierarchy]\label{def:cnbg}
For accident year $i = 1, \ldots, I$ and development year
$j = 1, \ldots, J$,
\begin{align}
N_{ij} \mid \lambda_i,\varepsilon_{ij}
        &\sim \mathrm{Poisson}(\lambda_i\,\varepsilon_{ij}\,w_j),
        \label{eq:cnbg_count}\\
    \lambda_i
        &\sim \mathrm{Gamma}\!\left(\kappa_r,\,\frac{\kappa_r}{\mu_i}\right),
        \label{eq:cnbg_frailty}\\
    \varepsilon_{ij}
        &\overset{\mathrm{iid}}{\sim} \mathrm{Gamma}(\kappa_c,\,\kappa_c),
        \label{eq:cnbg_cellshock}\\
    Y_k
        &\overset{\mathrm{iid}}{\sim} \mathrm{Gamma}(\alpha,\,\gamma),
        \label{eq:cnbg_severity}\\
    X_{ij}
        &= \sum_{k=1}^{N_{ij}} Y_k,
        \label{eq:cnbg_amount}
\end{align}
with $X_{ij} = 0$ when $N_{ij} = 0$, and $\lambda_i$, $(\varepsilon_{ij})_j$
and the severities mutually independent. The accident-year levels are
$\mu_i > 0$, the development weights $w_j > 0$ with $\sum_j w_j = 1$, the
row dispersion $\kappa_r > 0$, the cell dispersion $\kappa_c > 0$, the
severity shape $\alpha > 0$ and rate $\gamma > 0$.
\end{definition}

We call the marginal law of $X_{ij}$ the Compound Negative
Binomial--Gamma (CNBG) distribution. It is the compound Poisson--Gamma
model of \citet{JorgensenSouza1994} with two Gamma mixing layers on the
Poisson intensity; in the single-frailty case $\kappa_c = \infty$ it is
the compound negative binomial distribution with Gamma severities
\citep{KlugmanPanjerWillmot2019}. Here it is first the structural vehicle
through which the location of each source of uncertainty can be read, and
then, in Section~\ref{sec:cost}, the variance law of a reserving
predictive.
Write $U_i = \lambda_i/\mu_i$ for the normalised frailty,
$M_{ij} = U_i\varepsilon_{ij}$ for the total cell multiplier,
$\bar Y = \alpha/\gamma$ for the mean severity and
$\nu_{ij} = \mu_i w_j \bar Y$ for the expected cell amount. Small
$\kappa_r$ means high heterogeneity between accident years, small
$\kappa_c$ high heterogeneity between the cells of a row.

\paragraph{Notation.} We distinguish five dispersion quantities:
\begin{itemize}[nosep, leftmargin=*]
\item $\kappa_r$: the row dispersion, shared by the cells of an
      accident year;
\item $\kappa_c$: the cell-shock dispersion, specific to a cell;
\item $B = \kappa_r^{-1} + \kappa_c^{-1} + (\kappa_r\kappa_c)^{-1}$: the
      quadratic coefficient of the variance function, the heterogeneity
      composite;
\item $\kappa_B = 1/B$: the \emph{curvature dispersion}, the one scalar
      a second-moment fit sees;
\item $c$: the within-row concentration of the Dirichlet allocation
      model of Section~\ref{sec:dirichlet}, a different model.
\end{itemize}
\begin{equation}\label{eq:B}
B = \frac{1}{\kappa_r} + \frac{1}{\kappa_c} + \frac{1}{\kappa_r\kappa_c},
\qquad \kappa_B = 1/B.
\end{equation}
In the single-frailty case $\kappa_c = \infty$, which is the design of
every simulation reported below, $\kappa_B = \kappa_r$. Under free
levels the likelihood result of Section~\ref{sec:free} concerns
$\kappa_r$, while the second-moment working fit reads mainly $\kappa_c$
through $\kappa_B$. References to dispersion below name the quantity
involved.

\subsection{Mean, variance and within-row covariance}

\begin{proposition}[Moments \lvl{P}]\label{prop:moments}
Under Definition~\ref{def:cnbg}, with $A = (1 + 1/\alpha)\bar Y$:
\begin{enumerate}[label=(\roman*), nosep]
\item $\E[X_{ij}] = \nu_{ij}$;
\item $\Var(X_{ij}) = A\,\nu_{ij} + B\,\nu_{ij}^2$;
\item $\Cov(X_{ij}, X_{ij'}) = \nu_{ij}\nu_{ij'}/\kappa_r$ for $j \neq j'$:
      only the shared row frailty contributes;
\item $\Cov(X_{ij}, X_{i'j'}) = 0$ for $i \neq i'$.
\end{enumerate}
\end{proposition}

\begin{proof}
Conditionally on $M_{ij}$ the cell is compound Poisson--Gamma with
cumulants $\kappa_n(X_{ij} \mid M_{ij}) = (\nu_{ij}M_{ij}/\bar Y)\,\E[Y^n]$,
so $\E[X_{ij} \mid M_{ij}] = \nu_{ij}M_{ij}$ and
$\Var(X_{ij} \mid M_{ij}) = A\nu_{ij}M_{ij}$. Part (i) follows from
$\E[M_{ij}] = 1$. For (ii), the law of total variance gives
$A\nu_{ij} + \nu_{ij}^2\Var(M_{ij})$, and since $U_i$ and
$\varepsilon_{ij}$ are independent unit-mean Gammas,
$\Var(M_{ij}) = (1+\kappa_r^{-1})(1+\kappa_c^{-1}) - 1 = B$. For (iii),
cells of one row share $U_i$ and carry independent shocks, so
$\Cov(X_{ij}, X_{ij'}) = \nu_{ij}\nu_{ij'}(\E[U_i^2]-1) =
\nu_{ij}\nu_{ij'}/\kappa_r$. Part (iv) holds because different accident
years have independent frailties and shocks.
\end{proof}

The linear term $A\nu_{ij}$ is the severity component, the variability of
a sum of a random number of random claim sizes. The quadratic term
$B\nu_{ij}^2$ is the heterogeneity component; it is absent from the
Over-Dispersed Poisson (ODP) variance and can be imitated by a Tweedie
power variance only locally (Section~\ref{sec:tweedie}). Part (iii)
makes the cells of one accident year positively correlated, and it is the
only moment of order two that carries $\kappa_r$ alone. The third and
fourth moments are in Section~\ref{supp:moments}.

\subsection{The reserve variance}

\begin{proposition}[Reserve variance \lvl{P}]\label{prop:reserve_variance}
Write $\mathcal{D} = \{(i,j) : i+j \le I+1\}$ for the observed cells,
$S_i = \sum_{j : (i,j) \notin \mathcal{D}} \nu_{ij}$ for the expected
outstanding amount of accident year $i$ and $S = \sum_i S_i$. The
outstanding reserve $R = \sum_{(i,j) \notin \mathcal{D}} X_{ij}$ has
variance
\begin{equation}\label{eq:reserve_var}
\Var(R) \;=\;
  \underbrace{A\,S}_{\text{severity}}
  \;+\;
  \underbrace{\frac{1}{\kappa_r}\sum_{i} S_i^2}_{\text{row frailty}}
  \;+\;
  \underbrace{\frac{1}{\kappa_c}\Bigl(1+\frac{1}{\kappa_r}\Bigr)
  \sum_{(i,j)\notin\mathcal{D}} \nu_{ij}^2}_{\text{cell shock}}.
\end{equation}
\end{proposition}

\begin{proof}
Sum the variances of Proposition~\ref{prop:moments}(ii) and the
within-row covariances of (iii) over the future cells; cross-row
covariances vanish by (iv). The terms $\kappa_r^{-1}\nu_{ij}^2$ of the
cell variances and $\kappa_r^{-1}\nu_{ij}\nu_{ij'}$ of the covariances
combine to $\kappa_r^{-1}S_i^2$ for each row.
\end{proof}

For a row with $m_i$ future cells of comparable size the row-frailty term
grows as $m_i^2$ and the cell-shock term as $m_i$: the first does not
diversify within a row, the second does. At high heterogeneity the row
term dominates the reserve variance, and at low heterogeneity the
severity term does. The dispersion that governs the row term is
$\kappa_r$. Equation~\eqref{eq:reserve_var} is an unconditional variance,
the dispersion of the outstanding amount over triangles not yet observed.
A reserve is set on one observed triangle, and conditionally on it
$\kappa_r$ enters in one place only; the rest of the paper asks what a
triangle says about $\kappa_r$ and, through that place, when the answer
changes a prediction.

\begin{proposition}[Conditional reserve moments \lvl{P}]\label{prop:conditional}
Fix accident year $i$, write $\mathcal{D}_i$ for its observed cells,
$F_i = \sum_{j:(i,j)\in\mathcal{D}} w_j$ for its observed share of the
development pattern and $R_i = \sum_{j:(i,j)\notin\mathcal{D}} X_{ij}$ for
its outstanding amount. In the single-frailty case $\kappa_c = \infty$,
with the development weights and the severity known,
\begin{equation}\label{eq:cond_moments}
\begin{gathered}
\E[R_i \mid \mathcal{D}_i] = (1-F_i)\,\bar Y\,\E[\lambda_i \mid \mathcal{D}_i], \\
\Var(R_i \mid \mathcal{D}_i) =
\underbrace{(1-F_i)\,A\,\bar Y\,\E[\lambda_i \mid \mathcal{D}_i]}_{\text{process}}
+ \underbrace{(1-F_i)^2\,\bar Y^2\,\Var(\lambda_i \mid \mathcal{D}_i)}_{\text{level}},
\end{gathered}
\end{equation}
where the posterior of the realised intensity $\lambda_i$ is proportional
to $g(\lambda_i; \kappa_r, \kappa_r/\mu_i)\,f(\mathbf{x}_i \mid \lambda_i)$,
with $f$ the compound Poisson--Gamma likelihood of the observed cells.
The row dispersion enters the conditional law of $R_i$ only through this
posterior. For the count triangle the posterior is
$\mathrm{Gamma}(\kappa_r + N_i^{\mathrm{obs}}, \kappa_r/\mu_i + F_i)$, so that
\begin{equation}\label{eq:cond_level_count}
\Var(\lambda_i \mid N_i^{\mathrm{obs}})
= \frac{\kappa_r + N_i^{\mathrm{obs}}}{(\kappa_r/\mu_i + F_i)^2},
\end{equation}
which tends to $N_i^{\mathrm{obs}}/F_i^2$ as $\kappa_r \to 0$ and to $0$ as
$\kappa_r \to \infty$.
\end{proposition}

\begin{proof}
Given $\lambda_i$ the future cells of row $i$ are independent of its
observed cells and compound Poisson--Gamma with means $\lambda_i w_j \bar Y$
and variances $A\lambda_i w_j \bar Y$ (the conditional cumulants in the
proof of Proposition~\ref{prop:moments}), so $\E[R_i \mid \lambda_i] =
(1-F_i)\lambda_i\bar Y$ and $\Var(R_i \mid \lambda_i) = (1-F_i)A\lambda_i\bar Y$;
the laws of total expectation and variance over $\lambda_i \mid \mathcal{D}_i$
give \eqref{eq:cond_moments}. For counts, $N_i^{\mathrm{obs}} \mid \lambda_i
\sim \mathrm{Poisson}(\lambda_i F_i)$ by thinning, and Gamma--Poisson
conjugacy gives the posterior and \eqref{eq:cond_level_count}.
\end{proof}

\begin{remark}[A conditional predictive without $\kappa_r$]\label{rem:diffuse}
Let the level carry the scale prior $\pi(\mu_i) \propto 1/\mu_i$.
Integrating the frailty prior over it,
\[
\int_0^\infty g(\lambda_i; \kappa_r, \kappa_r/\mu_i)\,\frac{d\mu_i}{\mu_i}
= \frac{1}{\lambda_i}
\qquad\text{for every } \kappa_r > 0,
\]
by the substitution $t = \kappa_r\lambda_i/\mu_i$: the induced prior on
the realised intensity is flat in $\log\lambda_i$ whatever $\kappa_r$ is,
and the posterior of $\lambda_i$, hence the conditional law of $R_i$, does
not involve $\kappa_r$. The posterior is proper only on an informative
row: for the count triangle it is $\mathrm{Gamma}(N_i^{\mathrm{obs}},
F_i)$, improper when $N_i^{\mathrm{obs}} = 0$, and for amounts the same
condition is a positive observed row total, so an all-zero row has no
diffuse posterior and needs a proper anchor, from a prior or from an
exposure; what an implementation does with such rows is an operational
safeguard separate from this limit. This is the diffuse limit of the hierarchy and the
first-order target of the ODP bootstrap \citep{VanOirbeek2026cond}; its
level term for the count triangle is $N_i^{\mathrm{obs}}/F_i^2$, the
$\kappa_r \to 0$ value of \eqref{eq:cond_level_count}. The prior is
improper and the construction is not a recommendation; it shows that a
coherent conditional predictive need not import $\kappa_r$. What it does
instead is price the uncertainty about each realised row intensity at its
diffuse value, which exceeds the finite-$\kappa_r$ value
\eqref{eq:cond_level_count} if and only if $N_i^{\mathrm{obs}} \ge
\mu_i^2F_i^2/(\kappa_r + 2\mu_iF_i)$, a threshold below half the prior
expectation $\mu_iF_i$. A finite $\kappa_r$ sharpens the level term, most
on the thin latest accident years, and that $\kappa_r$ is what the
free-level fit does not supply (Theorem~\ref{thm:boundary}).
\end{remark}

\paragraph{When the row dispersion changes a conditional prediction.}
The two stances price the level term of \eqref{eq:cond_moments} by
\eqref{eq:cond_level_count} and by its diffuse limit, and their ratio
\[
\rho_i
 = \frac{(\kappa_r + N_i^{\mathrm{obs}})\,F_i^2}
        {(\kappa_r/\mu_i + F_i)^2\,N_i^{\mathrm{obs}}}
\]
measures what knowing $\kappa_r$ buys on row $i$. At the calibration of
Section~\ref{sec:illus:boundary}, with each row's count at its
expectation, $\rho_i$ is $0.77$ for the latest accident year ($F_i =
0.35$) and $0.86$ for a fully developed one at $\kappa_r = 5$, $0.62$ and
$0.75$ at $\kappa_r = 10$, and $0.25$ and $0.37$ at $\kappa_r = 50$: the
gain is largest where the heterogeneity is weakest, so that the anchor
binds, and on the thin latest years, and it concerns the level term
alone, which on the latest year is of the same order as the process
term. The question the paper asks is therefore conditional, what a
triangle supplies to a predictive of its own outstanding amount, and the
boundary of Section~\ref{sec:free} says that the row dispersion is not
among it: a free-level predictive either prices the levels diffusely,
with $\rho_i = 1$, or imports $\kappa_r$.

\paragraph{Scope.} The hierarchy is a model of claim emergence: claims
arrive in a cell and each pays an amount. It applies to incremental paid
triangles and not to incurred triangles, whose increments mix payments
with revisions of case reserves and IBNR provisions that have no
compound-sum representation and can be negative.

\subsection{The incumbent variance laws}\label{sec:tweedie}

The two variance laws in common use are the linear ODP law
$\phi\nu_{ij}$ and the Tweedie power law $\phi\nu_{ij}^p$ with
$p \in (1, 2)$. Neither is the law of Proposition~\ref{prop:moments}(ii),
and the relation between them is exact.

\begin{proposition}[Non-Tweedie marginal \lvl{P}]\label{prop:not_tweedie}
Under Definition~\ref{def:cnbg}:
\begin{enumerate}[label=(\roman*), nosep]
\item conditionally on $M_{ij}$, $X_{ij}$ is compound Poisson--Gamma and
      hence Tweedie with power $p_c = (\alpha+2)/(\alpha+1) \in (1,2)$;
\item no Tweedie family with a common power $p$ and dispersion $\phi$
      across cells reproduces the variance function of
      Proposition~\ref{prop:moments}(ii) at three or more distinct cell
      means, for any finite $\kappa_B$: the cells of a triangle are not a
      Tweedie GLM. (At two distinct means a power law can always be
      matched, since two parameters fit two points.)
\end{enumerate}
\end{proposition}

\begin{proof}
(i) is the standard compound Poisson--Gamma result
\citep{JorgensenSouza1994}. (ii) A Tweedie law with $p \in (1,2)$ has variance $\phi\nu^p$, so
$\log V = \log\phi + p\log\nu$ is affine in $t = \log\nu$. For
$V = A\nu + B\nu^2$ with $A, B > 0$, $\log V = t + \log(A + Be^{t})$,
whose second derivative in $t$ is $ABe^{t}/(A + Be^{t})^2 > 0$: a strictly
convex function of $t$, which agrees with an affine function at no three
distinct points. Two cells of different means can be matched, as with
$A = B = 1$, $\nu \in \{1, 2\}$, $p = \log 3/\log 2$ and $\phi = 2$;
three cannot. Whether a single marginal law is Tweedie for some
cell-specific $(p, \phi)$ is not decided by the variance function and is
not claimed.
\end{proof}

The Tweedie form holds conditionally because the compounding law is
Poisson; marginalising over the Gamma frailty turns that Poisson into a
negative binomial and adds the quadratic term, so a Tweedie GLM is a
conditional model fitted to marginal data. As $\kappa_B \to \infty$ the
quadratic term vanishes and the variance becomes the linear $A\nu_{ij}$:
the ODP law is the limit of no heterogeneity at all.

The marginal variance $A\nu + B\nu^2$ makes a fitted Tweedie power a
local, volume-dependent summary. The power matching the true variance function
around a mean $\nu_0$ is $1 + B\nu_0/(A + B\nu_0)$, which rises with the
claim volume and with either heterogeneity and falls with the severity
dispersion (Proposition~\ref{prop:local_p} and
Corollary~\ref{cor:p_not_structural}, Section~\ref{supp:moments}). Two
portfolios with the same structural parameters and different volumes show
different $\hat p$, and a test of $\hat p$ against $1$ or $2$ has power
against heterogeneity and against severity alternatives without telling
them apart. The structural pair $(B, \alpha)$ separates what $\hat p$
merges; what the triangle recovers of that pair is the subject of the
next three sections.

\section{Population Identification}\label{sec:population}

We first take the cell means $\nu_{ij}$ as known and study population
identification. Section~\ref{sec:free} then fits the means with free
accident-year levels. Theorem~\ref{thm:boundary} explains why this change
in the estimation problem changes which parameters are recovered.

\begin{proposition}[Identification from amounts \lvl{P}]\label{prop:identification}
The composites $A$ and $B$ are identified from the amount triangle
$\{X_{ij} : (i,j) \in \mathcal{D}\}$ alone, provided it contains at least
two distinct expected cell means. The row dispersion $\kappa_r$ is
identified by the within-row covariance, and with it the split of $B$
into $\kappa_r$ and $\kappa_c$.
\end{proposition}

\begin{proof}
By Proposition~\ref{prop:moments}(ii),
$\E[(X_{ij}-\nu_{ij})^2] = A\nu_{ij} + B\nu_{ij}^2$, a quadratic in
$\nu_{ij}$ without intercept; two distinct values of $\nu_{ij}$ give a
design of rank two and determine $(A, B)$, as for any quadratic variance
function \citep{DavidianCarroll1987}. By (iii),
$\E[(X_{ij}-\nu_{ij})(X_{ij'}-\nu_{ij'})] = \nu_{ij}\nu_{ij'}/\kappa_r$
for any two observed cells of one row, which determines $\kappa_r$; then
$\kappa_c$ follows from $B$ by \eqref{eq:B}.
\end{proof}

\begin{corollary}[The severity split \lvl{P}]\label{cor:cumulant}
The third central moment of $X_{ij}$ (Proposition~\ref{prop:higher_moments})
has leading coefficient $c_3 = (\alpha+1)(\alpha+2)\bar Y^2/\alpha^2$, with
$c_3/A^2 = (\alpha+2)/(\alpha+1) \in (1,2)$. In the population it
therefore identifies $\alpha$, hence $\bar Y$, which the variance
function leaves conflated in $A$.
\end{corollary}

Whether the third-moment channel is usable on a single triangle is not
settled here: the sampling variance of an empirical third moment involves
the sixth moment of $X_{ij}$, and we have not computed its precision at
$10 \times 10$. A second-moment fit leaves the channel unused, so a
procedure built on the first two moments carries no information on the
zero mass or the skewness of an individual future cell, both of which
depend on $(\alpha, \bar Y)$ separately; in practice the severity split
comes from the count triangle \citep{VanOirbeek2026nbcl}.

The marginal law of a single cell does not separate the two frailties:
$U_i$ and $\varepsilon_{ij}$ enter a cell only through their product
$M_{ij}$, so every single-cell moment depends on the cumulants of $M_{ij}$
alone (Section~\ref{supp:moments}). The separation in
Proposition~\ref{prop:identification} comes from the joint law of a row,
and that is the channel the free-level fit closes.

\section{What the Free-Level Fit Recovers}\label{sec:free}

\subsection{Two fits}\label{sec:free:fits}

Reserving fits the mean structure with one free level per accident year
and one free effect per development year, $\log\nu_{ij} = a_i + b_j$, the
parameterisation under which the quasi-Poisson fit reproduces the chain
ladder \citep{EnglandVerrall1999}. Two ways of reaching the dispersions
$\kappa_r$ and $\kappa_c$ from there are considered. The first is maximum likelihood under
Definition~\ref{def:cnbg} with the levels $\mu_i$ free. The second is the
two-stage second-moment fit: a quasi-likelihood fit of
$(a_i, b_j)$ with Poisson working variance, giving $\hat\nu_{ij}$, followed
by the weighted regression of the squared residuals on the variance
function,
\begin{equation}\label{eq:wls}
(\hat A, \hat B) = \arg\min_{A, B \ge 0}\sum_{(i,j)\in\mathcal{D}}
\hat\nu_{ij}^{-2}\bigl[(X_{ij}-\hat\nu_{ij})^2 - A\hat\nu_{ij} -
B\hat\nu_{ij}^2\bigr]^2,
\end{equation}
with $\hat\kappa_B = 1/\hat B$; the Bayesian version wraps the same
estimating equation in a working likelihood with priors
(Section~\ref{sec:predictive}).

\begin{lemma}[Absorption of the row frailty \lvl{S}]\label{lem:absorption}
Under the free-level quasi-likelihood fit the fitted row totals equal the
observed row totals, so the realised frailty $U_i$, constant across the
cells of row $i$, is carried by $\hat a_i$: the fitted means satisfy $\hat\nu_{ij} \approx \nu_{ij}U_i$ up to the sampling error of the development pattern and of the row total itself, whose compound noise the free level absorbs together with the frailty, and the first stage involves $\kappa_r$ nowhere. The lemma is a mechanism; the proposition that follows derives the target of the second stage with the first-stage error of the row total included. The
cell shock $\varepsilon_{ij}$, being neither row- nor column-constant,
cannot be absorbed by $a_i + b_j$ and remains in the residual.
\end{lemma}

The absorption mechanism is the classical one by which free fixed effects
absorb group-constant random effects \citep{neymanscott1948, Lancaster2000},
and no novelty is claimed for it. It is well known for count panels: with
one free effect per unit, the Poisson likelihood conditional on the unit
totals is multinomial and carries no information on the dispersion of the
unit effects \citep{HausmanHallGriliches1984}, and the conditional
``fixed-effects negative binomial'' of the same paper, which places the unit
effect in the dispersion rather than in the mean, is not a fixed-effects
model in the usual sense \citep{AllisonWaterman2002}. A paid triangle with free accident-year levels is such a panel, with the
accident years as units and the cells as repeated observations.

\begin{proposition}[Limit of the two-stage estimator under replicated free-level designs \lvl{S}]
\label{prop:two_stage_target}
Let the single-frailty hierarchy hold ($\kappa_c = \infty$), let $J$ be
fixed and let $I \to \infty$ with the cell population of the
$10 \times 10$ design replicated, so that the development pattern is
consistently estimated. Write $k_i$ for the number of observed cells of
row $i$, $\nu_{i+} = \sum_{j \le k_i}\nu_{ij}$ and $s_{ij} = \nu_{ij}/\nu_{i+}$
for the cell's share of its row's observed expectation.
\begin{enumerate}[label=(\roman*), nosep]
\item With the pattern at its limit, the free-level quasi-Poisson fit
      returns $\hat\nu_{ij} = \nu_{ij}U_i(1 + e_i)$ exactly, where
      $e_i = \sum_{j \le k_i} s_{ij}\varepsilon_{ij}$ is the relative error
      of the observed row total and $\varepsilon_{ij} = X_{ij}/(\nu_{ij}U_i) - 1$
      the relative error of the cell, with $\E[\varepsilon_{ij} \mid U_i] = 0$
      and $\Var(\varepsilon_{ij} \mid U_i) = A/(\nu_{ij}U_i)$.
\item The squared residual has conditional mean
      \[
      \E\bigl[(X_{ij} - \hat\nu_{ij})^2 \,\big|\, U_i\bigr]
      = A\,\nu_{ij}U_i\,(1 - s_{ij}),
      \]
      exactly in $U_i$ and free of $\kappa_r$: the row frailty enters the
      residual only through the level it has been absorbed into, and the
      first-stage error of the row total removes the share $s_{ij}$ of the
      cell's own variance.
\item \emph{(Leading order.)} Under the additional small-relative-noise
      approximation, with terms of relative order $A/(\nu_{i+}U_i)$
      neglected, the weighted regression \eqref{eq:wls} on the rows with
      a positive observed total is the least-squares fit of
      $A(1 - s_{ij})$ on $\nu_{ij}U_i$ with an intercept, whose slope is
      \[
      B^\ast = -A\,\frac{\Cov(s_{ij}, \nu_{ij}U_i)}{\Var(\nu_{ij}U_i)},
      \]
            the moments taken over the cells of the design and the frailty law.
      For the decreasing pattern and the replicated triangular design
            used below the covariance is positive, so $B^\ast < 0$ and the constrained estimator
      $\hat B \ge 0$ converges to the boundary $B = 0$: the cap is its
      target, not a failure of the optimiser. With the slope at the
      boundary the intercept is that of the constrained fit, the mean of
      $z_{ij}/\hat\nu_{ij}$, so $\hat A$ converges to $A(1 - \bar s)$ with
      $\bar s$ the mean share over the cells, and not to the unconstrained
      intercept $A^\ast = A(1 - \bar s) - B^\ast\E[\nu_{ij}U_i]$, which lies
      within a few percent of $A$; at the design below $\bar s = 10/55$ and
      the target is $0.82A$. The limits are those of the population
      least-squares problem of the implemented regression; the closed
      forms are their leading-order approximations, which the simulated
      estimates approach as the neglected relative order vanishes.
\end{enumerate}
\end{proposition}

\begin{proof}
(i) The quasi-Poisson score for the row effect $a_i$ is
$\sum_{j \le k_i}(X_{ij} - \hat\nu_{ij}) = 0$, so the fitted row total
equals the observed one; with $\hat b_j = b_j$ the fitted cells are the
observed total allocated by the true pattern, $\hat\nu_{ij} =
s_{ij}\sum_{l \le k_i}X_{il} = \nu_{ij}U_i(1 + e_i)$. The moments of
$\varepsilon_{ij}$ are those of a compound Poisson--Gamma cell given its
intensity (Proposition~\ref{prop:moments}). (ii) $X_{ij} - \hat\nu_{ij}
= \nu_{ij}U_i(\varepsilon_{ij} - e_i)$, and given $U_i$ the cells are
independent, so $\Cov(\varepsilon_{ij}, e_i \mid U_i) = s_{ij}A/(\nu_{ij}U_i)
= A/(\nu_{i+}U_i)$ and $\Var(e_i \mid U_i) = \sum_l s_{il}^2A/(\nu_{il}U_i)
= A/(\nu_{i+}U_i)$; hence $\E[(\varepsilon_{ij} - e_i)^2 \mid U_i] =
A/(\nu_{ij}U_i) - 2A/(\nu_{i+}U_i) + A/(\nu_{i+}U_i) = A(1 - s_{ij})/(\nu_{ij}U_i)$,
and multiplying by $\nu_{ij}^2U_i^2$ gives the display. (iii) Dividing
the squared residual by $\hat\nu_{ij}$ turns \eqref{eq:wls} into the
ordinary least-squares fit of $z_{ij}/\hat\nu_{ij}$ on $\hat\nu_{ij}$ with
intercept $A$ and slope $B$, and by (i)
\[
\frac{z_{ij}}{\hat\nu_{ij}}
 = \nu_{ij}U_i\,\frac{(\varepsilon_{ij} - e_i)^2}{1 + e_i},
\qquad
\hat\nu_{ij} = \nu_{ij}U_i(1 + e_i),
\]
defined on the rows whose observed total is positive, so that
$1 + e_i > 0$; a row with a zero observed total has $\hat\nu_{ij} = 0$,
no ratio, and is excluded from \eqref{eq:wls}, as the implementation
excludes cells without a positive fitted mean (at the calibration below
such rows have negligible probability). The squared residual and $e_i$
are dependent, so the expectation is taken given $U_i$ alone, with the
expansion $1/(1 + e_i) = 1 - e_i + e_i^2/(1 + e_i)$:
\[
\E\Bigl[\frac{z_{ij}}{\hat\nu_{ij}} \,\Big|\, U_i\Bigr]
 = A(1 - s_{ij})
   - \nu_{ij}U_i\,\E\bigl[(\varepsilon_{ij} - e_i)^2 e_i \mid U_i\bigr]
   + \nu_{ij}U_i\,\E\Bigl[\frac{(\varepsilon_{ij} - e_i)^2 e_i^2}{1 + e_i}
     \,\Big|\, U_i\Bigr],
\]
the first term by (ii). The second term is a third-order moment of the
cells: given $U_i$ the third cumulant of $\varepsilon_{ij}$ is
$c_3/(\nu_{ij}U_i)^2$ with $c_3$ the severity constant of
Section~\ref{supp:moments}, so the term is of order
$A^2/(\nu_{i+}U_i)$ up to a constant in $\alpha$; the third term is of
order $\Var(\varepsilon_{ij})\Var(e_i)\,\nu_{ij}U_i = A^2/(\nu_{i+}U_i)$ as
well. Both are of relative order $A/(\nu_{i+}U_i)$ against the leading
$A(1 - s_{ij})$, and the regressor $\nu_{ij}U_i(1 + e_i)$ differs from
$\nu_{ij}U_i$ by a factor whose variance is of the same relative order.
Neglecting them, the regression is that of $A(1 - s_{ij})$, which does not
depend on $U_i$, on $\nu_{ij}U_i$, which does, and the population slope
follows. At the calibration below the neglected relative order is
$A/(\nu_{i+}U_i) = (1 + 1/\alpha)/N_{i+}$, about $0.03$ to $0.05$ for rows of
$30$ to $47$ expected claims, which is why the simulated slopes stop
short of $B^\ast$ at every $I$: replication adds rows at fixed row volume
and leaves these terms in place.
Within a row $s_{ij}$ is proportional to $\nu_{ij}$, so the within-row
part of the covariance is positive; the between-row part depends on how
the row means of $s$ and of $\nu U$ co-vary with the number of observed
cells, and is not signed in general. For the design used below the total
is computed directly and is positive at every $\kappa_r$
(Table~\ref{tab:two_stage_target}).
\end{proof}

At the calibration of Section~\ref{sec:illus:boundary},
$B^\ast$ runs from $-0.026$ at $\kappa_r = 3$ to $-0.045$ at
$\kappa_r = 50$ against true values of $0.33$ to $0.02$; the unconstrained
intercept $A^\ast$ would lie within $7\%$ of $A$, but the constrained fit
returns the deflated $0.82A$ at every $\kappa_r$
(Table~\ref{tab:two_stage_target}). The unconstrained slope of the
implemented regression, on stacked copies of the $10 \times 10$ design,
has mean $-0.008$ at $I = 10$, $-0.017$ and $-0.027$ at $I = 50$ and
$-0.022$ and $-0.035$ at $I = 200$ at $\kappa_r = 3$ and $50$,
approaching but not reaching $B^\ast$, and the constrained estimate sits at the cap on $68\%$ of
triangles at $I = 10$, $90$ to $95\%$ at $I = 50$ and every triangle at
$I = 200$; its intercept averages $0.57A$ at $I = 10$, $0.74A$ at
$I = 50$ and $0.79A$ at $I = 200$, approaching $0.82A$ from below as the
leverage of the estimated pattern shrinks ($J - 1$ further parameters
fitted on $I$ rows); the gap that remains at $I = 200$ is of the order
of the neglected relative noise $A/(\nu_{i+}U_i)$, $3$ to $5\%$ here,
which replication at fixed row volume does not remove. The two thirds of
Section~\ref{sec:simulation} are therefore the finite-sample scatter of
an estimator whose limit is the boundary, and the second-moment fit
under free levels understates $A$ as well as capping $\kappa_B$; the
working fit of Section~\ref{sec:predictive} removes the deflation by the
leverage factor, and its adjusted interval covers $A$ on $90\%$ of
simulated triangles (Section~\ref{supp:estimation:bayes}). This recovery
is established numerically; no coverage result is proved.

\begin{table}[!htbp]
\centering
\caption{The target of the two-stage estimator under free levels at the
calibration of Section~\ref{sec:illus:boundary} ($\alpha = 3$): the true
$B = 1/\kappa_r$, the limit $B^\ast$ of the unconstrained slope
(Proposition~\ref{prop:two_stage_target}(iii)), the unconstrained
intercept $A^\ast$ relative to $A$, and the target of the implemented
constrained estimator $\hat A$ relative to $A$, which is the mean of
$1 - s_{ij}$ over the cells.}
\label{tab:two_stage_target}
\small
\begin{tabular}{@{}rcccc@{}}
\toprule
$\kappa_r$ & $B = 1/\kappa_r$ & $B^\ast$ & $A^\ast/A$ (unconstrained) & $\hat A$ at the cap, $/A$ \\
\midrule
3 & 0.333 & $-0.0263$ & 0.929 & 0.818 \\
5 & 0.200 & $-0.0321$ & 0.953 & 0.818 \\
10 & 0.100 & $-0.0383$ & 0.980 & 0.818 \\
20 & 0.050 & $-0.0425$ & 0.997 & 0.818 \\
50 & 0.020 & $-0.0454$ & 1.009 & 0.818 \\
\bottomrule
\end{tabular}

\end{table}

Lemma~\ref{lem:absorption} and Proposition~\ref{prop:two_stage_target}
are statements about the two-stage procedure. The stronger statement
concerns every estimator that maximises the free-level likelihood.

\subsection{The boundary}\label{sec:free:boundary}

The theorem below profiles the free level of each row. The step that
needs care is that the profile over the level is attained at a finite
value, so that the strict inequality between a mixture and the supremum
of what it mixes is inherited by the profile; the lemma isolates that
step and states the conditions on the row likelihood it uses.

\begin{lemma}[Attainment of the row profile]\label{lem:attain}
Fix $\theta$ and the observed cells $\mathbf{x}_i$ of row $i$, and write
$f(\lambda) = f(\mathbf{x}_i \mid \lambda, \theta)$ for the likelihood of
the row given its realised intensity. Assume
\begin{enumerate}[label=(A\arabic*), nosep]
\item $f$ is continuous on $(0, \infty)$ and not identically zero;
\item $f(\lambda) \to 0$ as $\lambda \downarrow 0$;
\item $f(\lambda) \to 0$ as $\lambda \uparrow \infty$.
\end{enumerate}
Then $f_i^\ast = \sup_\lambda f(\lambda)$ is finite and attained at an
interior point, and for every finite $\kappa_r > 0$ the mixture
$L_i(\mu_i, \kappa_r, \theta) = \int_0^\infty f(\lambda)\,
g(\lambda; \kappa_r, \kappa_r/\mu_i)\,d\lambda$ satisfies
\begin{enumerate}[label=(\roman*), nosep]
\item $L_i(\mu_i, \kappa_r, \theta) < f_i^\ast$ for every $\mu_i > 0$;
\item $\mu_i \mapsto L_i(\mu_i, \kappa_r, \theta)$ is continuous on
      $(0, \infty)$, positive, and tends to zero as $\mu_i \to 0$ and as
      $\mu_i \to \infty$;
\item consequently $\sup_{\mu_i} L_i(\mu_i, \kappa_r, \theta)$ is
      attained at a finite $\mu_i^\ast$ and is strictly below $f_i^\ast$.
\end{enumerate}
Under Definition~\ref{def:cnbg}, (A1)--(A3) hold for every row of the
amount triangle and of the count triangle that has at least one positive
observed cell, with or without the cell shock.
\end{lemma}

\begin{proof}
By (A2) and (A3) there is a compact $[a_0, b_0] \subset (0, \infty)$
outside which $f < \tfrac12\sup f$, and $f$ is continuous and positive
somewhere, so its supremum is finite and attained in $[a_0, b_0]$.
(i) The Gamma density has full support on $(0, \infty)$ and
$f \le f_i^\ast$ everywhere with $f < \tfrac12 f_i^\ast$ on the set
$(0, a_0) \cup (b_0, \infty)$ of positive Gamma probability, so
$\int f g < f_i^\ast \int g = f_i^\ast$. (ii) Positivity: $f \ge
\tfrac12 f_i^\ast$ on a neighbourhood of its maximiser, which has
positive Gamma probability. Continuity: if $\mu_n \to \mu$, the densities
$g_n = g(\cdot; \kappa_r, \kappa_r/\mu_n)$ converge pointwise to
$g = g(\cdot; \kappa_r, \kappa_r/\mu)$ and all integrate to one, so
$\int |g_n - g| \to 0$ by Scheff\'e's lemma and
$|L_i(\mu_n) - L_i(\mu)| \le f_i^\ast \int |g_n - g| \to 0$. Tails: given
$\varepsilon > 0$ choose $0 < a < b < \infty$ with $f < \varepsilon$
outside $[a, b]$, by (A2) and (A3), and split the integral,
\[
L_i(\mu_i, \kappa_r, \theta)
 \le \varepsilon \int_{(0,a) \cup (b,\infty)} g
   + f_i^\ast\, P_{\mu_i}\bigl([a, b]\bigr)
 \le \varepsilon + f_i^\ast\, P_{\mu_i}\bigl([a, b]\bigr),
\]
where $P_{\mu_i}$ is the $\mathrm{Gamma}(\kappa_r, \kappa_r/\mu_i)$ law.
With $\Gamma \sim \mathrm{Gamma}(\kappa_r, 1)$, $P_{\mu_i}([a, b]) \le
P(\Gamma \le b\kappa_r/\mu_i) \to 0$ as $\mu_i \to \infty$ and
$P_{\mu_i}([a, b]) \le P(\Gamma \ge a\kappa_r/\mu_i) \to 0$ as
$\mu_i \to 0$: the fixed compact interval loses all its mass when the
mean of the mixing law diverges or vanishes. Hence
$\limsup L_i \le \varepsilon$ at both ends, and $\varepsilon$ was
arbitrary. (iii) A continuous positive function on $(0, \infty)$ that
tends to zero at both ends attains its maximum at a finite point, and
(i) applies there.

For the last statement, consider first a single positive cell of the
amount triangle given its rate $r = \lambda\varepsilon_{ij}w_j$. Its
density at $x > 0$ is $h(x \mid r) = \sum_{n \ge 1} e^{-r}r^n/n!\,
\gamma_{n\alpha}(x)$ with $\gamma_{n\alpha}$ the
$\mathrm{Gamma}(n\alpha, \gamma)$ density. For fixed $x$ the sequence
$\gamma_{n\alpha}(x)$ is bounded, say by $c_x$, and tends to zero as
$n \to \infty$. Each term is continuous in $r$ and dominated locally
uniformly by $c_x r^n/n!$, so $h$ is continuous (A1);
$h(x \mid r) \le c_x(1 - e^{-r}) \to 0$ as $r \to 0$ (A2); and for any
$n_0$, $h(x \mid r) \le c_x P(N \le n_0 \mid r) + \sup_{n > n_0}
\gamma_{n\alpha}(x)$, whose second term is made small by $n_0$ and whose
first then vanishes as $r \to \infty$ (A3). A zero cell contributes
$e^{-r}$, continuous, bounded by one, and tending to zero as
$r \to \infty$. With the cell shock the cell factor is the average of
these over $\varepsilon_{ij} \sim \mathrm{Gamma}(\kappa_c, \kappa_c)$,
which preserves (A1)--(A3) by dominated convergence (the integrands are
bounded by $c_x$ or by one). The row likelihood is the product of its
cell factors, each bounded; it tends to zero at both ends because the
factor of a positive cell does, and it is continuous. For the count
triangle replace $h$ by the Poisson probability $e^{-r}r^n/n!$ of the
observed count, or by its Negative Binomial mixture over the cell shock,
which satisfy (A1)--(A3) for $n \ge 1$ by the same bounds.
\end{proof}

\begin{theorem}[Free-level profile likelihood of $\kappa_r$ \lvl{F}]
\label{thm:boundary}
Let the accident-year levels $\mu_1, \ldots, \mu_I$ be free parameters,
let $\theta$ collect the remaining parameters (development weights,
severity and cell dispersion), and let $f(\mathbf{x}_i \mid \lambda,
\theta)$ be the likelihood of the observed cells of row $i$ given the
realised intensity $\lambda_i = \lambda$. Under
Definition~\ref{def:cnbg} the marginal likelihood of row $i$ is
\[
  L_i(\mu_i, \kappa_r, \theta)
  = \int_0^\infty f(\mathbf{x}_i \mid \lambda, \theta)\,
    g(\lambda; \kappa_r, \kappa_r/\mu_i)\,d\lambda,
\]
with $g(\cdot; \kappa_r, \kappa_r/\mu_i)$ the $\mathrm{Gamma}(\kappa_r,
\kappa_r/\mu_i)$ density. Call row $i$ \emph{informative} if at least
one of its observed cells is positive. Then, for every triangle and
every $\theta$:
\begin{enumerate}[label=(\roman*), nosep]
\item for every informative row, $\sup_{\mu_i} L_i(\mu_i, \kappa_r,
      \theta) < \sup_\lambda f(\mathbf{x}_i \mid \lambda, \theta)$ for
      every finite $\kappa_r$, and the left-hand side converges to the
      right-hand side as $\kappa_r \to \infty$; for an all-zero row the
      two sides are equal for every $\kappa_r$, so such a row carries no
      information on $\kappa_r$;
\item consequently, whenever at least one row is informative, the
      profile likelihood $\prod_i \sup_{\mu_i} L_i$
      attains its supremum over $\kappa_r \in (0,\infty]$ only at
      $\kappa_r = \infty$: the free-level maximum-likelihood estimator of
      $\kappa_r$ exists at no finite value.
\end{enumerate}
The same holds for the count triangle $\{N_{ij}\}$. In the single-frailty
case ($\kappa_c = \infty$) $f$ is there the Poisson--multinomial row
likelihood and the row total $N_{i\cdot}$ is the only statistic that
carries $\lambda_i$; with a finite cell shock the counts given $\lambda_i$
are independent Negative Binomials, the within-row allocation carries
$\lambda_i$ as well, and the argument applies to that $f$ unchanged.
\end{theorem}

\begin{proof}
(i) An informative row satisfies (A1)--(A3) of Lemma~\ref{lem:attain},
whose part (iii) gives, for every finite $\kappa_r$,
\[
\sup_{\mu_i} L_i(\mu_i, \kappa_r, \theta) = L_i(\mu_i^\ast, \kappa_r, \theta)
< f_i^\ast := \sup_\lambda f(\mathbf{x}_i \mid \lambda, \theta),
\]
with $f_i^\ast$ attained at an interior $\lambda^\ast$.
For the limit, take $\mu_i = \lambda^\ast$; as $\kappa_r \to \infty$
the density $g(\cdot; \kappa_r, \kappa_r/\lambda^\ast)$ concentrates at
$\lambda^\ast$, and by continuity of $f$ in $\lambda$ and dominated
convergence $L_i(\lambda^\ast, \kappa_r, \theta) \to f_i^\ast$. For an
all-zero row, $f(\mathbf{x}_i \mid \lambda, \theta)$ is the
probability that every observed count of the row is zero given
$\lambda$: $\exp(-\lambda \sum_{j:(i,j)\in\mathcal{D}} w_j)$ in the
single-frailty case and $\prod_{j:(i,j)\in\mathcal{D}}
\bigl(\kappa_c/(\kappa_c + \lambda w_j)\bigr)^{\kappa_c}$ with a finite
cell shock. In both cases $f$ is continuous and strictly decreasing in
$\lambda$ with supremum one at the boundary $\lambda = 0$, so
$f_i^\ast = 1$; and since $g(\cdot; \kappa_r, \kappa_r/\mu_i)$
concentrates at $0$ as $\mu_i \to 0$, dominated convergence gives
$\sup_{\mu_i} L_i = \lim_{\mu_i \to 0} L_i = 1 = f_i^\ast$ for every
finite $\kappa_r$.
(ii) The product over independent rows of quantities each bounded by
$f_i^\ast$, with equality approached only as $\kappa_r \to \infty$ for
every informative row and attained for every $\kappa_r$ for the
all-zero rows, attains its supremum only in that limit as soon as one
row is informative.
\end{proof}

The argument uses nothing about the compound law beyond the fact that a
row's likelihood, as a function of the row intensity, vanishes at both
ends. A mixture over the intensity is an average of $f$, and an average
lies below the supremum it averages; a free level can always place the
mixture where $f$ peaks, and the only way to reach the peak is to let the
mixing distribution collapse. Theorem~\ref{thm:boundary} describes a
likelihood that varies with $\kappa_r$ but always prefers homogeneity
to a finite $\kappa_r$. A Bayesian posterior integrates the levels
under their priors. Its finiteness depends on those priors and the prior
on $\kappa_r$, while its departure from the prior depends on their
scales. Sections~\ref{sec:simulation} and~\ref{supp:evidence:sim}
measure that departure numerically.

These operations on the levels lead to three different analyses:
profiling in Theorem~\ref{thm:boundary}, integration under a prior in
Section~\ref{sec:cost}, and conditioning on fitted values in the
Schedule~P exact fit of Section~\ref{sec:illus:sp}. The latter two also
push $\kappa_r$ towards homogeneity because the fitted or integrated
levels carry realised frailty. Their behaviour is examined separately
from the profile-likelihood theorem.

The fixed-$\theta$ result in Theorem~\ref{thm:boundary} extends to joint
profiling as follows.

\begin{corollary}[Jointly profiled nuisance parameters \lvl{F}]
\label{cor:joint_profile}
Let $\theta$ range over a set $\Theta$ on which, for every finite
$\kappa_r$, the free-level likelihood $\theta \mapsto \prod_i
\sup_{\mu_i} L_i(\mu_i, \kappa_r, \theta)$ attains its supremum, for
instance $\Theta$ compact and the likelihood continuous in $\theta$, and
let at least one row be informative. Then the jointly profiled
likelihood $\kappa_r \mapsto \sup_{\theta \in \Theta} \prod_i
\sup_{\mu_i} L_i(\mu_i, \kappa_r, \theta)$, extended to $\kappa_r =
\infty$ by its limit $\sup_{\theta \in \Theta}\prod_i f_i^\ast(\theta)$,
attains its supremum over $(0, \infty]$ only at $\kappa_r = \infty$.
\end{corollary}

\begin{proof}
At a finite $\kappa_r$ the profile equals the free-level likelihood at
some $\theta_{\kappa_r} \in \Theta$, which by Theorem~\ref{thm:boundary}(i)
is strictly below $\prod_i f_i^\ast(\theta_{\kappa_r}) \le
\sup_{\theta \in \Theta}\prod_i f_i^\ast(\theta)$, the value at
$\kappa_r = \infty$; so no finite $\kappa_r$ attains the supremum, and
the value at $\infty$ does.
\end{proof}

Table~\ref{tab:boundary_joint} in the Supplement shows the same boundary
with $\theta$ estimated. The proof rests on an elementary property of
mixtures; its reserving consequence is Corollary~\ref{cor:blind}.
It covers every estimator that maximises the free-level likelihood.
The two-stage estimator requires the separate argument of
Proposition~\ref{prop:two_stage_target}; other estimators under free
levels remain outside the theorem's scope. Population identification
in Proposition~\ref{prop:identification} therefore coexists with failure
of the fits used in reserving.

\subsection{What the free-level fit estimates, and what anchoring adds}
\label{sec:free:anchoring}

\begin{proposition}[What the amount triangle estimates under free levels \lvl{F, S}]
\label{prop:two_kappa_id}
Under the free-level fit:
\begin{enumerate}[label=(\roman*), nosep]
\item \emph{$\kappa_r$ is not recovered.} The free-level likelihood is maximised at $\kappa_r = \infty$ (Theorem~\ref{thm:boundary}) and the second-moment regression's target for the quadratic coefficient is at or below zero (Proposition~\ref{prop:two_stage_target}). The count likelihood offers no wider
channel: in the single-frailty case a row frailty cancels from the
multinomial split of the counts and survives only in the row totals,
which the free levels absorb.
\item \emph{$\kappa_c$ survives, weakly.} The cell shock is not absorbed.
It remains in the within-row residual curvature,
\[
  \E\bigl[(X_{ij}-\nu_{ij}U_i)^2 \mid U_i\bigr]
  = A\,\nu_{ij}U_i + \frac{1}{\kappa_c}\,(\nu_{ij}U_i)^2,
\]
so the quantity $\hat\kappa_B$ returned by \eqref{eq:wls} under free levels
reads $\kappa_c$, not the composite $\kappa_B$, and it reads it only
where the cell shock is strong enough to outweigh the first-stage term
of Proposition~\ref{prop:two_stage_target} (Remark~\ref{rem:cellshock_target}).
At $I = J = 10$ and a data-rich calibration of $1{,}100$ to $1{,}700$
claims per accident year, where that term is small, the signal orders
$\kappa_c$ without pinning its magnitude: the posterior median of a
Bayesian fit tracks the prior mean with a log--log slope of $0.88$,
$0.37$ and $0.52$ at true $\kappa_c = 3$, $10$ and $50$
(Section~\ref{supp:evidence:priorsens}).
\end{enumerate}
\end{proposition}

\begin{remark}[The free-level second stage with a finite cell shock]
\label{rem:cellshock_target}
Propositions~\ref{prop:two_stage_target} and~\ref{prop:two_kappa_id}(ii)
combine. With $\varepsilon_{ij} \sim \mathrm{Gamma}(\kappa_c, \kappa_c)$
the cell variance given $U_i$ is $A\nu_{ij}U_i + (\nu_{ij}U_i)^2/\kappa_c$,
and the derivation of Proposition~\ref{prop:two_stage_target} goes
through with one extra term,
\[
\E\bigl[(X_{ij} - \hat\nu_{ij})^2 \,\big|\, U_i\bigr]
= A\,\nu_{ij}U_i\,(1 - s_{ij}) + \frac{(\nu_{ij}U_i)^2}{\kappa_c}\,q_{ij},
\qquad q_{ij} = 1 - 2s_{ij} + \sum_{l \le k_i}s_{il}^2 ,
\]
so that the population slope of the second stage is
\[
B^\ast + \frac{c_q}{\kappa_c}, \qquad
c_q = \frac{\Cov(\nu_{ij}U_i q_{ij}, \nu_{ij}U_i)}{\Var(\nu_{ij}U_i)},
\]
with $c_q$ between $0.39$ ($\kappa_r = 50$) and $0.46$ ($\kappa_r = 3$)
at the calibration of Section~\ref{sec:illus:boundary}. The second stage therefore reads a
cell dispersion only where $c_q/\kappa_c$ exceeds $|B^\ast|$, at this
calibration for $\kappa_c$ below about $10$ to $20$, and reads it with
an upward bias: the targets of $\hat\kappa_B$ are $8$ to $12$ at
$\kappa_c = 3$ and $15$ to $30$ at $\kappa_c = 5$, and the cap at
$\kappa_c = 20$ and $50$ for every $\kappa_r$
(Table~\ref{tab:two_stage_cellshock}). On stacked copies of the design
at $I = 50$ and $\kappa_r = 5$ the mean unconstrained slope is $0.052$,
$0.001$ and $-0.011$ at $\kappa_c = 5$, $20$ and $50$, against targets of
$0.056$, $-0.010$ and $-0.023$, with $6$, $49$ and $76\%$ of the fits at
the cap. The cell dispersion survives free levels in the population
sense of Proposition~\ref{prop:two_kappa_id}, but at practical claim
volumes the free-level second stage returns it only when the shock is
strong, and returns no dispersion at all when it is moderate; the
first-stage term shrinks with the claim volume, which is why the
data-rich calibration of Section~\ref{supp:evidence:priorsens} still
orders $\kappa_c$.
\end{remark}

\begin{table}[!htbp]
\centering
\caption{The free-level second stage with a finite cell shock at the
calibration of Section~\ref{sec:illus:boundary}: the single-frailty
first-stage term $B^\ast$ of Proposition~\ref{prop:two_stage_target} and
the target of $\hat\kappa_B = 1/(B^\ast + c_q/\kappa_c)$ where it is
positive, ``cap'' where the target of the constrained estimator is
$B = 0$.}
\label{tab:two_stage_cellshock}
\small
\begin{tabular}{@{}rccccccc@{}}
\toprule
$\kappa_r$ & $B^\ast$ & \multicolumn{5}{c}{target of $\hat\kappa_B$ at true $\kappa_c =$} \\
\cmidrule(lr){3-7}
 & & 3 & 5 & 10 & 20 & 50 \\
\midrule
3 & $-0.0263$ & 7.9 & 15.2 & 50.7 & cap & cap \\
5 & $-0.0321$ & 8.7 & 17.9 & 83.9 & cap & cap \\
10 & $-0.0383$ & 9.9 & 22.2 & 294.7 & cap & cap \\
20 & $-0.0425$ & 10.9 & 26.3 & cap & cap & cap \\
50 & $-0.0454$ & 11.7 & 30.4 & cap & cap & cap \\
\bottomrule
\end{tabular}

\end{table}

The boundary is a property of the parameterisation, not of the
information in the cells, and it is lifted by anchoring the levels: an
exposure offset $\mu_i = \pi_i\rho$ with a known exposure $\pi_i$ and one
common scale $\rho$, a common development rate, or a hierarchical prior
on the levels of credibility type \citep{BuhlmannGisler2005}. With $I$
free levels replaced by one, the row totals are no longer reproduced by
construction, and their dispersion around the anchored means carries
$\kappa_r$. The channel that opens is a channel of $I$ observations, one
per accident year: anchoring makes $\kappa_r$ estimable, not precise.
Count row totals carry the same channel without the severity noise of the
amount totals, and nothing more. Section~\ref{sec:illus:boundary}
measures what this channel delivers at $I = 10$.

\begin{corollary}[The free-level fit loses the reserve-dominant parameter \lvl{F, S}]
\label{cor:blind}
The free-level maximum-likelihood and two-stage fits do not recover
$\kappa_r$, which the population law identifies and which governs the
non-diversifying term $\kappa_r^{-1}\sum_i S_i^2$ of the unconditional
reserve variance \eqref{eq:reserve_var}. For a conditional predictive the
consequence is narrower (Proposition~\ref{prop:conditional}): $\kappa_r$
enters only through the posterior of each realised row intensity, so a
predictive built on the free-level fit must either price that posterior
at its diffuse value on its informative rows, which needs no $\kappa_r$ and over-states the level term of every row whose observed count exceeds the threshold $\mu_i^2F_i^2/(\kappa_r + 2\mu_iF_i)$ of Remark~\ref{rem:diffuse}, itself below half the prior expectation, or supply $\kappa_r$ from outside the triangle. The difference between the two is largest on the thin latest
accident years, which carry most of the reserve.
\end{corollary}

\section{A Dispersion that Survives Free Levels}\label{sec:dirichlet}

The boundary concerns a dispersion of the row levels. A dispersion of
the allocation within a row behaves differently, and a second paid-amount
model shows how. In the Dirichlet allocation model \citep{srirams2021},
the ultimate $S_i$ of accident year $i$ is a free level and its
allocation over development years is
\begin{equation}\label{eq:dirichlet_alloc}
\bigl(X_{i1}, \ldots, X_{iJ}\bigr)/S_i \sim
\mathrm{Dirichlet}\bigl(c\,w_1, \ldots, c\,w_J\bigr),
\end{equation}
independently of $S_i$, with $w$ the development pattern and $c > 0$ a
concentration: the larger $c$, the closer each row's development shares
are to $w$. Given the level, the cell variance
$S_i^2 w_j(1-w_j)/(c+1)$ is quadratic in it, as the heterogeneity term of
Proposition~\ref{prop:moments} is, and $c$ is a within-row dispersion,
the counterpart of $\kappa_c$. The model is the amount-side hierarchy of
the companion paper on the conditional predictive
\citep{VanOirbeek2026cond}, in which the level carries a Gamma prior;
here the level is free.

\begin{proposition}[The allocation concentration under free levels \lvl{P, F}]
\label{prop:dirichlet_c}
Under \eqref{eq:dirichlet_alloc}, with the levels $S_i$ free and
$k_i = \min(I-i+1, J)$ the number of observed cells of accident year $i$:
\begin{enumerate}[label=(\roman*), nosep]
\item The observed sub-composition $v_{ij} = X_{ij}/x_i$,
$j \le k_i$, where $x_i = \sum_{j \le k_i} X_{ij}$, is independent of
$S_i$, with law $\mathrm{Dirichlet}(c w_1, \ldots, c w_{k_i})$.
The log-likelihood of the observed sub-compositions,
\begin{equation}\label{eq:dirichlet_condlik}
\ell(c, w) = \sum_{i:\,k_i \ge 2}\Bigl[\log\Gamma(cF_{k_i})
 - \sum_{j \le k_i}\log\Gamma(c w_j)
 + \sum_{j \le k_i}(c w_j - 1)\log v_{ij}\Bigr],
\qquad F_k = \sum_{l \le k} w_l,
\end{equation}
is the likelihood of the triangle conditional on the observed row totals
$x_i$ and does not involve the levels. The concentration is identified,
and for fixed $J$ the maximiser of \eqref{eq:dirichlet_condlik} is
consistent as $I \to \infty$.
\item The partial-column proportions
$W_{ij}^{(k)} = X_{ij}/\sum_{l \le k} X_{il}$ of the accident years
observed beyond horizon $k$ are
$\mathrm{Beta}\bigl(cF_k\,w_j^{(k)},\, cF_k(1-w_j^{(k)})\bigr)$ with
$w_j^{(k)} = w_j/F_k$: their concentration is $cF_k$, not $c$. The moment
estimator built from their cross-row dispersion is consistent for $c$
only after division by $\hat F_k$; without it, it converges to $cF_{k^*}$
at the median contributing horizon $k^*$.
\item A row frailty that multiplies the level, $S_i \mapsto U_iS_i$,
leaves the laws in (i) and (ii) unchanged.
\end{enumerate}
\end{proposition}

The proof and the moment estimator in full are in
Section~\ref{supp:dirichlet}. Nothing is sent to a boundary here. The
incidental levels are removed by conditioning on the row totals, the
classical remedy for the incidental-parameter problem
\citep{neymanscott1948}, and what remains of the likelihood carries $c$ in
every row with two or more observed cells. The statement is about the
conditional likelihood and not about the profile over the free ultimates:
for a row observed to horizon $k_i$ the density of its observed cells
given $S_i$ carries the factor $(1 - x_i/S_i)^{c(1 - F_{k_i}) - 1}$ of the
unobserved remainder, which is unbounded as $S_i \downarrow x_i$ whenever
$c(1 - F_{k_i}) < 1$, so the likelihood profiled over free levels has no
finite supremum for such $c$ and free-level maximum likelihood in this
model has to be read as conditional maximum likelihood. Theorem~\ref{thm:boundary}
is a profile statement and Proposition~\ref{prop:dirichlet_c} a
conditional one; Table~\ref{tab:idmap} records the distinction. Part~(iii) is visible in
simulation: under a Gamma row frailty with $\Var(U_i)$ from $0$ to $1$ the
mean moment estimate stays between $65.6$ and $69.7$ at a true $c = 50$
($I = J = 10$, $500$ triangles per frailty level), whereas the free-level
likelihood of Theorem~\ref{thm:boundary} sends the row dispersion to the
boundary on every triangle. The two models make the same point from two
sides: a row frailty and a within-row dispersion live in different
channels of the triangle, and the free levels close only the first.

\section{Precision at Practical Triangle Sizes}\label{sec:precision}

We now turn from identification and recovery to precision. Three
channels survive in some form, each weak at $I = J = 10$.

\subsection{The curvature channel}\label{sec:precision:curvature}

A second-moment fit reaches a dispersion only through the quadratic term
of the variance function. Among the estimating equations linear in the squared residuals that treat the cells as independent, the one with the largest information is the inverse-variance-weighted one,
\[
\sum_{ij}\frac{\partial\mu_z^{ij}}{\partial\kappa_B}\,
\frac{z_{ij}-\mu_z^{ij}}{\Var(z_{ij})} = 0
\]
\citep{Godambe1960}; it is the score of the working likelihood with cell-level shapes of
Section~\ref{sec:predictive}, and under working independence its
information bounds the precision of the WLS \eqref{eq:wls}, which weights
by $\hat\nu_{ij}^{-2}$ instead. The following result gives that
information for the diagonally weighted equation with the cell means
and $A$ known. Section~\ref{supp:precision} specifies the estimators
covered by this benchmark and explains why it cannot serve as a general
information boundary for paid triangles.

\begin{proposition}[Information of the working-independence curvature channel \lvl{N}]
\label{prop:godambe}
Let $\mu_z^{ij} = A\nu_{ij} + \nu_{ij}^2/\kappa_B$, let
$\mathrm{CV}^2(z_{ij}) = \E[(X_{ij}-\nu_{ij})^4]/(\mu_z^{ij})^2 - 1$ be the
squared coefficient of variation of the squared residual
$z_{ij} = (X_{ij}-\nu_{ij})^2$ (Proposition~\ref{prop:higher_moments}),
and let the cell means be known.
\begin{enumerate}[label=(\roman*), nosep]
\item With $A$ held at its true value, the working-independence
information for $\kappa_B$ carried by the optimally weighted equation is
\begin{equation}\label{eq:fisher_kappa}
  \mathcal{I}_{\mathrm{diag}}(\kappa_B)
  \;=\;
  \sum_{(i,j)\in\mathcal{D}}
  \frac{(\nu_{ij}^2/\kappa_B^2)^2}
       {\mathrm{CV}^2(z_{ij}) \cdot (\mu_z^{ij})^2},
\end{equation}
In the heterogeneity-dominated limit $\kappa_B \ll \nu_{ij}/A$ the
per-cell term is, in the single-frailty case, $1/(2\kappa_B^2 + 6\kappa_B)$,
since there $\mu_z^{ij} \approx \nu_{ij}^2/\kappa_B$ and
$\mathrm{CV}^2(z_{ij}) \to 2 + 6/\kappa_B$; in the opposite limit it is
$O(\kappa_B^{-4})$.
\item The shared frailty makes the scores of cells in one row positively
correlated. With $\Delta(\kappa_B)$ the sum of the within-row score
cross-covariances, the precision of the working-independence estimator
is the Godambe information
\begin{equation}\label{eq:godambe}
  G(\kappa_B) \;=\; \frac{\mathcal{I}_{\mathrm{diag}}(\kappa_B)^2}
  {\mathcal{I}_{\mathrm{diag}}(\kappa_B) + \Delta(\kappa_B)}
  \;\le\; \mathcal{I}_{\mathrm{diag}}(\kappa_B),
\end{equation}
and its asymptotic variance is $1/G(\kappa_B)$.
\end{enumerate}
\end{proposition}

\noindent The proof is in Section~\ref{supp:precision}. Under free levels
the same channel carries $\kappa_c$ (Proposition~\ref{prop:two_kappa_id}(ii)).
The per-cell information \eqref{eq:fisher_kappa} is then the same function
of the cell moments with $\kappa_c$ in place of $\kappa_B$, but the
within-row correction is not: for a cell shock with the means known the
cell scores are uncorrelated, $\Delta = 0$ and $G = \mathcal{I}_{\mathrm{diag}}$,
a coefficient of variation of $0.6$ in place of $0.8$ at $\kappa = 5$,
while the free-level fit adds the first-stage dependence of
Proposition~\ref{prop:two_stage_target} and the threshold of
Remark~\ref{rem:cellshock_target}, neither of which the table carries.
Table~\ref{tab:godambe} is therefore the row-frailty benchmark at known
means; the precision of $\hat\kappa_c$ after free-level fitting is not
derived here. The table evaluates $G$ at the calibration of the simulation of
Section~\ref{sec:illus:boundary} and converts it into the asymptotic
coefficient of variation of the estimate, $1/(\kappa_B\sqrt{G})$.

\begin{table}[!htbp]
\centering
\caption{Godambe information of the optimally weighted curvature channel
on a $10 \times 10$ triangle (Proposition~\ref{prop:godambe}; $\bar Y = 5{,}000$, $30$--$47$
expected claims per accident year, the development pattern of
Section~\ref{sec:illus:boundary}, information averaged over
$\alpha \in \{1, 3, 10\}$, cell means known) and the implied asymptotic
coefficient of variation of the estimate of $\kappa_B$.}
\label{tab:godambe}
\small
\begin{tabular}{@{}rrr@{}}
\toprule
$\kappa_B$ & $G(\kappa_B)$ & asymptotic $\mathrm{SD}(\hat\kappa_B)/\kappa_B$ \\
\midrule
3 & 0.164 & 0.82 \\
5 & 0.061 & 0.81 \\
10 & 0.012 & 0.91 \\
20 & 0.0017 & 1.23 \\
50 & $7.5 \times 10^{-5}$ & 2.30 \\
\bottomrule
\end{tabular}

\end{table}

The asymptotic coefficient of variation is $0.8$ at the smallest
dispersions, where the heterogeneity signal is strongest, and grows to
$2.3$ at $\kappa_B = 50$, where the information decays as
$\kappa_B^{-4}$. Information is additive over independent rows, so these
describe a $10 \times 10$ triangle. Independent rows add information,
and the estimator is consistent (Proposition~\ref{prop:sandwich}). At large $\kappa_B$ the weakly determined
quantity is also the inconsequential one, since the quadratic term
vanishes and the reserve variance is dominated by the severity term.
Section~\ref{supp:precision} sets $G$ against two summaries of the
precision of the reference prior used by the Bayesian version of the fit;
both are of the same order as $G$ at $\kappa_B \le 10$, and the
empirical prior sensitivity of Proposition~\ref{prop:two_kappa_id}(ii),
not that comparison, carries the operational statement. Allowing the optimal row-block weighting improves the coefficient of
variation by at most $11\%$ at this calibration, and the WLS weighting
$\hat\nu_{ij}^{-2}$ of \eqref{eq:wls} loses little to the optimal diagonal
one (Tables~\ref{tab:godambe_block} and~\ref{tab:godambe_wls},
Section~\ref{supp:precision}). Changing the weighting therefore offers
little relief from the weak curvature signal at this calibration.

\subsection{The row-total channel under anchoring}\label{sec:precision:row}

Once the levels are anchored, $\kappa_r$ is read from the dispersion of
$I$ row totals around their anchored means. Each row total is itself a
compound sum, so the frailty signal competes with severity and count
noise, and there are only $I$ of them. The channel is therefore weak at
$I = 10$ for a reason that a larger triangle in the development direction
does not repair: more development years add cells to each row but no
rows. Section~\ref{sec:illus:boundary} quantifies the resulting uncertainty.

\subsection{The allocation channel}\label{sec:precision:allocation}

The conditional likelihood \eqref{eq:dirichlet_condlik} carries $c$ in
every row with two or more observed cells, but on a ten-row triangle both
its maximiser and the moment estimator overstate $c$ by a quarter or more
(Section~\ref{sec:illus:dirichlet}), and at $I = 10$ the variance of the median aggregate of the moment estimator is $2.8$ times a per-cell asymptotic reference whose own limit for the aggregate is $0.73$ of itself, so that the aggregate's finite-sample variance is about four times its large-$I$ value (Section~\ref{supp:dirichlet}): like the Godambe information above, the asymptotics describe a regime that single triangles do not reach.

\subsection{The information map}\label{sec:precision:map}

Table~\ref{tab:idmap} collects the results.

\begin{table}[!htbp]
\centering
\caption{The information map of a paid triangle. For each parameter: the
feature of the triangle that carries it; whether it is identified in the
population \lvl{P}, recovered by free-level maximum likelihood \lvl{F},
the profile likelihood unless the entry says conditional,
and by the two-stage second-moment fit \lvl{S}; how precisely a
$10 \times 10$ triangle determines it \lvl{N}; and what supplies it when
the triangle does not. $\checkmark$: yes; $\times$: no; ``not
studied'': not examined in this paper; ``not applicable'': the level
has no meaning for that parameter.}
\label{tab:idmap}
\scriptsize
\renewcommand{\arraystretch}{1.15}
\setlength{\tabcolsep}{4pt}
\newcolumntype{L}[1]{>{\raggedright\arraybackslash}p{#1}}
\newcolumntype{C}[1]{>{\centering\arraybackslash}p{#1}}
\begin{tabular}{@{}L{2.0cm}L{2.8cm}C{0.45cm}L{1.6cm}L{1.8cm}L{2.6cm}L{2.5cm}@{}}
\toprule
Parameter & Carried by & \lvl{P} & \lvl{F} & \lvl{S} & \lvl{N} at $10 \times 10$
  & Supplied by, if not by the triangle \\
\midrule
Mean structure $\mu_i$, $w_j$; the fit returns the realised $\lambda_i = \mu_iU_i$, not $\mu_i$ & first moments of the cells (up to the scale normalisation $\sum_j w_j = 1$)
  & $\checkmark$ & $\checkmark$ for $\lambda_i$, $w_j$ (quasi-Poisson chain ladder; the exact amount MLE need not equal it; Lemma~\ref{lem:absorption}) & $\checkmark$ for $\lambda_i$, $w_j$ (chain ladder)
  & error of the order of the process variance; exposed to calendar-year change
  & not applicable \\
Severity composite $A$ & linear term of the variance
    & $\checkmark$ & not studied & $\checkmark$ up to the first-stage deflation $1 - \bar s$ of Prop.~\ref{prop:two_stage_target}, $0.82$ at the design studied & usable with measured under-coverage: the working posterior's adjusted $95\%$ interval covers $A$ on $90\%$ of simulated triangles (Section~\ref{sec:predictive}) & not applicable \\
Row dispersion $\kappa_r$ & within-row covariance; row totals under anchoring
  & $\checkmark$ & $\times$ (Thm.~\ref{thm:boundary})
  & $\times$ (Prop.~\ref{prop:two_stage_target})
  & low even when anchored (Table~\ref{tab:boundary})
  & anchored levels, or a prior \\
Cell dispersion $\kappa_c$ & curvature of the within-row residuals
  & $\checkmark$ & not established (only \lvl{S} studied) & $\checkmark$ ordinal, only below the threshold of Remark~\ref{rem:cellshock_target} ($\kappa_c \lesssim 10$--$20$ at the calibration); at the cap above it
    & low, prior-sensitive (Table~\ref{tab:godambe} is the row-frailty benchmark at known means; the free-level precision of $\kappa_c$ is not derived; Section~\ref{supp:evidence:priorsens})
  & count triangle \\
Severity split $\alpha$, $\bar Y$ & third cumulant
  & $\checkmark$ & not studied & $\times$ (second moments only)
  & not established & count triangle \\
Dirichlet concentration $c$ & within-row compositions given the row totals
        & $\checkmark$ & $\checkmark$ by the conditional likelihood (Prop.~\ref{prop:dirichlet_c}); the profile over free ultimates is unbounded where $c(1 - F_{k_i}) < 1$ & not applicable (no second-moment fit is defined for this model)
  & low: $+23$ to $31\%$ at $I = 10$ (Table~\ref{tab:dirichlet_c})
  & integration over $c$, in part \\
\bottomrule
\end{tabular}
\end{table}

\FloatBarrier
\section{Numerical and Empirical Illustrations}\label{sec:illus}

\subsection{The boundary and the anchored channel in simulation}\label{sec:illus:boundary}

Triangles are drawn from the single-frailty hierarchy at the calibration of the simulation of Section~\ref{sec:simulation}: $I = J = 10$,
expected claim counts $\mu_i = 30\,e^{0.05(i-1)}$ ($30$ to $47$ claims per
accident year), development weights
$(0.35, 0.25, 0.15, 0.10, 0.07, 0.04, 0.02, 0.01, 0.005, 0.005)$, mean
severity $\bar Y = 5{,}000$ and severity shape $\alpha = 3$, with
$\kappa_r \in \{3, 5, 10, 20, 50\}$ and $500$ triangles each. On every
triangle the exact likelihood of Definition~\ref{def:cnbg} (compound
Poisson--Gamma cells, the row intensity integrated numerically) is
maximised under three parameterisations of the levels, with the
development weights and the severity held at their true values:
\emph{free levels}, the profile likelihood evaluated on a grid of
$\kappa_r$ and at $\kappa_r = \infty$; levels \emph{anchored up to a common
scale}, $\mu_i = \rho\pi_i$ with $\pi_i$ the true relative levels and $\rho$
free, an exposure offset with an exact exposure; and \emph{levels known}.
Holding the other parameters at the truth favours the anchored fits and
leaves the free fit unaffected, since Theorem~\ref{thm:boundary} holds for
every $\theta$, so the comparison isolates the parameterisation of the
levels; a second run, reported below, estimates them jointly. The experiment is repeated with $I = 40$ rows spanning the same
range of levels, to separate what more rows buy from what more claims buy.
Table~\ref{tab:boundary} reports the results; the computation is in
Section~\ref{supp:boundary}.

\begin{table}[!htbp]
\centering
\caption{Maximum likelihood for the row dispersion under three
parameterisations of the accident-year levels: the share of triangles on
which the estimate is $\kappa_r = \infty$, and the median and $10\%$ and
$90\%$ quantiles of $\hat\kappa_r/\kappa_r$ (triangles at the boundary
counted at $\infty$). Exact likelihood of the single-frailty hierarchy,
development weights and severity at their true values, $\alpha = 3$,
$J = 10$, $500$ triangles per row of the table. The free-level profile
likelihood was non-decreasing in $\kappa_r$ on every triangle.}
\label{tab:boundary}
\footnotesize
\setlength{\tabcolsep}{4pt}
\begin{tabular}{@{}rrrrrrrrr@{}}
\toprule
& & \multicolumn{1}{c}{Free levels} & \multicolumn{3}{c}{Anchored up to a common scale} & \multicolumn{3}{c}{Levels known} \\
\cmidrule(lr){3-3}\cmidrule(lr){4-6}\cmidrule(lr){7-9}
$I$ & $\kappa_r$ & \% at $\infty$ & \% at $\infty$ & median $\hat\kappa_r/\kappa_r$ & [q10, q90] & \% at $\infty$ & median $\hat\kappa_r/\kappa_r$ & [q10, q90] \\
\midrule
$10$ & 3 & 100.0 & 0.4 & 1.21 & [0.62, 2.88] & 0.2 & 1.05 & [0.59, 2.18] \\
 & 5 & 100.0 & 1.0 & 1.23 & [0.60, 3.61] & 0.4 & 1.05 & [0.58, 2.85] \\
 & 10 & 100.0 & 3.8 & 1.33 & [0.57, 7.95] & 2.0 & 1.07 & [0.53, 3.88] \\
 & 20 & 100.0 & 13.0 & 1.64 & [0.55, $\infty$] & 8.2 & 1.25 & [0.48, 18.68] \\
 & 50 & 100.0 & 36.8 & 2.75 & [0.37, $\infty$] & 26.6 & 1.41 & [0.32, $\infty$] \\
\midrule
$40$ & 3 & 100.0 & 0.0 & 1.03 & [0.79, 1.44] & 0.0 & 1.00 & [0.77, 1.39] \\
 & 5 & 100.0 & 0.0 & 1.04 & [0.77, 1.53] & 0.0 & 1.01 & [0.75, 1.47] \\
 & 10 & 100.0 & 0.0 & 1.06 & [0.74, 1.76] & 0.0 & 1.02 & [0.73, 1.67] \\
 & 20 & 100.0 & 0.2 & 1.14 & [0.70, 2.29] & 0.0 & 1.08 & [0.68, 2.04] \\
 & 50 & 100.0 & 5.4 & 1.14 & [0.55, 5.78] & 4.2 & 1.05 & [0.52, 4.55] \\
\bottomrule
\end{tabular}

\end{table}

The free-level profile likelihood is maximised at $\kappa_r = \infty$ on
every one of the $5{,}000$ triangles, and it is non-decreasing over the
whole grid on every triangle: Theorem~\ref{thm:boundary} is visible
without exception, at $\kappa_r = 3$ as at $\kappa_r = 50$, with ten rows
as with forty. Replacing the free levels by one common scale opens the
channel. At $I = 10$ and $\kappa_r = 3$ the estimate is finite on $99.6\%$
of triangles, but its central $80\%$ runs from $0.62$ to $2.88$ times the
truth, a factor of $4.7$, and its median lies $21\%$ above the truth; the
factor grows with $\kappa_r$, and from $\kappa_r = 20$ on, $13$ to $37\%$
of triangles return no finite estimate. Knowing the levels exactly removes
most of the median bias (median ratios $1.05$ to $1.41$) but little of the
spread, a factor of $3.7$ at $\kappa_r = 3$. With forty rows the same
anchored estimator is nearly unbiased up to $\kappa_r = 10$ and its central
$80\%$ narrows to a factor of $1.8$ at $\kappa_r = 3$, while the free-level
fit is unchanged. The precision of the anchored channel is set by the
number of accident years, as Section~\ref{sec:precision:row} argued:
anchoring turns $\kappa_r$ from unrecoverable into estimable, and at
$I = 10$ it delivers an estimate whose uncertainty is of the order of the
parameter itself.

Holding the development weights and the severity at their true values
favours the anchored fits, so the common-scale arm was repeated with them
estimated jointly by maximum likelihood, on the same triangles ($200$ per
configuration at $I = 10$, $100$ at $I = 40$; Table~\ref{tab:boundary_joint}).
The picture does not change. At $I = 10$ and $\kappa_r = 3$ the central
$80\%$ of the estimates spans a factor of $4.8$, against $4.7$ with
$\theta$ known; the share of triangles at $\kappa_r = \infty$ is of the same
order; and with forty rows the estimate is again within $18\%$ of the truth
at the median. Where the two runs differ, at large $\kappa_r$ with ten rows,
the joint fit is less biased upward; it also overstates the severity shape
(median $\hat\alpha = 4.4$ against the true $3$), which leaves less of the
dispersion of the row totals to severity and more to the frailty. Under
free levels the joint fit reaches $\kappa_r = \infty$ on all $50$ triangles
on which the profile was computed, as Theorem~\ref{thm:boundary} requires,
and refitting from the true parameters reproduces the estimate on $98$ of
the $100$ triangles checked (Section~\ref{supp:boundary}).

\subsection{The allocation concentration in simulation}\label{sec:illus:dirichlet}

\begin{table}[!htbp]
\centering
\caption{The allocation concentration estimated from a single paid
triangle (Dirichlet allocation, $J = 5$,
$w = (0.45, 0.25, 0.15, 0.10, 0.05)$, true $c = 50$, $2{,}000$ triangles
per column). Mean and standard deviation of the moment estimator and of
the conditional maximum-likelihood estimator
\eqref{eq:dirichlet_condlik} with the chain-ladder pattern, divided by
$c$; and the ratio of the measured variance of the moment estimator, a median over correlated cell estimators with the pattern estimated, to the per-cell asymptotic reference of Proposition~\ref{prop:c_variance}, which holds $F_k$ known ($10{,}000$ triangles per column); the two are unlike objects, and the ratio's own limit is $0.73$. At $c = 20$ and $100$ the ratios agree with these
to within $0.05$ for the means; Table~\ref{tab:sigma_c} gives the
variance ratio at all three.}
\label{tab:dirichlet_c}
\small
\begin{tabular}{@{}lccccc@{}}
\toprule
$I$ & 7 & 10 & 15 & 20 & 50 \\
\midrule
\multicolumn{6}{@{}l}{\textit{Moment estimator}} \\
\quad mean $/\,c$ & 1.90 & 1.31 & 1.14 & 1.10 & 1.03 \\
\quad s.d.\ $/\,c$ & 1.90 & 0.74 & 0.43 & 0.34 & 0.18 \\
\quad $I\,\widehat{\Var}(\hat c)/\bar\sigma_c^2$ & 14.00 & 2.77 & 1.47 & 1.14 & 0.79 \\
\midrule
\multicolumn{6}{@{}l}{\textit{Conditional maximum likelihood}} \\
\quad mean $/\,c$ & 1.52 & 1.23 & 1.13 & 1.09 & 1.03 \\
\quad s.d.\ $/\,c$ & 0.70 & 0.36 & 0.24 & 0.19 & 0.11 \\
\bottomrule
\end{tabular}

\end{table}

Table~\ref{tab:dirichlet_c} reports both estimators. At $I = 10$ the
moment estimator overstates $c$ by $31\%$ on average with a standard
deviation of $0.74c$; the conditional maximum-likelihood estimate halves
the spread, to $0.36c$, but not the bias, which is $23\%$. At $I = 7$ the
two overstate $c$ by $90\%$ and $52\%$, and both approach $c$ only slowly,
to a bias of $3\%$ at $I = 50$. On the Taylor--Ashe triangle the two
estimates are $189.1$ and $112.4$, on Mortgage $152.0$ and $66.0$. The
allocation concentration is therefore the case in which the free-level
fit loses nothing to the boundary and a ten-row triangle still determines
the concentration weakly, the position of $\kappa_c$ in
Proposition~\ref{prop:two_kappa_id}(ii). What substituting the estimate
costs a predictive interval is reported in Section~\ref{supp:dirichlet}.

\subsection{Free and anchored levels on Schedule~P}\label{sec:illus:sp}

The CAS Schedule~P database \citep{Meyers2011, CAS2024} holds ten-year
paid triangles of U.S.\ insurers in six lines of business, accident
years 1988--1997. Two fits on its eligible portfolios bear on the
boundary. Both are computed on the outcome-free panel of Section~\ref{sec:scheduleP}: the $452$ portfolios with a complete
triangle, a negative incremental share of at most $5\%$ and a
chain-ladder reserve above \$$100$k, selected without reference to the
realised outcome, with zero cells kept and negative cells excluded
(Table~\ref{tab:eligibility_chain}). Every denominator in the paper
descends from one chain: $779$ complete $10 \times 10$ triangles in the
six lines, $631$ after the negative-increment screen, and $452$ after
the outcome-free reserve threshold; the method-specific denominators on
that panel and their reconciliation are in
Table~\ref{tab:eligibility_fits} and Section~\ref{supp:scheduleP}. The
panel is broader than the $200$
groups of \citet{Meyers2015} because we include the wider range of
triangles a practitioner might encounter.

The first fit has the row dispersion $\kappa_r$ itself as its parameter. It
is the exact frailty-augmented posterior of the single-frailty model with
the cell means at their free-level fitted values, $\alpha = 3$
and a $\mathrm{Gamma}(2, 0.4)$ prior on $\kappa_r$ (median $4.2$), fitted
on the $392$ portfolios on which it is computationally feasible. It
conditions on fitted levels, the third operation distinguished after
Theorem~\ref{thm:boundary}. The fitted levels carry each row's realised frailty, so the
realised intensities sit at their fitted means, the frailty likelihood
reads them as homogeneous, and $\kappa_r$ is pushed upward, held only by
the prior. Across the $363$ portfolios that also have a
working fit, the median of the posterior medians is $12.4$ against the
prior's $4.2$, and the lower $2.5\%$ posterior quantile lies above the
prior median on $66\%$ of them
(Section~\ref{supp:evidence:real}).

The second fit concerns a different parameter from $\kappa_r$. It is the
Bayesian second-moment fit, whose single dispersion parameter is the
curvature dispersion $\kappa_B$ of \eqref{eq:B}, which pools every source
of quadratic variance into one scalar; it is run once with free
levels and once with the levels anchored to net earned premium,
$\mu_i = \pi_i \cdot \mathrm{ELR}$ with a common expected loss ratio,
on the $281$ portfolios with a positive premium in every accident year
and both fits usable (Table~\ref{tab:kappa_sp}).

\begin{table}[!htbp]
\centering
\caption{The curvature dispersion $\kappa_B$ of the working fit, not the row
dispersion $\kappa_r$, on Schedule~P under free and
premium-anchored accident-year levels: Bayesian second-moment fit
(working likelihood, reference prior $\mathrm{Gamma}(2, 0.4)$ with median
$4.2$ and $95\%$ interval width $13.3$), $281$ portfolios with both fits
usable. Quantiles across portfolios of the posterior median, the median
ratio of the posterior to the prior $95\%$ interval width, and the share
of portfolios whose $95\%$ posterior interval excludes the prior median.}
\label{tab:kappa_sp}
\small
\begin{tabular}{@{}lrrrrrr@{}}
\toprule
& & \multicolumn{3}{c}{Posterior median across portfolios} & Width / prior & Excludes prior \\
\cmidrule(lr){3-5}
Accident-year levels & $n$ & q05 & median & q95 & width (median) & median (\%) \\
\midrule
Free & $281$ & 3.7 & 7.6 & 23.6 & 1.06 & 38.4 \\
Premium-anchored & $281$ & 3.6 & 6.7 & 20.5 & 1.02 & 33.8 \\
\midrule
\multicolumn{7}{@{}l}{Spearman correlation of the two posterior medians across portfolios: $0.93$} \\
\bottomrule
\end{tabular}

\end{table}

Anchoring restores the row-level channel in principle. In the working
second-moment fit, however, the combined scalar dispersion changes little
when the channel of ten row totals is added to the within-row curvature
the fit already reads: the two posteriors have nearly the same
cross-portfolio distribution (medians $7.6$ and $6.7$), nearly the same
width relative to the prior, and rank the portfolios alike (Spearman
$0.93$). Both update the prior. Several explanations for their similarity
remain possible: the shared prior, error in the premium anchor, and weak
information in each channel at $I = 10$. This comparison cannot separate
them. On real triangles the free-level working
fit reads the within-row curvature, which carries the cell dispersion,
where it exceeds the first-stage term of Remark~\ref{rem:cellshock_target},
and whatever cell-level misspecification the data hold
(Proposition~\ref{prop:two_kappa_id}(ii)); the anchored fit reads the same
curvature plus ten row totals, and a channel of ten observations moves the
result little, as the simulation of Section~\ref{sec:illus:boundary}
predicts. Table~\ref{tab:kappa_sp} concerns what the one-dispersion fit
displays at $I = 10$, leaving identification of $\kappa_r$ under
anchoring intact. Curvature-channel strength, anchor error and weakness
of both channels remain unresolved explanations. The exact likelihood
with free levels sends $\kappa_r$ to the boundary; the working fit's
$\kappa_B$ shows little of what anchoring adds at $I = 10$.

\FloatBarrier
\section{What the Imports Cost a Predictive}\label{sec:cost}

The information map gives an amount-only predictive three inputs: the
severity composite $A$ from the triangle, the row dispersion $\kappa_r$
from anchored levels or a prior, and a fitted mean structure whose
estimation error on a $10 \times 10$ triangle is of the order of the
process variance. We examine their consequences for interval coverage.
How much does propagating mean-structure estimation error change
calibration? Can a variance law of the correct form help when the fit
has absorbed the signal for its dispersion?

We organise the procedures by variance law and treatment of
mean-structure estimation error. Since several components change between
procedures, this is a comparison of complete implementations, not a
factorial experiment. A separate ablation replaces individual inputs of
one benchmark with their true values. Together these analyses support
the comparisons below, but cannot causally decompose coverage into a
variance-law effect and a mean-structure effect. The benchmark is used
to examine these inputs, without recommending it as a standard.

``Misspecified'' and ``correct'' on the variance-law axis refer to the
marginal variance function of Proposition~\ref{prop:moments}. Conditional
on realised row intensity, cell variance is linear. The quadratic term
represents uncertainty about that intensity, supplied through the
treatment of the levels.

\subsection{Predictive constructions}\label{sec:predictive}

A predictive interval for the reserve has a variance law and a treatment
of the estimation error of the mean structure. We use three treatments.

\paragraph{Plug-in.} The mean structure is held at $\hat\theta$ and only
process variance is simulated, the conditional target
$p(R \mid \mathcal{D}, \hat\theta)$. ODP-cond and Tweedie-cond use the
linear and power variance functions; WLS-cond uses the correct marginal cell
variance of the CNBG law at the point estimate $(\hat\kappa_B, \hat A)$
of \eqref{eq:wls}, without the shared-frailty dependence, and draws cells
independently (Algorithm~\ref{alg:bootstrap}). WLS-cond provides the
lower-edge ablation in the design; it is not proposed for use.

\paragraph{Refit.} The England--Verrall bootstrap resamples residuals,
refits the mean structure on each pseudo-triangle and simulates future
cells: with the linear variance (ODP), the power variance (Tweedie) or
the CNBG-form variance law with its dispersion re-estimated by
\eqref{eq:wls} from the free-level residuals of each pseudo-triangle
(Frequentist CNBG). Under free accident-year levels the refit reproduces,
to first order, the conditional predictive variance under a diffuse prior
on the levels and the pattern, a correspondence established for the count
hierarchy by \citet{VanOirbeek2026cond}.

\paragraph{Draws: the modular benchmark.}\label{sec:estimation:bayes}
The third treatment holds the observed triangle fixed and draws every
input from a stated source. The severity composite is drawn from a
working posterior: the point estimator \eqref{eq:wls} saturates its cap on
about two thirds of correctly specified triangles
(Section~\ref{supp:evidence:sim}), so the variance equation is instead
wrapped in a Gamma working density for the squared residual
$z_{ij} = (X_{ij}-\hat\nu_{ij})^2$,
\begin{equation}\label{eq:working_lik}
  z_{ij} \;\overset{\mathrm{ind}}{\sim}\;
  \mathrm{Gamma}\!\left(s_{ij},\;\frac{s_{ij}}{\mu_z^{ij}}\right),
  \qquad
  \mu_z^{ij} = A\hat\nu_{ij} + \frac{\hat\nu_{ij}^2}{\kappa_B},
\end{equation}
a composite-likelihood Bayes fit \citep{Pauli2011, Ribatet2012} with
priors on $(\kappa_B, A)$. The shape does not move the location of the
posterior, which concentrates on the target of the estimating equation
whatever $s_{ij}$ is (Proposition~\ref{prop:sandwich}), but it sets the
curvature; the squared coefficient of variation of $z_{ij}$ is at least
$2$ in every cell (Proposition~\ref{prop:fourth_moment}), so the shapes are
set cell by cell from the fourth moment and the log-likelihood is rescaled
by the curvature adjustment of \citet{Ribatet2012}. Under free levels the
$A$ posterior comes from a fit whose working mean is the absorbed variance
$A\hat\nu_{ij}$, since the quadratic term is then at zero
(Lemma~\ref{lem:absorption}); across the simulation grid its adjusted
$95\%$ interval covers the true $A$ on $90.1\%$ of replications. This
five-point under-coverage remains in the construction
(Section~\ref{supp:estimation:bayes}). The prior
is $\kappa_B \sim \mathrm{Gamma}(2, 0.4)$, with $95\%$ of its mass on
$(0.6, 13.9)$ and median $4.2$, and
$A \sim \mathrm{Gamma}(2, 2/\hat A_{\mathrm{emp}})$ with
$\hat A_{\mathrm{emp}}$ the mean squared residual divided by the mean
fitted amount (Section~\ref{supp:estimation:bayes}). With $\{A^{(b)}\}$ the
posterior draws of $A$ and $\pi(\kappa_r)$ the prior, each predictive draw
$s = 1, \ldots, B$ proceeds as follows.
\begin{enumerate}[label=\arabic*., nosep]
\item Draw $A^{(s)}$ from $\{A^{(b)}\}$ and $\kappa_r^{(s)} \sim \pi$,
      with $\kappa_r^{(s)} \ge 0.5$: the triangle supplies $A$ and does not
      supply $\kappa_r$.
\item Draw the mean-structure coefficients
      $\theta^{(s)} \sim \mathcal{N}(\hat\theta, \widehat{\Var}(\hat\theta))$,
      with the quasi-Poisson covariance scaled by the Pearson dispersion,
      and set $\nu_{ij}^{(s)} = \exp\bigl(x_{ij}^\top\theta^{(s)} -
      \tfrac12 x_{ij}^\top\widehat{\Var}(\hat\theta)x_{ij}\bigr)$, capped at
      $10\,\hat\nu_{ij}$; the half-variance term is the log-normal
      retransformation correction.
\item Draw each row frailty from its conjugate update under a Poisson
      working law for $X_{ij}/A$,
      \begin{equation}\label{eq:U_posterior}
        U_i^{(s)} \mid \mathcal{D}_i \;\sim\;
        \mathrm{Gamma}\!\left(
          \kappa_r^{(s)} + \frac{\sum_{j:(i,j)\in\mathcal{D}} X_{ij}}{A^{(s)}},\;\;
          \kappa_r^{(s)} + \frac{\sum_{j:(i,j)\in\mathcal{D}} \nu_{ij}^{(s)}}{A^{(s)}}
        \right),
      \end{equation}
      so that the realised level the triangle does carry is used.
\item Draw each future cell from the Gamma law matched to the first two
      conditional moments,
      \begin{equation}\label{eq:cell_draw}
        X_{ij}^{(s)} \sim
        \mathrm{Gamma}\!\left(\frac{\nu_{ij}^{(s)}U_i^{(s)}}{A^{(s)}},\;
          \frac{1}{A^{(s)}}\right),
        \qquad (i,j)\notin\mathcal{D},
      \end{equation}
      with all future cells of a row sharing $U_i^{(s)}$, and set
      $R^{(s)} = \sum_{(i,j)\notin\mathcal{D}} X_{ij}^{(s)}$.
\end{enumerate}
We call this construction CNBG-Mod, the modular CNBG predictive. Its
prior draw, sampling-distribution draw, working posterior and approximate
conjugate update do not form the posterior predictive of a single joint
model. Steps 2 and 3 may count the latest accident years' level uncertainty
twice, and step 4 approximates the compound law by two moments.
Section~\ref{sec:simulation:oracle} tests the consequences by ablation.
Sections~\ref{supp:estimation} and~\ref{supp:predictive} give the
step-by-step justification, the exact frailty-augmented posterior used
as a reference, and WLS-cond.

\subsection{Simulation: two axes and an ablation}\label{sec:simulation}
\label{sec:simulation:design}

Triangles are drawn from the single-frailty case of
Definition~\ref{def:cnbg} ($\kappa_c = \infty$), the calibration of
Section~\ref{sec:illus:boundary}: $\mu_i = 30\, e^{0.05(i-1)}$ expected
claims, the ten-column development pattern, $\bar Y = 5{,}000$, an expected
outstanding reserve of about \$$355$k, $\kappa_r \in \{3, 5, 10, 20, 50\}$
and $\alpha \in \{1, 3, 10\}$, $15$ configurations with $M = 500$ triangles
each and $B = 500$ predictive draws per triangle, seed $2026$. Of the
$22{,}500$ Bayesian fits, $64$ failed at initialisation and none reached
the wall-clock limit. The Monte Carlo standard error of a coverage near
$92\%$ is about $1.2$ points per configuration and $0.3$ points for the
means over configurations. Widths are medians of the interval width as a
fraction of the realised reserve $R_{\mathrm{true}}$, because the refit
procedures occasionally return a degenerate interval on a single
replication that destroys a mean and leaves the median and the coverage
unaffected.

\label{sec:simulation:headline}
Table~\ref{tab:two_axis} places every procedure by its variance law and
its treatment of mean-structure uncertainty and reports the means over the
$15$ configurations; the per-configuration results are in
Table~\ref{tab:simulation}.

\begin{table}[!htbp]
\centering
\caption{The procedures by variance law and treatment of the estimation
error of the mean structure (held at $\hat\theta$, refitted by bootstrap,
or drawn). Simulation: means over the $15$ configurations of the coverage
of nominal $95\%$ intervals (C, \%) and of the median width (W, \% of
$R_{\mathrm{true}}$); $M = 500$ per configuration; Monte Carlo standard
error $\sim 0.3$ points. Schedule~P: the outcome-free panel of
Section~\ref{sec:scheduleP}, $452$ portfolios; C on the procedure's own
denominator $n$, C all on the full panel with portfolios lacking an
interval scored as misses, W the median width (\% of
$R_{\mathrm{actual}}$).}
\label{tab:two_axis}
\small
\begin{tabular}{@{}lllrrrrrr@{}}
\toprule
& Mean & & \multicolumn{2}{c}{Simulation} & \multicolumn{4}{c}{Schedule~P} \\
\cmidrule(lr){4-5}\cmidrule(lr){6-9}
Variance law & structure & Procedure & C & W & $n$ & C & C all & W \\
\midrule
Linear (ODP) & held at $\hat\theta$ & ODP-cond & 67.1 & 52.0 & $440$ & 28.0 & 27.2 & 40.7 \\
 & refit & ODP bootstrap & 91.5 & 95.3 & $440$ & 50.7 & 49.3 & 122.1 \\
\addlinespace[2pt]
Power (Tweedie) & held at $\hat\theta$ & Tweedie-cond & 68.2 & 53.4 & $437$ & 32.3 & 31.2 & 43.9 \\
 & refit & Tweedie bootstrap & 93.2 & 102.2 & $429$ & 55.2 & 52.4 & 118.8 \\
\addlinespace[2pt]
Linear + quadratic & held at $\hat\theta$ & WLS-cond & 62.2 & 46.5 & $440$ & 30.9 & 30.1 & 59.2 \\
(CNBG) & refit & Frequentist CNBG & 84.9 & 79.0 & $437$ & 51.7 & 50.0 & 102.9 \\
 & drawn & CNBG-Mod & 92.6 & 99.0 & $371$ & 68.5 & 56.2 & 108.2 \\
\bottomrule
\end{tabular}

\end{table}

At this calibration, treatment of mean-structure uncertainty separated
the implementations most. The controlled ablation in
Section~\ref{sec:simulation:oracle} supports this reading: switching off
the modular construction's mean-structure draw lowers coverage from the
$91$ to $94\%$ range to $82$ to $83\%$, with the frailty update
overlapping it. Within each variance law, propagating the estimation error of the mean
structure raises mean coverage from $67.1$ to $91.5\%$ for the linear law,
from $68.2$ to $93.2\%$ for the power law, and from $62.2\%$ to $84.9\%$ by
refitting and to $92.6\%$ by drawing for the CNBG law. Within a treatment,
the variance law moves coverage by at most $6.0$ points among the plug-in
procedures and $8.3$ among the refit bootstraps. That the estimation
error of the mean structure is of the order of the process error on a
$10 \times 10$ triangle has been known since \citet{mack1993} and
\citet{england2002}; what the table adds is that it dominates the choice
among plausible variance laws, and that the variance-law axis runs through
the dispersion each law is given. Paired over the same triangles, CNBG-Mod minus the
ODP bootstrap is $+1.1$ points (standard error $0.2$), minus the Tweedie
bootstrap $-0.6$ ($0.2$), minus the Frequentist CNBG bootstrap $+7.7$
($0.3$) and minus the plug-in procedures $+24$ to $+30$ ($0.5$). The pooled
$92.6\%$ of CNBG-Mod, with a Monte Carlo standard error of $0.3$ points,
is systematic under-coverage of nominal $95\%$, as is the ODP bootstrap's
$91.5\%$; only the Tweedie bootstrap and the oracles are within two points
of nominal. The Frequentist CNBG bootstrap has the correct
variance law and the refit mechanism, yet covers $6.6$ points less than
the ODP bootstrap at a narrower width: its point estimate of $\kappa_B$ saturates the cap on about two thirds of replications, the finite-sample scatter of an estimator whose target is the cap (Proposition~\ref{prop:two_stage_target}; Theorem~\ref{thm:boundary} concerns the profile likelihood, not this estimator), so its quadratic term is nearly zero, and its linear coefficient is the deflated intercept of Proposition~\ref{prop:two_stage_target}(iii), about $0.57A$ at $I = 10$, whereas the Pearson dispersion of the ODP bootstrap divides by $n - q$ and undoes the leverage deflation: the law it draws from is narrower than ODP's through the linear coefficient as much as through the missing quadratic term. WLS-cond, which adds
the same saturation to a plug-in, is the lowest-coverage procedure.
The free-level second-moment fit deflates the linear coefficient as well
as capping the dispersion, and the two defects separate: with the linear
coefficient corrected by the leverage factor and nothing else changed,
the refit bootstrap's coverage rises by $9$ to $11$ points to the ODP
bootstrap's on the same triangles, at a $6$ to $7\%$ greater width
(Section~\ref{supp:simulation:leverage}). The deficit of the Frequentist
CNBG bootstrap is therefore its deflated linear coefficient, not its
variance law: with the dispersion at the cap the refitted CNBG law and
the linear law cover alike. In these implementations a variance law of
the correct form added coverage only when its dispersion was supplied by
something other than the free-level fit, and then by about a point:
CNBG-Mod carries
$\kappa_r$ on the prior, and its coverage stays within $91$ to $94\%$ over
the $17$-fold range of the true $\kappa_r$, with a prior of mean $8$ in
place of $5$ moving pooled coverage by $0.4$ points
(Section~\ref{supp:simulation}).

\begin{figure}[!htbp]
\centering
\includegraphics[width=\textwidth]{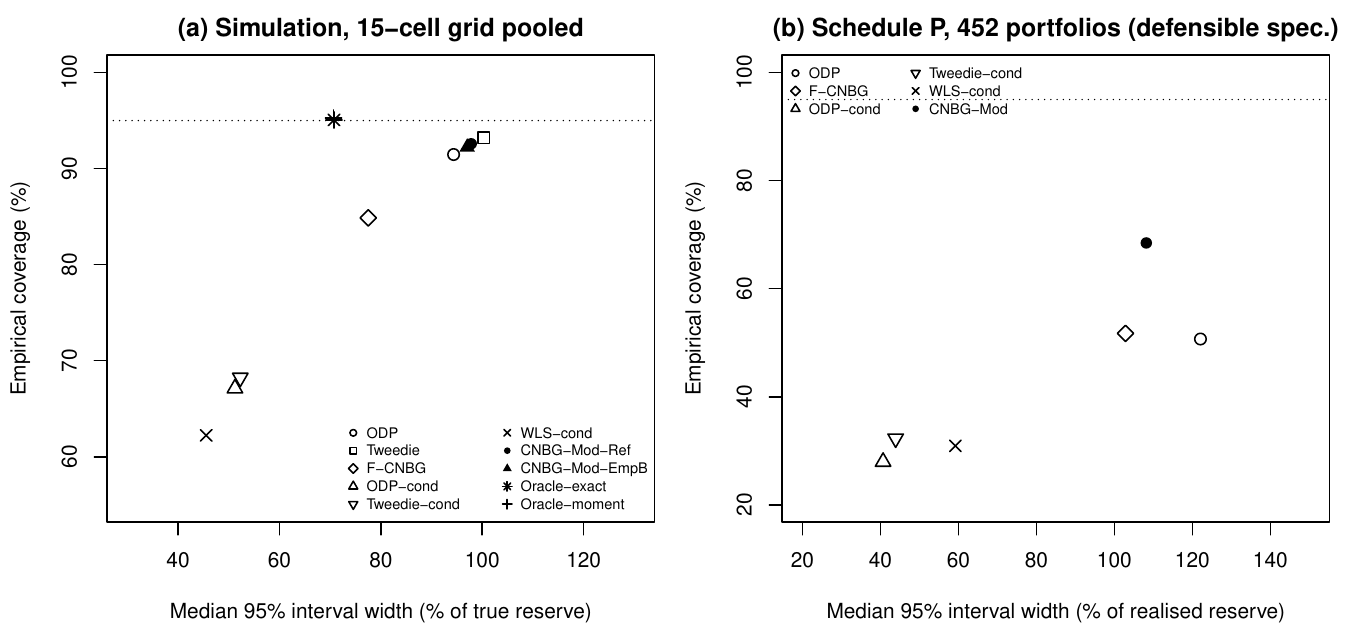}
\caption{Empirical $95\%$ coverage against median relative interval
width. (a) Simulation, pooled over the $15$-cell grid ($7{,}500$
replications; widths are pooled medians, which differ by up to two
points from the means of per-cell medians in
Table~\ref{tab:two_axis}); the oracle points average the three
configurations of Table~\ref{tab:oracle}. (b) Schedule~P, outcome-free
panel of Section~\ref{sec:scheduleP}, $452$ portfolios (CNBG-Mod on its
$371$ usable fits; each procedure on its own denominator). The dotted
line is nominal coverage.}
\label{fig:coverage_width}
\end{figure}

\label{sec:simulation:oracle}
The attribution question is answered by holding the construction fixed
and varying what is known. At three configurations
($(\kappa_r, \alpha) = (3, 3)$, $(10, 1)$ and $(50, 3)$; $M = 500$ each)
three reference procedures are scored alongside CNBG-Mod:
\emph{Oracle-exact}, the true parameters, the row frailty drawn from its
exact conditional law given the observed row by importance resampling
and future cells from the exact compound law, which is
$p(R \mid \mathcal{D}, \theta_0)$ and covers at the nominal rate by
construction; \emph{Oracle-moment}, as before with future cells drawn
from \eqref{eq:cell_draw}, so that the difference is the cost of the
two-moment cell draw; and the \emph{exact posterior} with the levels
anchored at their true values, so that $\kappa_r$ is estimable, with
future cells from the exact compound law given posterior draws
($M = 100$).

\begin{table}[!htbp]
\centering
\caption{Oracle decomposition at three configurations ($M = 500$;
$M = 100$ for the anchored exact posterior; $I = J = 10$): coverage of
nominal $95\%$ intervals, median relative width and median signed centre
error for the modular benchmark and the three reference procedures.}
\label{tab:oracle}
\small
\renewcommand{\arraystretch}{1.15}
{\scriptsize\setlength{\tabcolsep}{3pt}%
\begin{tabular}{@{}llrrrrrrrrrrrr@{}}
\toprule
& & \multicolumn{3}{c}{CNBG-Mod-Ref} & \multicolumn{3}{c}{CNBG-Mod-Ref, no step 2}
  & \multicolumn{3}{c}{Oracle-moment} & \multicolumn{3}{c}{Oracle-exact} \\
\cmidrule(lr){3-5}\cmidrule(lr){6-8}\cmidrule(lr){9-11}\cmidrule(lr){12-14}
$\kappa_r$ & $\alpha$ & C & W & E & C & W & E & C & W & E & C & W & E \\
\midrule
 3 & 3 & 92 & 95.1 & $-2.1$ & 83 & 69.6 & $-0.7$ & 94 & 71.1 & $-0.0$ & 94 & 71.4 & $+0.5$ \\
10 & 1 & 94 & 113.5 & $-3.2$ & 83 & 87.2 & $-0.4$ & 97 & 81.2 & $+2.5$ & 97 & 80.8 & $+2.6$ \\
50 & 3 & 91 & 94.4 & $-1.6$ & 82 & 71.0 & $-0.7$ & 95 & 59.7 & $+1.3$ & 95 & 60.0 & $+1.3$ \\
\midrule
\multicolumn{14}{l}{\textit{Exact-posterior (anchored), $M = 100$: C, W and E in the first three columns}} \\
 3 & 3 & 95 & 71.8 & $+4.3$ & & & & & & & & & \\
10 & 1 & 93 & 83.3 & $+5.0$ & & & & & & & & & \\
50 & 3 & 97 & 63.6 & $-0.6$ & & & & & & & & & \\
\bottomrule
\end{tabular}}

\end{table}

The oracles cover $93.6$ to $96.8\%$, at nominal up to Monte Carlo error
(probability integral transforms uniform, Kolmogorov--Smirnov
$p \ge 0.055$ in every configuration), and the moment-matched cell draw
costs nothing measurable at this calibration (Table~\ref{tab:oracle}).
CNBG-Mod covers $91.8$, $93.6$ and $91.4\%$, within two to three points
of the oracles, at $24$ to $35$ points more width and with its interval
centre $1.6$ to $3.2\%$ below the realised reserve at the median; the
anchored exact posterior covers $95.0$, $93.0$ and $97.0\%$ at
oracle-like width. Table~\ref{tab:components} removes or replaces each component of the
construction in turn on the same triangles, and Table~\ref{tab:steps}
switches the two draws off in every combination. The two draws
interact. With both on, coverage is $92$, $94$ and $91\%$ at the three
configurations; with the frailty update off (step 3) it is $93$, $95$
and $92\%$; with the mean-structure draw off (step 2) it is $83$, $83$
and $82\%$; with both off, $71$, $72$ and $70\%$. The frailty update
therefore adds about twelve points when the mean-structure draw is
absent and takes about one point away when it is present: the two steps
overlap in representing the level uncertainty of the accident years, the
double counting that Section~\ref{sec:predictive} flagged as possible,
and removing one while the other is kept does not show the first to be
inert. What the ablation establishes is narrower than harmlessness: that
the mean-structure draw is the component the construction's coverage
rests on, that the frailty update is redundant with it at this
calibration, and that the moment-matched cell draw and the prior's
location change coverage by at most $1.2$ points. The coverage of the
construction is empirical and configuration-specific, and the width above
the oracles, $24$ to $35$ points, is attributable to the mean-structure
draw and its overlap with the frailty update jointly, not to the
estimation error of the mean structure alone. The coherent joint model is
the anchored exact posterior, in which $\kappa_r$ is estimable because the
levels are anchored; it receives the true accident-year levels, which no
free-level procedure possesses. It therefore serves as a reference with
more information: it covers at $95.0$, $93.0$ and $97.0\%$ at
oracle-like width. Under free levels the profile likelihood supplies no finite estimate
of $\kappa_r$; a Bayesian predictive can nevertheless be defined, and
what it infers about $\kappa_r$ then depends on the priors placed on the
levels and on $\kappa_r$. The exact model with free levels, the levels
carrying a lognormal prior of log-scale $2$ centred on each row's
crude count scale, fitted on ten replicates per configuration, stays
close to its prior for $\kappa_r$ at every true value (posterior median
$4.7$ to $5.1$ against the prior's $4.2$, interval $[1.2, 14.4]$ against
$[0.6, 13.9]$). Integrating the levels under this prior moves
$\kappa_r$ little, consistent with the mechanism behind
Theorem~\ref{thm:boundary}. The size of the movement depends on the
level prior's scale: at
$(\kappa_r, \alpha) = (3, 3)$ a log-scale of $0.5$, which holds each level
near its own realised value, moves the posterior median of $\kappa_r$ to
$7.5$ with interval $[2.7, 18.1]$, above the prior and towards the
boundary, as the Schedule~P fit that conditions on the fitted levels
does, while log-scales of $2$ and $4$ give $5.0$ and $4.5$
(Section~\ref{supp:evidence:sim}). These posterior changes quantify the
prior-scale dependence discussed after the theorem. The variant of
Remark~\ref{rem:diffuse}, step 1 with $\kappa_r = 0$ so that step 3 is the
diffuse conjugate update and the predictive uses no $\kappa_r$ at all,
covers $91.6$, $94.4$ and $92.2\%$ against the benchmark's $91.8$, $93.6$
and $91.4\%$ on the same triangles (paired differences $-0.2$, $+0.8$ and
$+0.8$ points, standard errors $0.4$ to $0.7$), at widths $3$ to $6$ points
greater. On these triangles, whose rows carry $30$ to $47$ claims, a
diffuse-update variant of the modular construction, which uses no
$\kappa_r$, is as well calibrated as the variant that imports it; the
variant keeps the sampling-distribution draw of the mean structure and
the working posterior of $A$, so it is not the coherent predictive of
Remark~\ref{rem:diffuse}, only its update rule. The prior on $\kappa_r$
matters where the rows are thin, and the diffuse treatment pays for its freedom from $\kappa_r$ in
width.

\begin{table}[!htbp]
\centering
\caption{What each component of the modular benchmark contributes:
change in coverage of nominal $95\%$ intervals and in median relative width
(percentage points) when the component is removed or replaced, at the three
oracle configurations ($M = 500$ each, same triangles). $^\ast$The cell
draw is compared at true parameters (Oracle-exact minus Oracle-moment).
The prior row compares the two priors on the full simulation at the same
configurations.}
\label{tab:components}
\footnotesize
\setlength{\tabcolsep}{4pt}
\begin{tabular}{@{}llrrr@{}}
\toprule
& & \multicolumn{3}{c}{Change in coverage / width (points), $(\kappa_r, \alpha) =$} \\
\cmidrule(lr){3-5}
Component & Removed or replaced by & $(3, 3)$ & $(10, 1)$ & $(50, 3)$ \\
\midrule
Mean-structure draw (step 2) & held at $\hat\theta$ & $-8.8$ / $-25.5$ & $-10.2$ / $-26.3$ & $-9.8$ / $-23.4$ \\
Frailty update (step 3) & $U_i = 1$ & $+1.2$ / $+3.0$ & $+1.2$ / $+4.2$ & $+0.2$ / $+2.4$ \\
Steps 2 and 3 & both removed & $-21.0$ / $-41.3$ & $-21.8$ / $-46.0$ & $-21.6$ / $-40.4$ \\
Prior on $\kappa_r$ (mean $5$) & prior with mean $8$ & $-0.6$ / $+0.2$ & $-0.8$ / $-0.8$ & $-0.6$ / $+0.6$ \\
Moment-matched cell draw$^\ast$ & exact compound law & $-0.4$ / $+0.3$ & $+0.0$ / $-0.4$ & $-0.2$ / $+0.3$ \\
Prior on $\kappa_r$ (mean $5$) & none: $\kappa_r = 0$, diffuse update & $-0.2$ / $+3.3$ & $+0.8$ / $+5.8$ & $+0.8$ / $+4.3$ \\
\bottomrule
\end{tabular}

\end{table}

The information map still applies: under free levels the row dispersion
is imported or priced diffusely, the severity split remains unrecovered,
and the predictive uses two moments. At $10 \times 10$, the ablation
locates the measurable coverage cost on correctly specified data in
mean-structure estimation error. The construction propagates this error
with an overlapping contribution from the frailty update. At ten times the claim volume every width falls
by a factor of about three and the gap to the oracles narrows to $8$--$9$
points (Section~\ref{supp:simulation}). 

The ordering survives a finite cell shock, although every procedure
loses coverage because none models that shock as such: on data with
$(\kappa_r, \kappa_c) = (5, 5)$, $(5, 20)$ and $(50, 5)$ the propagating
procedures cover $79$ to $90\%$ and the plug-in procedures $54$ to
$69\%$, with gaps of $22$ to $33$ points between the treatments, and the
lead of the modular benchmark over the ODP bootstrap is $2.0$ to $2.5$
points (Table~\ref{tab:cellshock}, Section~\ref{supp:simulation}).

\paragraph{A misspecified mean structure.} The simulation so far has the
mean structure correctly specified, and Schedule~P suggests that on real
triangles the binding problem is location. A bridging experiment
multiplies the claim rate by $\exp\{-\delta\max(0, t - 7)\}$, with
$t = i + j - 1$ the calendar year and $\delta \ge 0$, so that a downward
change the chain-ladder mean structure cannot represent begins in the
eighth calendar year, inside the last three observed diagonals, and
continues into the future cells; $\kappa_r = 5$, $\alpha = 3$,
$\delta \in \{0, 0.05, 0.10, 0.20\}$ per year and $M = 200$ triangles per
value (Table~\ref{tab:drift}, Section~\ref{supp:simulation:drift}).
Because the multiplicative chain-ladder fit is informed largely by
diagonals preceding the change, it overstates subsequent development and
places the predictive centres above the realised reserves, the direction
observed on the Schedule~P panel. Nothing in an amount-only fit repairs
the location: at $\delta = 0.10$ the interval centres of every procedure
sit $27$ to $35\%$ above the realised reserve at the median, and at
$\delta = 0.20$ $57$ to $69\%$ above it. Coverage falls with the
drift and the misses are below the interval, $31\%$ for the ODP
bootstrap and $52\%$ for ODP-cond at $\delta = 0.20$ against $3$ and
$19\%$ without drift, the pattern of Section~\ref{sec:scheduleP}. The
treatment of the estimation error of the mean structure keeps its weight
throughout: the ODP bootstrap and CNBG-Mod cover $23$ to $33$ points more
than the plug-in ODP-cond at every $\delta$, because their wider
intervals absorb part of the displacement, with the Frequentist CNBG
bootstrap between them, and the lead of CNBG-Mod over the ODP bootstrap
grows from $1.5$ to $9.5$ points as the drift strengthens. Propagating
estimation uncertainty improves coverage under a location error without
repairing it, which is what Section~\ref{sec:scheduleP} finds on real
triangles.

\subsection{Schedule~P: where the binding constraint lies}\label{sec:scheduleP}
\label{sec:scheduleP:data}

The same procedures are run on the outcome-free Schedule~P panel of
Section~\ref{sec:illus:sp}, $452$ portfolios in six lines, with zero cells
kept, negative cells excluded for every procedure, the reference prior on
every portfolio and $B = 500$ draws. Negative increments lie outside the
support of the model, and one rule is applied to every procedure alike:
portfolios with more than $5\%$ negative observed increments are
excluded from the panel, the remaining negative cells are excluded from
every fit, and zero cells are retained; the preprocessing changes the
estimand to the reserve of a triangle without recoveries, and is a
limitation. The target that scores every procedure is the realised
reserve as recorded, the sum of the future incremental payments of the
lower triangle with any negative increments included, so the evaluation
retains the recoveries that the fitted object excludes and a procedure
is penalised for recoveries it could not model; the $17$ non-positive
realised reserves of the panel are of that origin. Four quantities are
kept apart in every table: portfolios on which a procedure returns no
interval ($n$); fits that fail the inference-quality rule of
Section~\ref{supp:estimation:bayes}, four of the $371$ CNBG-Mod fits,
retained and reported, whose exclusion moves its coverage from $68.5$ to
$68.4\%$ on $367$; coverage conditional on an interval (C); and
operational coverage over the eligible panel with missing intervals
scored as misses (C all). Restricting the panel further to the $255$ portfolios with
no negative cell at all lowers every procedure's coverage, because that
subset is the sparser part of the panel (the ODP bootstrap covers
$42.0\%$ on $255$ and CNBG-Mod $64.9\%$ on $191$, with a paired lead of
$12$ points on $191$), and leaves the ordering and the one-sided misses
($54\%$ below and $4\%$ above the ODP bootstrap's interval) unchanged
(Section~\ref{supp:scheduleP}). The first stage fails to converge on
$11$ portfolios and one fit exceeds a five-minute cap; the
working-likelihood sampler returns no usable posterior on $81$ portfolios
($18\%$), so CNBG-Mod is reported on its own denominator of $371$ and on
the full panel with the portfolios without an interval scored as misses.
An earlier implementation that screened on the realised reserve flattered
every procedure by three to five points uniformly
(Section~\ref{supp:scheduleP}). Paired differences between procedures
carry a $95\%$ interval from $2{,}000$ bootstrap resamples of insurers,
all portfolios of a company resampled together.

\label{sec:scheduleP:defensible}
The Schedule~P columns of Table~\ref{tab:two_axis} give pooled coverage
and width. No procedure approaches nominal coverage. CNBG-Mod covers
$68.5\%$ on its usable fits and $56.2\%$ on the full panel, the refit
bootstraps $50.7$ to $55.2\%$, and the plug-in procedures $28.0$ to
$32.3\%$. The ordering by treatment of mean-structure uncertainty is the
one the simulation predicts, with the plug-in procedures $30$ to $36$
points below CNBG-Mod on the common portfolios; within the propagating
procedures CNBG-Mod leads the ODP bootstrap by $9.4$ points on the $361$
portfolios with both intervals once the ten degenerate refits are left
out (interval $4.9$ to $13.8$) with a lower interval score
(Table~\ref{tab:defensible}). The misses are one-sided: the realised
reserve falls below the ODP bootstrap's interval on $44\%$ of portfolios
and above it on $5\%$, where a calibrated predictive would give $2.5\%$
on each side, and for CNBG-Mod the shares are $21\%$ and $11\%$. Such a
strongly one-sided miss pattern is difficult to attribute to predictive
skewness alone and points to a systematic displacement of the predictive
location relative to the outcomes, whose possible sources
Section~\ref{sec:scheduleP:bias} takes up. Every score in this
section has a stated denominator. Coverage uses none. Widths and relative interval scores
divided by the realised reserve are reported on the portfolios where it
is positive ($17$ of the $452$ realised reserves are zero or negative, on
which such a ratio is undefined or reversed), and Table~\ref{tab:scoring}
in the Supplement repeats them divided by the chain-ladder reserve that
defines the panel, which is positive and outcome-free on every portfolio,
together with the unnormalised Winkler score in dollars, a proper score,
and paired denominator-free comparisons; ten numerically degenerate
refits, identified without the outcome, are left out of every score, and
the explosive intervals are counted and kept in the dollar means, which
they dominate; the medians beside those means are the robust descriptive
summaries, and no score is trimmed on its realised value. The orderings are the same
under every denominator: CNBG-Mod leads the ODP bootstrap by a mean log
ratio of interval scores of $-0.31$ $[-0.46, -0.17]$ and the plug-in
procedures by $-0.55$ to $-0.70$, and a median outcome-normalised score,
which is not itself a proper scoring rule, is kept only beside those.

\label{sec:scheduleP:positive}
That lead is largely a property of near-empty triangles. The outcome-free
rule admits them: a quarter of the panel has $16$ or fewer positive cells
of $55$, and the sampler fails mostly there. A rule that requires at least
half of the observed cells to be positive uses the observed triangle only
and keeps $303$ portfolios (Table~\ref{tab:sp_positive}). On them the
sampler returns an interval on every portfolio whose first stage converges
($293$), coverage rises for every procedure, the ordering by treatment of
mean-structure uncertainty is unchanged ($38.6$ to $41.6\%$ for the plug-in
procedures against $61.1$ to $71.0\%$ for the rest), and the lead of
CNBG-Mod narrows to $5.1$ points over the ODP bootstrap (interval $1.6$ to
$8.7$) and $1.0$ point over the Tweedie bootstrap ($-2.0$ to $4.2$).
Thresholds of $20$ and $40$ positive cells give the same picture, and the
misses stay one-sided ($28\%$ below against $6\%$ above for the ODP
bootstrap). On triangles with enough positive cells the modular benchmark
and the refit bootstraps are close.

\label{sec:scheduleP:bias}
The location of the chain-ladder forecast relative to the outcome limits
every procedure. We measure it by the ratio of realised reserve to chain-ladder reserve
(Table~\ref{tab:cl_bias}; on an outcome-free panel the denominator has to
be the forecast, since $17$ of the $452$ realised reserves are zero or
less). The median portfolio realises $0.70$ of its chain-ladder reserve and
$78\%$ of portfolios realise less; summed over portfolios, the realised
reserve is $0.85$ of the chain-ladder reserve with the one numerically
degenerate fit excluded. The central $80\%$ of the ratio runs from $0.09$
to $1.26$, and the median portfolio's realised reserve differs from its
chain-ladder reserve by $44\%$ of the latter. An interval calibrated
correctly on the variance side still misses when its centre is off by that
much and its half-width is $20$ to $60\%$ of the reserve, so the dispersion
of the forecast errors is large enough to dominate the coverage
comparisons for every procedure. Realised error includes process
variation, so its size and asymmetry support, without isolating, a
mean-structure explanation. The displacement is consistent
with the late-1990s calendar-year changes in claim frequency and severity
that violate the calendar-year invariance of the chain ladder
\citep{TaylorMcGuire2004, Shapland2019, ClarkDingZhou2022}, and
\citet{aslan2026} finds that calendar-year effects relocate cohort
coefficients on the same database. Structural mean error, estimation
error and process variation remain inseparable in this design.
The outcome-free eligibility rule also conditions on a large
chain-ladder forecast. All procedures exclude negative observed
increments, whereas the realised reserves used for scoring retain
negative future increments. On the $255$ portfolios without any negative
cell, coverages and their ordering are unchanged
(Section~\ref{supp:estimation:wls}).

The simulated calendar-year change in Section~\ref{sec:simulation}
produces one-sided misses of this size at the same displacement of the
centres. Establishing it as the cause on Schedule~P would require a direct
comparison with a calendar-aware mean structure and testing on the
independent 1998--2007 window of \citet{cas2025}; both remain future work.
At the median, the chain-ladder point estimate is prudent. The backtest
calls into question the confidence attached to the interval around it.

\begin{table}[H]
\centering
\caption{The realised reserve as a fraction of the chain-ladder reserve
across portfolios: quantiles, the share of portfolios below one (the chain
ladder over-reserved), and the median absolute deviation from one.
Outcome-free: the $452$ portfolios with a chain-ladder reserve above
\$$100$k; historical: the $398$ portfolios with a realised reserve above
\$$100$k, shown for comparison.}
\label{tab:cl_bias}
\small
\begin{tabular}{@{}lrrrrrrrr@{}}
\toprule
& & \multicolumn{5}{c}{Quantiles of $R_{\mathrm{actual}}/R_{\mathrm{CL}}$} & Below & Median \\
\cmidrule(lr){3-7}
Panel & $n$ & 10\% & 25\% & 50\% & 75\% & 90\% & one (\%) & $|1 - \text{ratio}|$ (\%) \\
\midrule
Outcome-free & $452$ & 0.09 & 0.29 & 0.70 & 0.97 & 1.26 & 77.9 & 43.5 \\
Historical & $398$ & 0.24 & 0.48 & 0.80 & 1.07 & 1.67 & 70.4 & 35.2 \\
\bottomrule
\end{tabular}

\end{table}

\label{sec:scheduleP:anchored}
Anchoring the accident-year levels to net earned premium, the remedy that
Section~\ref{sec:free:anchoring} prescribes for the row dispersion, does
not repair location. On the $286$ portfolios with a positive premium in
every accident year and both intervals, it narrows the intervals from a
median $97.0\%$ to $91.4\%$ of the realised reserve and lowers pooled
coverage from $72.0\%$ to $68.2\%$ (Table~\ref{tab:anchored}); a
premium-anchored mean structure is as exposed to the chain-ladder location
error as the free one, and anchoring takes part of the prior out of the
width without moving the centre.

\subsection{What the comparison establishes}\label{sec:scheduleP:established}

The comparisons cover one development pattern, one claim-volume range
and single-frailty Gamma data in the main grid. Procedures differ in
dependence, estimation and inference as well as variance law, so the
results apply to these implementations and calibrations.

Treatment of mean-structure estimation error separated the amount-only
intervals by several times more than the choice among plausible variance
laws. Within the modular construction, switching that treatment off
confirmed its importance for coverage. A correct variance law improved
coverage only when its dispersion was supplied by a prior. Refit on free
levels, it covered less than the linear law; correcting the deflated
linear coefficient removed that deficit
(Section~\ref{supp:simulation:leverage}). With dispersion at the cap it
behaved as the linear law, whereas with a prior-supplied dispersion it
matched the oracles' coverage to within three points.

On real paid triangles, the displacement of the chain-ladder forecast
relative to the outcome limits every variance law. The simulated
calendar-year change produces the same difficulty. Addressing this
shortfall requires work on the mean structure or information beyond the
triangle: a calendar-year-aware mean structure, still untested here,
richer data scopes or external information on the ultimate. Changing the
amount-only variance law alone does not resolve it.

\FloatBarrier
\section{Implications for Macro-Level Reserving}\label{sec:implications}

\paragraph{What must be imported.} A conditional predictive built on the
free-level fit has to take a stance on the realised intensity of every
accident year, and $\kappa_r$ enters only there
(Proposition~\ref{prop:conditional}). Two stances are coherent. The
diffuse one, which the ODP bootstrap takes to first order
\citep{VanOirbeek2026cond}, needs no $\kappa_r$ on any row with a positive observed count and prices the level uncertainty of a row at
$N_i^{\mathrm{obs}}/F_i^2$, which over-states it whenever
$N_i^{\mathrm{obs}} \ge \mu_i^2F_i^2/(\kappa_r + 2\mu_iF_i)$, a threshold
below half the prior expectation $\mu_iF_i$, and under-states it only
below that threshold (Remark~\ref{rem:diffuse}); the statement is exact
for the count triangle and a leading-order one for amounts. The proper one needs a
$\kappa_r$ that the triangle does not supply under free levels, from
anchored levels, from peers or from a prior. The difference between the two
is largest on the thin latest accident years, which carry most of the
reserve, and an informative prior on the levels changes the conditional
law of the reserve itself, its location as well as its width
\citep{VanOirbeek2026cond}.
Section~\ref{sec:cost} measures both stances on simulated and on
Schedule~P data.

\paragraph{What counts add.} The count triangle separates frequency from
severity, supplies the severity split that the second-moment fit leaves
unused, and estimates the cell dispersion with the full cell signal of
the count likelihood \citep{VanOirbeek2026nbcl}. It does not remove the
boundary: with free accident-year parameters it sends $\kappa_r$ to
infinity exactly as the amount fit does (Theorem~\ref{thm:boundary}).

\paragraph{What anchoring adds.} An exposure offset, a common development
rate or a hierarchical prior on the levels opens the row-total channel and
makes $\kappa_r$ estimable. At $I = 10$ that channel is narrow
(Table~\ref{tab:boundary}); on Schedule~P it barely moves the curvature
dispersion $\kappa_B$ of a working fit, to which ten new observations add
little beside the within-row curvature (Table~\ref{tab:kappa_sp};
the separate channel contributions remain unresolved), so a
fit that is to show what anchoring adds needs $\kappa_r$ as a parameter of
its own. Anchoring is the first requirement, without ensuring useful
precision by itself. Peer information on $\kappa_r$, through a prior
fitted across portfolios, is the natural complement.

\paragraph{What to report.} We propose that every amount-only interval be
reported with three statements: (i) its treatment of the estimation error
of the mean structure, held fixed, refitted or drawn; (ii) for every
dispersion it uses, which level of Table~\ref{tab:idmap} that dispersion
reaches, estimated from the triangle, read from anchored levels, imported
from counts or peers, or carried on a prior; and (iii) the conditions under
which its coverage has been checked, triangle size and panel. Readers
need this record to interpret the width: the contributions learned from
the triangle and those assumed are of the same order at $10 \times 10$.

\FloatBarrier
\section{Conclusion}\label{sec:conclusion}

Population identification is not fit recovery. Given the law of the paid
increments, the first moments determine the mean structure, the variance
function determines the severity and heterogeneity composites, and the
covariance of cells in one accident year determines the row dispersion
that governs the non-diversifying part of the reserve variance. With one
free level per accident year that last parameter is lost: the levels
absorb the realised frailty, the profile likelihood prefers homogeneity
on every triangle, in amounts and in counts, and the two-stage
second-moment fit converges to the cap under the replicated free-level
designs of Proposition~\ref{prop:two_stage_target}.

Where a dispersion lives decides whether free levels absorb it. A
dispersion of the row levels enters through the row levels, which the
free fit reproduces and the profile removes with them (for counts with
the pattern known the row total is sufficient for the level; for amounts
the allocation within a row also carries information on the latent
counts, which Theorem~\ref{thm:boundary} shows the profile removes all the
same); a dispersion of the allocation within a row is
carried by the likelihood conditional on the row totals, which the free
levels leave intact. A fitted model cannot estimate a component whose
signal its own parameterisation absorbs.

Even the surviving channels are weak on practical triangles. With cell
means known, the curvature channel gives an asymptotic coefficient of
variation of $0.8$ or more for its dispersion estimate; under free levels,
it recovers cell dispersion only below a threshold. Anchoring provides
$I$ observations for the row channel. The allocation concentration is
overstated by a quarter on ten rows.

Prediction therefore requires explicit information accounting. A
conditional predictive built on the free-level fit either prices the
realised row intensities diffusely, as the ODP bootstrap does, or
imports the row dispersion, and should state which. On a $10 \times 10$
triangle, treatment of mean-structure estimation error separated the
compared procedures more than any other choice. A variance law of the
correct form improved coverage only with dispersion supplied from outside
the free-level fit. On real paid triangles, outcomes sit systematically
below fitted predictive centres. Alongside its width, a reserve interval
needs an information ledger: which components came from the triangle,
which survived the fitted parameterisation, and which were imported.

\FloatBarrier
\section*{Declarations}

\paragraph{Competing interests.} The author declares none.

\paragraph{Use of AI-assisted tools.} AI-assisted coding and language
tools (Anthropic's Claude) were used to support code development,
numerical cross-checking and language editing. All modelling choices,
derivations, interpretations and conclusions are the author's, and the
author takes full responsibility for the manuscript.

\paragraph{Data and code availability.} The CAS Schedule~P loss triangles
are publicly available from the Casualty Actuarial Society. The
procedures of this paper are implemented in the \texttt{hgr} R package
(\url{https://github.com/robin-vo/hgr}) as a port of the scripts behind
its tables. The replication material itself, the R code, seeds, stored
result files and notes recording the R, Stan and package versions and
the script behind each result, from which one command
(\texttt{Rscript R/reproduce\_tables.R}) regenerates every table, is too
large for the package repository and is available from the author on
request.


\section*{Author contact}
Robin Van Oirbeek, University of Antwerp, Department of Mathematics,
Antwerp, Belgium. Email: \texttt{robin.vanoirbeek@gmail.com}.
\clearpage
\setcounter{section}{0}
\renewcommand{\thesection}{S\arabic{section}}
\renewcommand{\theHsection}{S\arabic{section}}
\setcounter{table}{0}
\renewcommand{\thetable}{S\arabic{table}}
\renewcommand{\theHtable}{S\arabic{table}}
\setcounter{equation}{0}
\renewcommand{\theequation}{S\arabic{equation}}
\renewcommand{\theHequation}{S\arabic{equation}}
\setcounter{proposition}{0}
\setcounter{footnote}{0}

\begin{center}
{\LARGE Supplementary Material}\\[6pt]
{\large What Paid Triangles Identify about Reserve Uncertainty}\\[4pt]
Robin Van Oirbeek
\end{center}
\bigskip

\noindent This Supplement holds the higher moments of the CNBG
distribution with their proof (Section~\ref{supp:moments}), the proof of
the precision result with its comparison against the prior and two
further consequences of the identification map
(Section~\ref{supp:precision}), the allocation concentration of a
Dirichlet paid triangle in full (Section~\ref{supp:dirichlet}), further
evidence on the free-level fit from simulation and from Schedule~P
(Section~\ref{supp:evidence}), the computational details of the boundary
simulation (Section~\ref{supp:boundary}), the estimators, the working
likelihood and the exact posterior (Section~\ref{supp:estimation}), the
modular benchmark and the conditional WLS bootstrap in full
(Section~\ref{supp:predictive}), the full simulation comparison
(Section~\ref{supp:simulation}), the Schedule~P details
(Section~\ref{supp:scheduleP}) and further limitations
(Section~\ref{supp:limitations}). Numbering carries the prefix S;
references without it point to the main text.

\section{Higher moments of the CNBG distribution}\label{supp:moments}

\begin{proposition}[Higher central moments \lvl{P}]\label{prop:higher_moments}
Under Definition~\ref{def:cnbg}, with $\nu_{ij} = \mu_{ij}\bar{Y}$,
$A = (1 + 1/\alpha)\bar Y$ and $\bar{Y} = \alpha/\gamma$, the third
and fourth central moments of $X_{ij}$ are
\begin{align}
\E[(X_{ij}-\nu_{ij})^3]
    &= c_3\,\nu_{ij}
       \;+\; 3AB\,\nu_{ij}^{2}
       \;+\; \kappa_3(M)\,\nu_{ij}^{3},
       \qquad
       c_3 = \frac{(\alpha+1)(\alpha+2)}{\alpha^2}\,\bar{Y}^2,
       \label{eq:third_moment}\\[2pt]
  \E[(X_{ij}-\nu_{ij})^4]
    &= 3\bigl(A\nu_{ij} + B\nu_{ij}^{2}\bigr)^{2}
       \;+\; r(\nu_{ij}),
       \label{eq:fourth_moment}
\end{align}
where $M=U_i\varepsilon_{ij}$ is the total cell multiplier, with
$\E[M]=1$ and cumulants
$B \equiv \kappa_2(M)=\kappa_r^{-1}+\kappa_c^{-1}+(\kappa_r\kappa_c)^{-1}$
and
$\kappa_3(M)=2\kappa_r^{-2}+2\kappa_c^{-2}+6(\kappa_r\kappa_c)^{-1}
+6\kappa_r^{-1}\kappa_c^{-2}+6\kappa_r^{-2}\kappa_c^{-1}+4\kappa_r^{-2}\kappa_c^{-2}$
(here $\kappa_n(\cdot)$ denotes a cumulant, not a dispersion), and
$r(\nu_{ij})$ collects strictly positive terms of orders $\nu_{ij}^1$ to
$\nu_{ij}^4$, given in full in \eqref{eq:fourth_full_expanded}. Both
reduce to the single-frailty cumulants
$\kappa_n(U)=(n-1)!\,\kappa_r^{-(n-1)}$ as $\kappa_c\to\infty$.
Conditionally on $M$ the cell is compound Poisson--Gamma, so every
single-cell central moment is the one-level expression with $\kappa_n(U)$
replaced by $\kappa_n(M)$: the row frailty and the cell shock enter a
single cell only through their product and cannot be separated from
single-cell moments of order two, though the cumulants of order three and
four depend on $\kappa_r$ and $\kappa_c$ separately. In particular, every
coefficient of \eqref{eq:fourth_moment} is positive, so
$\E[(X_{ij}-\nu_{ij})^4] \ge 3(A\nu_{ij} + B\nu_{ij}^2)^2$ and the squared
coefficient of variation of the squared residual used in
Proposition~\ref{prop:godambe} satisfies $\mathrm{CV}^2(z_{ij}) \ge 2$.
\end{proposition}

\subsection{Proof of Proposition~\ref{prop:higher_moments}}
\label{app:cnbg_moments}

We derive the third and fourth central marginal moments of $X_{ij}$
under the CNBG generative model of Definition~\ref{def:cnbg}. The
derivation uses iterated cumulant calculations through the hierarchy
$(\lambda_i, N_{ij}, Y_{ijk})$. Throughout we suppress the subscripts
$(i,j)$ on $X$, $N$, and $\nu$ where they are not needed.

\subsubsection{Setup and conditional cumulants}\label{app:moments:setup}

Recall the structural hierarchy:
\begin{itemize}[itemsep=0pt]
\item $U = \lambda_i/\mu_i \sim \mathrm{Gamma}(\kappa_r, \kappa_r)$
(row frailty, shared across the row) and
$\varepsilon \sim \mathrm{Gamma}(\kappa_c, \kappa_c)$ (cell shock,
independent across cells), with $\mathbb{E}[U]=\mathbb{E}[\varepsilon]=1$.
Write $M = U\varepsilon$ for the total cell multiplier; $\mathbb{E}[M]=1$.
\item $N \mid U,\varepsilon \sim \mathrm{Poisson}(\mu w U\varepsilon)
= \mathrm{Poisson}(\nu M / \bar{Y})$ since $\nu = \mu w \bar{Y}$.
\item $Y_k \stackrel{\mathrm{iid}}{\sim} \mathrm{Gamma}(\alpha,
\gamma)$ independent of $N$, with $\bar{Y} = \alpha/\gamma$.
\item For a \emph{single} cell, $U$ and $\varepsilon$ enter only through
their product $M$; the split between $\kappa_r$ and $\kappa_c$ is invisible
to the single-cell \emph{second} moment, which depends on them only
through $B$, but not to the higher single-cell cumulants
\eqref{eq:kappa3_M}--\eqref{eq:kappa4_M}, which depend on $a$ and $b$
separately, nor to the cross-cell covariance, where
$U$ is shared and $\varepsilon$ is not.
\end{itemize}

The moments of the Gamma severity (shape $\alpha$, rate $\gamma$)
are
\begin{equation}\label{eq:gamma_moments}
\mathbb{E}[Y] = \bar{Y}, \quad
\mathbb{E}[Y^2] = \bar{Y}^2 \cdot \frac{\alpha+1}{\alpha}, \quad
\mathbb{E}[Y^3] = \bar{Y}^3 \cdot \frac{(\alpha+1)(\alpha+2)}{\alpha^2}.
\end{equation}
The central moments of $Y$ are $\mathbb{E}[(Y-\bar{Y})^2] =
\bar{Y}^2/\alpha$, $\mathbb{E}[(Y-\bar{Y})^3] = 2\bar{Y}^3/\alpha^2$,
$\mathbb{E}[(Y-\bar{Y})^4] = 3\bar{Y}^4(\alpha+2)/\alpha^3$.

\subsubsection{The compound Poisson cumulants}\label{app:moments:cp}

Conditional on $M$, $X = \sum_{k=1}^{N} Y_k$ is a compound Poisson
sum with Poisson rate $\nu M/\bar{Y}$ and Gamma severity, so
\begin{equation}\label{eq:cp_cumulants}
\kappa_n(X \mid M) \;=\; \frac{\nu M}{\bar{Y}}\, \mathbb{E}[Y^n].
\end{equation}
Using \eqref{eq:gamma_moments}, the first four conditional cumulants of
$X$ given $M$ are
\begin{align}
\kappa_1(X \mid M) &\;=\; \nu M, \label{eq:k1_cond}\\
\kappa_2(X \mid M) &\;=\; A\, \nu M, \label{eq:k2_cond}\\
\kappa_3(X \mid M) &\;=\; c_3\, \nu M, \label{eq:k3_cond}\\
\kappa_4(X \mid M) &\;=\; c_4\, \nu M, \label{eq:k4_cond}
\end{align}
with $A = (1 + 1/\alpha)\bar Y$,
$c_3 = \bar{Y}^2(\alpha+1)(\alpha+2)/\alpha^2$, and
$c_4 = \bar{Y}^3(\alpha+1)(\alpha+2)(\alpha+3)/\alpha^3$. The severity
constants are unchanged from the single-frailty case; only the mixing
variable is now $M$ rather than $U$.

\subsubsection{Marginal cumulants via the law of total cumulants}
\label{app:moments:lotc}

To pass from conditional cumulants given $M$ to marginal cumulants
of $X$, we use the law of total cumulants. For $n = 2$ this is the
familiar law of total variance:
\begin{equation*}
\kappa_2(X) \;=\; \mathbb{E}[\kappa_2(X \mid M)]
\;+\; \Var(\kappa_1(X \mid M)).
\end{equation*}
The corresponding identities for $n = 3, 4$ are well-established
\citep{BrillingerCumulants1969}:
\begin{align}
\kappa_3(X) &\;=\; \mathbb{E}[\kappa_3(X \mid M)]
  \;+\; 3\,\Cov(\kappa_1(X \mid M), \kappa_2(X \mid M))
  \;+\; \kappa_3(\kappa_1(X \mid M)),
\label{eq:total_kappa3}\\
\kappa_4(X) &\;=\; \mathbb{E}[\kappa_4(X \mid M)]
  \;+\; 4\,\Cov(\kappa_1(X \mid M), \kappa_3(X \mid M))
  \;+\; 3\,\Var(\kappa_2(X \mid M))\nonumber\\
  &\qquad+ \; 6\,\Cov_2(\kappa_1, \kappa_2)
  \;+\; \kappa_4(\kappa_1(X \mid M)),
\label{eq:total_kappa4}
\end{align}
where in \eqref{eq:total_kappa4} the term $\Cov_2(\kappa_1, \kappa_2)$
is the joint third cumulant $\kappa(\kappa_1, \kappa_1, \kappa_2)$ of
the conditional cumulants. Because all conditional cumulants
\eqref{eq:k1_cond}--\eqref{eq:k4_cond} are linear in $M$, it reduces to
a multiple of $\kappa_3(M)$ and does \emph{not} vanish; it is evaluated
in Section~\ref{app:moments:fourth}. (The identities
\eqref{eq:total_kappa3}--\eqref{eq:total_kappa4} are the
Fa\`a di Bruno expansion of $K_X(t) = K_M(h(t))$ with
$h(t) = (\nu/\bar Y)(M_Y(t)-1)$, which gives
$\kappa_4(X) = h'''' + \kappa_2(M)[4h'h''' + 3(h'')^2]
+ 6\kappa_3(M)(h')^2 h'' + \kappa_4(M)(h')^4$ at $t = 0$.)

\subsubsection{Cumulants of the cell multiplier $M = U\varepsilon$}
\label{app:moments:M_cumulants}

Since $U$ and $\varepsilon$ are independent unit-mean Gammas, the raw
moments of $M$ factorise. For a unit-mean $\mathrm{Gamma}(k,k)$ variate,
$\mathbb{E}[V^j] = \prod_{l=1}^{j-1}(1 + l/k) \equiv g_j(k)$, so
\begin{equation}\label{eq:M_raw_moments}
  \mathbb{E}[M^j] \;=\; g_j(\kappa_r)\,g_j(\kappa_c),
  \qquad
  g_2(k)=1+\tfrac1k,\;\;
  g_3(k)=\bigl(1+\tfrac1k\bigr)\bigl(1+\tfrac2k\bigr),\;\;
  g_4(k)=\bigl(1+\tfrac1k\bigr)\bigl(1+\tfrac2k\bigr)\bigl(1+\tfrac3k\bigr).
\end{equation}
Converting raw to central moments via
$\kappa_2 = \mathbb{E}[M^2]-1$,
$\kappa_3 = \mathbb{E}[M^3]-3\mathbb{E}[M^2]+2$, and
$\kappa_4 = \mathbb{E}[M^4]-4\mathbb{E}[M^3]-3\mathbb{E}[M^2]^2
+12\mathbb{E}[M^2]-6$, and writing $a = 1/\kappa_r$, $b = 1/\kappa_c$,
\begin{align}
\kappa_2(M) &= a + b + ab \;\equiv\; B, \label{eq:kappa2_M}\\
\kappa_3(M) &= 2a^2 + 2b^2 + 6ab + 6ab^2 + 6a^2b + 4a^2b^2,
  \label{eq:kappa3_M}\\
\kappa_4(M) &= 6a^3 + 6b^3 + 36a^2b + 36ab^2 + 102a^2b^2
  + 36a^3b + 36ab^3 + 66a^3b^2 + 66a^2b^3 + 36a^3b^3.
  \label{eq:kappa4_M}
\end{align}
As $\kappa_c \to \infty$ ($b \to 0$) these collapse to the single-frailty
cumulants $\kappa_n(U) = (n-1)!\,a^{\,n-1}$, namely
$\kappa_2 \to a$, $\kappa_3 \to 2a^2$, $\kappa_4 \to 6a^3$, recovering the
row-only model; the symmetric limit $a\to 0$ gives the pure cell-shock
model. We will need \eqref{eq:kappa2_M}--\eqref{eq:kappa4_M} in
\eqref{eq:total_kappa3}--\eqref{eq:total_kappa4}.

\subsubsection{Third marginal cumulant}\label{app:moments:third}

All conditional cumulants \eqref{eq:k1_cond}--\eqref{eq:k4_cond}
are of the form $c_n \cdot \nu M$. Hence
\begin{align*}
\mathbb{E}[\kappa_3(X \mid M)] &\;=\; c_3\nu, \\
\Cov(\kappa_1(X \mid M), \kappa_2(X \mid M))
  &\;=\; A\nu^2 \,\Var(M) \;=\; A B\,\nu^2, \\
\kappa_3(\kappa_1(X \mid M)) &\;=\; \nu^3\,\kappa_3(M).
\end{align*}
Substituting into \eqref{eq:total_kappa3}:
\begin{equation}\label{eq:third_central_full}
\mathbb{E}[(X - \nu)^3]
  \;=\; \kappa_3(X)
\;=\; c_3\nu \;+\; 3AB\,\nu^2 \;+\; \kappa_3(M)\,\nu^3,
\end{equation}
with $B$ and $\kappa_3(M)$ from \eqref{eq:kappa2_M}--\eqref{eq:kappa3_M}.
At $\kappa_c\to\infty$ this is the single-frailty
$c_3\nu + 3A\kappa_r^{-1}\nu^2 + 2\kappa_r^{-2}\nu^3$.

\subsubsection{Fourth marginal moment}\label{app:moments:fourth}

For the fourth moment, we work with the central moment directly
(the fourth cumulant differs by $3 \kappa_2^2$):
\begin{equation}\label{eq:fourth_central_decomp}
\mathbb{E}[(X-\nu)^4] \;=\; \kappa_4(X) \;+\; 3 \kappa_2(X)^2.
\end{equation}
We need both terms. For the $3\kappa_2(X)^2$ term, Proposition
\ref{prop:moments} gives $\kappa_2(X) = A\nu + B\nu^2$, so
\begin{equation}\label{eq:three_var_squared}
3 \kappa_2(X)^2 \;=\; 3 (A\nu + B\nu^2)^2.
\end{equation}
For the $\kappa_4(X)$ term, applying
\eqref{eq:total_kappa4}:
\begin{align*}
\mathbb{E}[\kappa_4(X \mid M)] &\;=\; c_4\nu,\\
\Cov(\kappa_1(X \mid M), \kappa_3(X \mid M))
  &\;=\; c_3\nu^2 \,\Var(M) \;=\; c_3 B\,\nu^2,\\
\Var(\kappa_2(X \mid M)) &\;=\; (A\nu)^2 \Var(M)
  \;=\; A^2 B\,\nu^2,\\
\kappa_4(\kappa_1(X \mid M)) &\;=\; \nu^4\,\kappa_4(M).
\end{align*}
The $\Cov_2$ term is the third cumulant of the jointly-linear pair
$\kappa_1(X \mid M) = \nu M$, $\kappa_2(X \mid M) = A\nu M$:
\begin{equation*}
\Cov_2(\kappa_1, \kappa_2) \;=\; A\nu^3 \,\kappa_3(M).
\end{equation*}
Combining all terms:
\begin{equation}\label{eq:fourth_cumulant_full}
\kappa_4(X) \;=\; c_4\nu
  \;+\; (4c_3 + 3A^2)\,B\,\nu^2
  \;+\; 6A\,\kappa_3(M)\,\nu^3
  \;+\; \kappa_4(M)\,\nu^4.
\end{equation}
Combining these, substituting
\eqref{eq:three_var_squared} and \eqref{eq:fourth_cumulant_full}
into \eqref{eq:fourth_central_decomp} gives
\begin{equation}\label{eq:fourth_full_expanded}
\mathbb{E}[(X-\nu)^4]
\;=\; 3\bigl(A\nu + B\nu^2\bigr)^2
  \;+\; c_4\nu \;+\; (4c_3 + 3A^2)\,B\,\nu^2
  \;+\; 6A\,\kappa_3(M)\,\nu^3 \;+\; \kappa_4(M)\,\nu^4,
\end{equation}
with $B,\kappa_3(M),\kappa_4(M)$ from
\eqref{eq:kappa2_M}--\eqref{eq:kappa4_M}. Every coefficient is strictly
positive, so $\mathbb{E}[(X-\nu)^4] \ge 3(A\nu+B\nu^2)^2 = 3\mu_z^2$,
giving $\mathrm{CV}^2(z) \ge 2$, a floor that the cell-shock layer
reinforces rather than weakens.
The term $3(A\nu + B\nu^2)^2$ is the Gaussian-kurtosis square of the
variance; the remaining terms are $O(\nu) + O(\nu^2) + O(\nu^3) +
O(\nu^4)$ with strictly positive coefficients. Two of these are
genuinely lower-order in $\nu$ and vanish in relative terms as $\nu$
grows: the $c_4\nu$ severity term and the $(4c_3+3A^2)B\nu^2$ term are
dominated by $3A^2\nu^2$ and $3B^2\nu^4$ respectively for large cells.
The $\nu^4$ terms, however, are the \emph{same} order in $\nu$ and must
be compared by coefficient: the squared-variance contribution
$3(B\nu^2)^2 = 3B^2\nu^4$ exceeds the kurtosis contribution
$\kappa_4(M)\nu^4$ if and only if $3B^2 \ge \kappa_4(M)$. This holds for
mild-to-moderate heterogeneity ($3B^2/\kappa_4(M)$ exceeds one for
$\kappa_r,\kappa_c \gtrsim 8$ on the diagonal, and across most of the
$\{3,5,10,20,50\}^2$ grid) but \emph{fails} in the strong
double-heterogeneity corner: at $\kappa_r=\kappa_c=3$ the ratio is
$0.31$, so the kurtosis term is roughly three times the squared-variance
term. The fourth moment is therefore not dominated by a single term
throughout the parameter space, and the compact form
\eqref{eq:fourth_moment} reports $3(A\nu+B\nu^2)^2$ as the leading term
only in the sense of the variance-square anchor, with $r(\nu_{ij})$
collecting strictly positive corrections that are themselves
$O(\nu^4)$-competitive at small $\kappa_B$. This does not affect the
floor $\mathrm{CV}^2(z) \ge 2$,
which needs only $\E[(X-\nu)^4] \ge 3\mu_z^2$ and holds with
all corrections positive regardless of which $\nu^4$ term is larger. The compact form
\eqref{eq:fourth_moment} of Proposition~\ref{prop:higher_moments}
reports the variance-square anchor plus the remainder $r(\nu)$, which
collects the strictly positive terms of orders $\nu^1$ through $\nu^4$
displayed in \eqref{eq:fourth_full_expanded}, completing the proof.
\hfill$\Box$

\subsection{What a fitted Tweedie power measures}\label{supp:moments:tweedie}

The elasticity of the true variance function gives the power a Tweedie
fit tracks.

\begin{proposition}[Locally optimal Tweedie power \lvl{P}]\label{prop:local_p}
The power of the Tweedie variance function that matches the variance
function $A\nu + B\nu^2$ locally around a mean $\nu_0$ is
\begin{equation}\label{eq:p_local}
    p^*(\nu_0) = 1 + \frac{B\nu_0}{A + B\nu_0} \in (1,2).
\end{equation}
\end{proposition}

\begin{proof}
The locally optimal power is the elasticity $p^*(\nu) = \nu V'(\nu)/V(\nu)$;
for $V(\nu) = A\nu + B\nu^2$ it is $(A + 2B\nu)/(A + B\nu)
= 1 + B\nu/(A + B\nu)$.
\end{proof}

``Locally optimal'' is a statement about the variance function alone,
the power whose log-slope matches the true one at $\nu_0$; it is not the
power that a Tweedie likelihood selects on a triangle, which depends on
the spread of the cell means and on the distribution of the cells and
not only on their variance. The corollary below reads $p^*$ as the
quantity a fitted power tracks, not as its expectation.

\begin{corollary}[$\hat p$ is not a structural constant]\label{cor:p_not_structural}
$p^*$ increases with the claim volume $\mu_0 = \nu_0/\bar Y$ at fixed
$(A, B)$, and is invariant to the monetary unit, since a rescaling
multiplies $\nu_0$ and $A$ alike and leaves $B$ unchanged; it increases
with $B$, so more heterogeneity of either kind pushes it towards $2$; it
increases with $\alpha$ at fixed $(\bar Y, \nu_0, B)$, so heavier severity
dispersion pushes it towards $1$; and it tends to $1$ as
$\kappa_B \to \infty$ and to $2$ as $\kappa_B \to 0$.
\end{corollary}

Two portfolios with the same structural parameters and different claim
volumes can therefore show different $\hat p$, since the local elasticity
moves with the cell means and what a fitted power settles on depends on
the whole triangle, and two portfolios with the
same $\hat p$ can have entirely different $(\kappa_B, \alpha)$. Comparing
$\hat p$ across portfolios or lines without accounting for volume is
uninformative, and a test of $\hat p$ against $1$ or $2$ has power against
heterogeneity and against severity alternatives without telling them
apart. The structural pair $(B, \alpha)$ separates what $\hat p$ merges;
what the triangle recovers of that pair is the subject of the next three
sections.

\section{Precision of the curvature channel: proof, the prior, and two consequences}
\label{supp:precision}

\subsection{Proof of Proposition~\ref{prop:godambe}}
\begin{proof}[Proof of Proposition~\ref{prop:godambe}]
\textit{(i)} The score for $\kappa_B$ of the squared-residual estimating
equation is $\sum_{ij}(\partial\mu_z^{ij}/\partial\kappa_B)
(z_{ij}-\mu_z^{ij})/\Var(z_{ij})$, where
$\partial\mu_z^{ij}/\partial\kappa_B = -\nu_{ij}^2/\kappa_B^2$ and
$\Var(z_{ij}) = \mathrm{CV}^2(z_{ij})\,(\mu_z^{ij})^2$. The
working-independence information is the sum of the cell-wise score
variances, which is \eqref{eq:fisher_kappa}. The $O(\kappa_B^{-2})$ rate is
the heterogeneity-dominated limit $\kappa_B \ll \nu_{ij}/A$, in which
$\mu_z^{ij} \approx \nu_{ij}^2/\kappa_B$; the rate steepens towards
$\kappa_B^{-4}$ in the opposite limit $\kappa_B \gg \nu_{ij}/A$, where the
heterogeneity term of the reserve variance is negligible
(Proposition~\ref{prop:reserve_variance}). At the calibration of
Table~\ref{tab:godambe} the observed cells straddle the two regimes
($\nu_{ij}/A$ has median $3.0$ and range $0.1$--$12$ over the $55$ cells at
$\alpha = 3$), and over $\kappa_B \in [3, 50]$ the information
$\mathcal{I}_{\mathrm{diag}}$ decays with a fitted log--log slope between
$-3.0$ and $-3.2$ across $\alpha \in \{1, 3, 10\}$ (between $-2.6$ and
$-3.0$ for $G$).

\textit{(ii)} Cells in different accident years are independent. Within
accident year $i$, cells $(i,j)$ and $(i,j')$ share the frailty $\lambda_i$,
which induces a positive within-row covariance of their scores. Numerical
evaluation at the same calibration gives
$\Delta(\kappa_B)/\mathcal{I}_{\mathrm{diag}}(\kappa_B) \in [0.03, 1.67]$
across the $(\kappa_B, \alpha)$ grid: the cross terms range from negligible
to exceeding the diagonal information itself. The Godambe form
\eqref{eq:godambe} is the standard sandwich precision of an
estimating-equation estimator \citep{Lindsay1988, VarinReidFirth2011}. When
$A$ is estimated jointly, the relevant quantity is the $\kappa_B$ entry of
the $2 \times 2$ sandwich, which is no larger.
\end{proof}

\paragraph{Scope of the benchmark.} Proposition~\ref{prop:godambe}
gives the precision of the diagonally weighted equation with cell means
and $A$ known. With correlated squared residuals, the optimal linear
equation uses the full row-block covariance. Part (ii) therefore gives
the diagonal equation's own sandwich precision, rather than an upper
bound for all correlated equations. Table~\ref{tab:godambe_block}
computes the optimum and finds it within $11\%$. The full likelihood has
additional dispersion information in zero cells, higher cumulants and
within-row dependence. The operational procedure also estimates the
means and $A$, both held fixed in this calculation.

\paragraph{The weighting.} The WLS \eqref{eq:wls}, which weights by $\hat\nu_{ij}^{-2}$, solves a
different equation and is less precise, but not by much: under working
independence at $\alpha = 3$ its asymptotic coefficient of variation is
$0.59$ against $0.52$ for the optimally weighted equation at
$\kappa_B = 3$, and the two coincide from $\kappa_B = 20$
(Table~\ref{tab:godambe_wls}). The diagonal weighting is itself not
optimal once the within-row correlation of the squared residuals is
admitted: among all estimating equations linear in the squared residuals
the optimal one weights each accident year's residual vector by the
inverse of its full covariance matrix, $\sum_i D_i^\top V_i^{-1}(z_i -
\mu_z^i) = 0$ with $D_i = \partial\mu_z^i/\partial\kappa_B$ and $V_i =
\Var(z_i)$, and its information $\mathcal{I}_{\mathrm{block}} = \sum_i
D_i^\top V_i^{-1}D_i$ is the largest attainable by such an equation and
exceeds the Godambe information $G$ of the diagonal one
\citep{Godambe1960, Lindsay1988}. Table~\ref{tab:godambe_block} computes
both at the calibration of Table~\ref{tab:godambe}, with $V_i$ from
$2 \times 10^5$ simulated rows per accident year; the row-block optimum
improves the coefficient of variation from $0.81$ to $0.77$ at
$\kappa_r = 3$, by $3\%$ at $\kappa_r = 10$ and by nothing from
$\kappa_r = 20$, so the diagonal equation, with its sandwich precision, is
within $11\%$ of the best linear equation in the squared residuals.
The weak curvature signal persists under the better weighting.

\begin{table}[H]
\centering
\caption{The optimally weighted squared-residual equation against the WLS
\eqref{eq:wls} with weights $\hat\nu_{ij}^{-2}$: working-independence
information of the former and the asymptotic coefficient of variation of
the estimate of $\kappa_B$ under each, at the calibration of
Table~\ref{tab:godambe} with $\alpha = 3$ and the cell means known.}
\label{tab:godambe_wls}
\small
\begin{tabular}{@{}rrrr@{}}
\toprule
$\kappa_B$ & $\mathcal{I}_{\mathrm{diag}}(\kappa_B)$ & $\mathrm{SD}/\kappa_B$, optimal weights & $\mathrm{SD}/\kappa_B$, WLS \eqref{eq:wls} \\
\midrule
3 & 0.415 & 0.52 & 0.59 \\
5 & 0.123 & 0.57 & 0.62 \\
10 & 0.018 & 0.74 & 0.76 \\
20 & 0.002 & 1.10 & 1.11 \\
50 & $8.3 \times 10^{-5}$ & 2.20 & 2.20 \\
\bottomrule
\end{tabular}

\end{table}

\begin{table}[H]
\centering
\caption{The optimal row-block estimating equation against the diagonal
one at the calibration of Table~\ref{tab:godambe} ($\alpha = 3$, cell
means and $A$ known): the ratio of the within-row score cross-covariances
to the diagonal information, the Godambe information $G$ of the diagonal
equation, the information $\mathcal{I}_{\mathrm{block}}$ of the row-block
equation, and the asymptotic coefficients of variation of the two
estimates of $\kappa_B$; within-row covariances of the squared residuals
by Monte Carlo, $2 \times 10^5$ rows per accident year.}
\label{tab:godambe_block}
\small
\begin{tabular}{@{}rccccc@{}}
\toprule
$\kappa_r$ & $\Delta/\mathcal{I}_{\mathrm{diag}}$ & $G$ (diagonal) & $\mathcal{I}_{\mathrm{block}}$ & CV, diagonal & CV, row-block \\
\midrule
3 & 1.45 & 1.70e-01 & 1.90e-01 & 0.81 & 0.77 \\
5 & 0.93 & 6.39e-02 & 6.84e-02 & 0.79 & 0.76 \\
10 & 0.44 & 1.29e-02 & 1.33e-02 & 0.88 & 0.87 \\
20 & 0.17 & 1.77e-03 & 1.78e-03 & 1.19 & 1.18 \\
50 & 0.05 & 7.89e-05 & 7.90e-05 & 2.25 & 2.25 \\
\bottomrule
\end{tabular}

\end{table}

\subsection{The curvature channel against the prior}

The working likelihood of Section~\ref{sec:predictive} places a $\mathrm{Gamma}(a, b) = \mathrm{Gamma}(2, 0.4)$ reference prior on
$\kappa_B$. Two scalar summaries of the prior's precision can be set
against $G(\kappa_B)$. The local curvature of the log prior at the true
value is $-\partial^2\log\pi/\partial\kappa_B^2 = (a-1)/\kappa_B^2 =
1/\kappa_B^2$. The expected squared prior score,
$\mathcal{I}_\pi = \E_\pi[((a-1)/\kappa_B - b)^2] = b^2/(a-2)$ for $a > 2$,
diverges at $a = 2$; under the fit-stage truncation $\kappa_B \ge 0.1$ of the
implementation it is $\mathcal{I}_\pi \approx 0.28$, a value sensitive to
the truncation at the second decimal. Table~\ref{tab:godambe_prior} reports
the ratio $\mathcal{I}/(G + \mathcal{I})$ for both.

\begin{table}[ht]
\centering
\caption{The Godambe information of Table~\ref{tab:godambe} against two
summaries of the precision of the reference prior $\mathrm{Gamma}(2,
0.4)$: its local curvature $1/\kappa_B^2$ at the true value and its expected
squared score under the truncation $\kappa_B \ge 0.1$,
$\mathcal{I}_\pi = 0.28$. Ratio $= \mathcal{I}/(G + \mathcal{I})$.}
\label{tab:godambe_prior}
\small
\begin{tabular}{@{}rrrrr@{}}
\toprule
$\kappa_B$ & $G$ (Godambe) & $1/\kappa_B^2$ & ratio (local curvature) & ratio ($\mathcal{I}_\pi = 0.28$) \\
\midrule
3 & 0.164 & 0.111 & 40\% & 63\% \\
5 & 0.061 & 0.040 & 40\% & 82\% \\
10 & 0.012 & 0.010 & 45\% & 96\% \\
20 & 0.0017 & 0.0025 & 60\% & 99.4\% \\
50 & $7.5 \times 10^{-5}$ & 0.0004 & 84\% & 99.97\% \\
\bottomrule
\end{tabular}

\end{table}

The ratio compares information magnitudes. Interpreting it as a share
of posterior precision would require a Laplace calculation using local
posterior curvature in $\kappa_B$ with $A$ as a nuisance parameter.
The two summaries disagree by up to a factor of $30$. Read with
the local curvature, the data match the prior at roughly $0.7$ triangles'
worth of cells at $\kappa_B = 3$ and $5$ triangles' worth at
$\kappa_B = 50$; read with the expected squared score, at $1.7$ and
$3{,}700$. A van Trees bound cannot be obtained from $G$: it requires
the full likelihood's Fisher information averaged over the prior,
whereas this Godambe information is evaluated pointwise for one estimating
equation. The reference prior also violates the boundary condition.
The practical evidence on prior dependence comes from
Section~\ref{supp:evidence:priorsens}.

\subsection{Importing a cell dispersion from counts, and frequency--severity dependence}

If a count triangle is available, the cell dispersion $\kappa_c$ can be
estimated from it with the Negative Binomial chain ladder
\citep{VanOirbeek2026nbcl}, typically with higher efficiency, since the
count likelihood carries $\kappa_c$ directly rather than through the second
moment of a compound sum. The estimate can be imported into the amount
analysis as a fixed constraint, reducing the second-stage regression to
one free parameter; under free levels this is coherent, since the
recoverable part of $B$ is then $1/\kappa_c$ in any case
(Proposition~\ref{prop:two_kappa_id}(ii)). It does not supply the row
dispersion $\kappa_r$, which free accident-year parameters send to the
boundary in the count fit as in the amount fit, and which is read, under
anchored levels, from the row totals of whichever triangle is used.

The CNBG model takes individual claim sizes $Y_k \overset{iid}{\sim}
\mathrm{Gamma}(\alpha, \gamma)$ independent of the accident-year
frailty. This independence is embedded in the model structure and cannot
be relaxed within the CNBG framework. Suppose instead that the mean claim
size of accident year $i$ scales with the normalised frailty as
$U_i^{\eta}$ for some $\eta \ge 0$, so that the cell mean given $U_i$ is
$m_{ij}(U_i) = \nu_{ij}U_i^{1+\eta}$ and $\eta = 0$ recovers CNBG.
Integrating out the cell shock, the cell variance given $U_i$ is
$A\,U_i^{\eta}\,m_{ij}(U_i) + m_{ij}(U_i)^2/\kappa_c$: the quadratic
coefficient given the row is still $1/\kappa_c$, and the dependence makes
the linear coefficient row-dependent and the realised row level
$U_i^{1+\eta}$ rather than $U_i$. Under free levels the row level absorbs
$U_i^{1+\eta}$ as it absorbs $U_i$, but a second stage with one common
linear coefficient folds the row-dependent term $A\,U_i^{\eta}$ into its
slope: with the realised means known, the population slope of the
regression of $z_{ij}/m_{ij}$ on $m_{ij}$ is
$1/\kappa_c + A\,\Cov(U_i^{\eta}, m_{ij})/\Var(m_{ij})$, which at
$\eta = 1$, $\kappa_r = 5$, $\kappa_c = 5$ and $A = 1$ on cells of mean
$10U_i^2$ is $0.238$ against $0.2$, an apparent cell dispersion of $4.2$
for a true $5$. The import of the count estimate is therefore coherent
only if the amount variance model carries the row-dependent linear
coefficient. Under anchored levels the row-total channel reads the
dispersion of $U_i^{1+\eta}$, whose squared coefficient of variation is
about $(1+\eta)^2/\kappa_r$ for small $1/\kappa_r$, and would report a row
dispersion near $\kappa_r/(1+\eta)^2$. The second-moment procedure
identifies neither $\eta$ nor $\alpha$ separately (the population law
identifies $\alpha$ through the third cumulant,
Corollary~\ref{cor:cumulant}); a dependence of this kind needs individual
claim data or a joint frequency--severity model, and no diagnostic for it
is offered here.

\subsection{Testable implication: the variance linearity check}

Proposition~\ref{prop:identification} implies a testable structural
restriction. Given estimates $(\hat{A}, \hat{B})$, define the
\emph{adjusted squared residual} for cell $(i,j)$ as
\begin{equation}\label{eq:adj_resid}
    r_{ij} = \frac{(X_{ij} - \hat{\nu}_{ij})^2}{1 - h_{ij}} - \hat{B}\,\hat{\nu}_{ij}^2,
\end{equation}
with $h_{ij}$ the leverage of the cell in the Stage-1 fit. Under the CNBG
model, $\E[r_{ij}] \approx A\nu_{ij}$, the leverage factor undoing the
first-stage deflation of Proposition~\ref{prop:two_stage_target}(ii):
without it the slope is deflated by $1 - \bar s$, $0.82$ with the pattern
known at the design of Section~\ref{sec:illus:boundary}, and by about
$1 - q/n = 0.65$ on a $10 \times 10$ triangle with both effects
estimated. A plot of $r_{ij}$ against $\hat{\nu}_{ij}$ should then be
linear through the origin with positive slope $A$. Systematic curvature, a negative estimated slope or a non-zero
intercept signal that the variance function is not of the form
$A\nu + B\nu^2$: that the heterogeneity is not multiplicative with a
constant dispersion, or that the severity variance is not proportional
to the mean. The check tests the second-moment structure only. It cannot
identify the Gamma assumptions of Definition~\ref{def:cnbg}, since any
severity law with the same first two moments and any unit-mean frailty
law with the same variance give the same variance function and pass it.

\section{The allocation concentration of a Dirichlet paid triangle}
\label{supp:dirichlet}

This section holds the proof of Proposition~\ref{prop:dirichlet_c}, the
moment estimator of the concentration with its asymptotic variance and
finite-sample behaviour, the study behind Table~\ref{tab:dirichlet_c},
the integration over $c$ and the cost of substituting it, and what the
estimate reports about a portfolio, including its limit under compound
Poisson data. The numbers come from the scripts in \texttt{multinom/R}
of the replication package, and the tables are written by
\texttt{R/make\_tables\_multinom.R}.

\subsection{Aggregation, sub-composition and the proof of Proposition~\ref{prop:dirichlet_c}}

\begin{lemma}[Aggregation and sub-composition]\label{lem:dirichlet_subcomp}
Let $F_k = \sum_{l \le k} w_l$, and let $V = (V_1, \ldots, V_J)$ have
the law $\mathrm{Dirichlet}(c w_1, \ldots, c w_J)$.
\begin{enumerate}[label=(\roman*), nosep]
\item $\sum_{l \le k} V_l \sim \mathrm{Beta}\bigl(cF_k,\, c(1-F_k)\bigr)$.
\item The sub-composition $(V_1, \ldots, V_k)/\sum_{l \le k} V_l$ is
$\mathrm{Dirichlet}(c w_1, \ldots, c w_k)$ and independent of
$\sum_{l \le k} V_l$. In particular each component is
$\mathrm{Beta}\bigl(cF_k w_j^{(k)},\, cF_k(1 - w_j^{(k)})\bigr)$ with
$w_j^{(k)} = w_j/F_k$: its concentration is $cF_k$, not $c$.
\end{enumerate}
\end{lemma}

\begin{proof}
Both are standard properties of the Dirichlet distribution
\citep{aitchison1982}. Write $V_l = G_l/\sum_m G_m$ with independent
$G_l \sim \mathrm{Gamma}(c w_l, 1)$, and $G_{\le k}$ and $G_{>k}$ for the
sums over $l \le k$ and $l > k$, independent Gamma variables of shapes
$cF_k$ and $c(1-F_k)$. Then
\[
\sum_{l \le k} V_l = \frac{G_{\le k}}{G_{\le k} + G_{>k}},
\qquad
\frac{(V_1, \ldots, V_k)}{\sum_{l \le k} V_l}
 = \frac{(G_1, \ldots, G_k)}{G_{\le k}} .
\]
The first is $\mathrm{Beta}(cF_k, c(1-F_k))$, which is (i). The second is
$\mathrm{Dirichlet}(cw_1, \ldots, cw_k)$ and independent of $G_{\le k}$ by
the Gamma--Dirichlet factorisation, and as a function of
$(G_1, \ldots, G_k)$ it is independent of $G_{>k}$; it is therefore
independent of the first, which is (ii). The Beta marginal is (i) applied
within the sub-composition, whose total concentration is $cF_k$.
\end{proof}

\begin{proof}[Proof of Proposition~\ref{prop:dirichlet_c}]
(i) The allocation vector of row $i$ is independent of $S_i$, and its
observed part, rescaled to its own total, is the sub-composition of
Lemma~\ref{lem:dirichlet_subcomp}(ii) with $k = k_i$; it is
$\mathrm{Dirichlet}(c w_1, \ldots, c w_{k_i})$ and independent of $S_i$ and
of the observed share $x_i/S_i$. The observed row is in one-to-one
correspondence with $(x_i, v_i)$, so the density of the triangle factors
into the density of the sub-compositions, whose logarithm is
\eqref{eq:dirichlet_condlik}, times the density of the row totals, which
alone involves the levels. Identification: the first accident year is
fully developed when $J \le I$, and its composition
$\mathrm{Dirichlet}(c w_1, \ldots, c w_J)$ identifies $c w$, hence $c$ since
$\sum_j w_j = 1$. Each term of \eqref{eq:dirichlet_condlik} is the
log-density of a regular exponential family in $(cw_1, \ldots, cw_{k_i})$,
and for fixed $J$ the number of accident years observed through every
horizon grows with $I$, so the maximiser is consistent by the standard
argument for M-estimators of independent, regular exponential-family
terms.
(ii) The proportions are the components of the horizon-$k$
sub-composition, so Lemma~\ref{lem:dirichlet_subcomp}(ii) gives their law.
The moment estimator of Definition~\ref{def:c_moment} below reads the
cross-row variance $w^{(k)}(1-w^{(k)})/(cF_k + 1)$ as that of a
concentration $a$, so its bracketed quantity estimates $cF_k$; division
by $\hat F_k$ makes every cell estimator converge to $c$, and the median
of homogeneous estimators converges to the common limit. Without the
division the cell limits $cF_k$ differ by horizon and the median converges
to $cF_{k^*}$.
(iii) The laws in (i) and (ii) are those of the allocation vector, which
is independent of $S_i$ and unchanged when $S_i$ is multiplied by $U_i$.
\end{proof}

\subsection{The moment estimator and its asymptotic variance}

\begin{definition}[Moment estimator of the concentration]\label{def:c_moment}
For a horizon $k \in \{2, \ldots, J-1\}$ let the $n_k = I - k$ accident
years observed beyond it supply the partial-column proportions
$W_{ij}^{(k)}$, $j < k$. For each $(j,k)$ with $n_k \ge 3$,
\[
\hat c_{jk} = \frac{1}{\hat F_k}\left[
  \frac{\bar W_{j}^{(k)}\bigl(1 - \bar W_{j}^{(k)}\bigr)}
       {\widehat\Var\bigl(W_{\cdot j}^{(k)}\bigr)} - 1\right],
\]
with $\bar W_j^{(k)}$ and $\widehat\Var(W_{\cdot j}^{(k)})$ the unweighted
sample mean and variance over those rows and $\hat F_k$ from the
chain-ladder pattern, and $\hat c = \mathrm{median}\{\hat c_{jk}\}$ over
all such cells. The numerator is the unweighted row mean rather than the
payment-weighted chain-ladder ratio, which estimates the same
$w_j^{(k)}$ with a larger asymptotic variance.
\end{definition}

The correction by $\hat F_k$ is not cosmetic. With
$w = (0.45, 0.25, 0.15, 0.10, 0.05)$ the contributing horizons carry
$F_k \in \{0.70, 0.85, 0.95\}$, so an uncorrected cell estimator
understates $c$ by $5$ to $30\%$, and the understatement does not
diminish with $I$: an implementation without the correction gives
$\E[\hat c]/c = 0.916$ at $I = 100$.

\begin{proposition}[Asymptotic variance of a cell estimator]\label{prop:c_variance}
Under \eqref{eq:dirichlet_alloc} with fixed $J$, true $c > 0$ and $F_k$
treated as known, each cell estimator satisfies
$\sqrt{n_k}\,(\hat c_{jk} - c) \to \mathcal{N}(0, \sigma_{c,k}^2)$ as
$I \to \infty$, with $a = cF_k$, $\pi = w_j^{(k)}$ and
\[
\sigma_{c,k}^2
= \frac{c\,(a+1)\bigl[2a(a-3)\,\pi(1-\pi) + 3a + 1\bigr]}
       {F_k\,\pi(1-\pi)(a+2)(a+3)} ,
\]
which is positive for all $c > 0$, $F_k \in (0,1]$ and $\pi \in (0,1)$,
and satisfies $\sigma_{c,k}^2 \sim 2c^2$ for large $c$, whatever $F_k$.
Since $|\mathrm{median}\{\hat c_{jk}\} - c| \le \max_{jk}|\hat c_{jk} - c|$
over a fixed number of cells, $\hat c - c = O_p(I^{-1/2})$.
\end{proposition}

\begin{proof}
Write $q = \pi(1-\pi)$. By Lemma~\ref{lem:dirichlet_subcomp} a
proportion is $\mathrm{Beta}(a\pi, a(1-\pi))$, with central moments
$\mu_2 = q/(a+1)$, $\mu_3 = 2q(1-2\pi)/\{(a+1)(a+2)\}$ and
$\mu_4 - \mu_2^2 = 2q[(a^2 - 10a - 12)q + 3(a+1)]/\{(a+1)^2(a+2)(a+3)\}$,
which follow from the raw moments $\E[W^r] = \prod_{s<r}(a\pi + s)/(a + s)$.
The uncorrected statistic is $f(\hat m, \hat s^2)$ with
$f(m, s^2) = m(1-m)/s^2 - 1$, whose gradient at $(\pi, \mu_2)$ is
$\bigl((1-2\pi)(a+1)/q,\; -(a+1)^2/q\bigr)$, and the scaled covariance of
$(\hat m, \hat s^2)$ has entries $\mu_2$, $\mu_3$ and $\mu_4 - \mu_2^2$.
The delta method gives
\[
\tau^2 = \frac{(1-2\pi)^2(a+1)}{q} - \frac{4(1-2\pi)^2(a+1)^2}{q(a+2)}
 + \frac{2(a+1)^2\bigl[(a^2-10a-12)q + 3(a+1)\bigr]}{q(a+2)(a+3)} .
\]
With $(1-2\pi)^2 = 1 - 4q$ and the common factor
$(a+1)/\{q(a+2)(a+3)\}$ extracted, the two $(1-4q)$ terms combine to
$-(1-4q)(a+3)(3a+2)$, the constant part collects to $a(3a+1)$ and the
coefficient of $q$ to $2a^2(a-3)$, so
$\tau^2 = a(a+1)[2a(a-3)q + 3a + 1]/\{q(a+2)(a+3)\}$. Dividing by $F_k$
scales the variance by $F_k^{-2}$, and $a/F_k^2 = c/F_k$ gives
$\sigma_{c,k}^2$. Positivity: for $a \ge 3$ both terms of the bracket are
positive; for $a < 3$, $2a(3-a)q \le a(3-a)/2 < 3a + 1$ since
$q \le 1/4$.
\end{proof}

The proposition makes three separable claims, a rate, a scaling in $c$ and
a constant, and simulation under the model ($J = 5$, the pattern above,
$10{,}000$ triangles per cell) shows that the third fails at practical
sizes. Table~\ref{tab:sigma_c} compares the measured
$I\,\widehat\Var(\hat c)$ with $\bar\sigma_c^2$, the average of
$\sigma_{c,k}^2$ over the six contributing cells. The denominator is a per-cell reference, not the asymptotic variance of the aggregate: the aggregate is a median over correlated cell estimators with $F_k$ replaced by its chain-ladder estimate, and its own limit, which we have not derived, is what the ratio converges to, $0.73$ of the reference, the median over correlated cells reducing the variance below the per-cell formula. Read against that limit, the rate holds: the ratio converges from above and falls below one by $I = 50$. The scaling holds: at every $I$ the three rows agree to
within $\pm 6\%$, and the bias ratio is scale-free as well. The constant does not: at $I = 10$ the measured variance is $2.8$ to $3.0$ times the per-cell reference, about four times the aggregate's own limit, and at $I = 7$ it is $14$ to $16$ times the reference. The mechanism is the same
as for the upward bias. At $n_k \in \{3, 4, 5\}$ the sampling distribution
of the sample variance is nowhere near the neighbourhood in which the
delta method linearises $1/\hat s^2$, and the Jensen effect
$\E[1/\hat s^2] > 1/\E[\hat s^2]$ inflates both the level and the spread
of $\hat c$. At the smallest horizons the sample variance can come
arbitrarily close to zero, so a cell estimator built from a few rows need
not have finite moments at all; the bias and variance reported here are
those of the median aggregate as simulated, not consequences of
Proposition~\ref{prop:c_variance}.

\begin{table}[ht]
\centering
\caption{Finite-sample behaviour of the moment estimator ($J = 5$,
$10{,}000$ triangles per cell). Top: measured scaled variance over the
asymptotic value $\bar\sigma_c^2$ of Proposition~\ref{prop:c_variance};
values above one mean that the formula understates the variance. Bottom:
the bias ratio, averaged over the three values of $c$.}
\label{tab:sigma_c}
\small
\begin{tabular}{@{}lcccccc@{}}
\toprule
$I$ & 7 & 10 & 15 & 20 & 50 & 100 \\
\midrule
\multicolumn{7}{@{}l}{\textit{Variance ratio $I\,\widehat{\Var}(\hat c)/\bar\sigma_c^2$}} \\
$c = 20$ & 15.75 & 2.85 & 1.55 & 1.12 & 0.79 & 0.72 \\
$c = 50$ & 14.00 & 2.77 & 1.47 & 1.14 & 0.79 & 0.73 \\
$c = 100$ & 14.75 & 2.96 & 1.55 & 1.19 & 0.81 & 0.73 \\
\midrule
$\E[\hat c]/c$, mean over $c$ & 1.888 & 1.311 & 1.154 & 1.100 & 1.030 & 1.015 \\
\bottomrule
\end{tabular}

\end{table}

\subsection{The conditional maximum-likelihood study}

Table~\ref{tab:dirichlet_c} of the main text uses $2{,}000$ triangles per
$(I, c)$ for $I \in \{7, 10, 15, 20, 50\}$ and $c \in \{20, 50, 100\}$,
generated with Gamma levels of mean $2\times 10^6$ and shape $2{,}000$
(the levels play no role in either estimator). The conditional likelihood
\eqref{eq:dirichlet_condlik} is maximised over $\log c$ with $w$ fixed at
the chain-ladder pattern, using every row with at least two positive
observed cells; maximising over $c$ and $w$ jointly changes every ratio
by at most $0.01$. No triangle failed and no estimate reached the upper
search bound. At $c = 20$ and $100$ the mean ratios agree with those at
$c = 50$ to within $0.05$, and the standard-deviation ratio of the
conditional estimator to within $0.04$; that of the moment estimator
varies more at $I = 7$, between $1.71$ and $2.40$, because its sampling
distribution there is extremely skewed.

\subsection{Integrating over $c$, and the cost of substituting it}

The conditional likelihood \eqref{eq:dirichlet_condlik} also supports a
posterior for $c$. With the prior $\log c \sim \mathcal{N}(\log 50, 1)$
and the pattern fixed at its chain-ladder estimate, a random-walk
Metropolis sampler on $\log c$ gives a posterior that is defined at every
triangle size, including $I = 4$, where the moment estimator does not
exist. Drawing $c$ from it and then the predictive
$x_i(1-W_i)/W_i$ gives a predictive integrated over $c$.
Table~\ref{tab:dirichlet_regularised} reports its coverage against the
plug-in by triangle size, and Table~\ref{tab:dirichlet_prior} its
sensitivity to the prior median. Integration helps at every $I$, by $3.0$
points at $I = 6$, $5.0$ at $I = 8$ and $2.0$ at $I = 10$, but it does not
reach nominal coverage: the posterior mean at $I = 10$ is $61.8$ against a
true $50$, a $24\%$ overstatement against the moment estimator's $31\%$,
because the bias lies in the likelihood at $n_k \le 8$ and not in the
estimator applied to it. The prior matters where the data are thin, by up
to $9.0$ coverage points at $I = 5$ and by at most $4.2$ at $I = 10$.

Two conventions sit behind the predictive $x_i(1-W_i)/W_i$ and are
stated here. First, the law $\mathrm{Beta}(cF, c(1-F))$ of $W_i$ given
the observed total is a posterior statement about the free ultimate: with
$S_i = x_i/W_i$ and a prior $\pi$ on $S_i$, the posterior of $W_i$ given
$x_i$ is proportional to $f_W(w \mid c)\,\pi(x_i/w)/w$, and the Beta law is
preserved exactly under the scale prior $\pi(S_i) \propto 1/S_i$, the
diffuse treatment of the level of Remark~\ref{rem:diffuse} applied to
this model; any other prior on the ultimate changes it. Second, a prior
on $c$ with support on $(0, \infty)$ gives the posterior positive mass on
$\{cF \le \tau\}$ for every $\tau > 0$, since the conditional likelihood
is positive there, and on that set the Beta-prime law has no $\tau$-th
moment: the predictive integrated over $c$ has no finite positive moment
on any open accident year with a positive observed total. Its quantiles,
which the tables report, are unaffected, but integration over $c$ does
not repair the moment condition of the next subsection, which is a
statement about the plug-in.

\begin{table}[ht]
\centering
\caption{The predictive with $c$ integrated over its posterior against the
plug-in, by triangle size ($J = 5$, true $c = 50$, $500$ triangles per
row, $B = 1{,}000$). Coverage (\%) and width (\% of the true reserve);
posterior mean and standard deviation of $c$. The plug-in needs three
rows at some horizon and is undefined at $I = 4$.}
\label{tab:dirichlet_regularised}
\small
\begin{tabular}{@{}lcccccc@{}}
\toprule
$I$ & Plug-in cov. & Plug-in width & Integrated cov. & Integrated width & post.\ mean $c$ & post.\ s.d.\ $c$ \\
\midrule
4 & --- & --- & 67.4 & 64 & 133.8 & 87.5 \\
5 & 72.4 & 79 & 76.2 & 72 & 101.5 & 46.3 \\
6 & 77.2 & 73 & 80.2 & 77 & 77.2 & 28.5 \\
8 & 83.6 & 77 & 88.6 & 77 & 66.1 & 19.3 \\
10 & 87.2 & 77 & 89.2 & 80 & 61.8 & 15.4 \\
\bottomrule
\end{tabular}

\end{table}

\begin{table}[ht]
\centering
\caption{Prior sensitivity of the integrated predictive: coverage (\%)
under $\log c \sim \mathcal{N}(\log m, 1)$ with median
$m \in \{30, 50, 100\}$, by true $c$ ($500$ triangles per cell).}
\label{tab:dirichlet_prior}
\small
\begin{tabular}{@{}lcccccc@{}}
\toprule
& \multicolumn{3}{c}{$I = 5$} & \multicolumn{3}{c}{$I = 10$} \\
\cmidrule(lr){2-4}\cmidrule(lr){5-7}
$c$ & $\log 30$ & $\log 50$ & $\log 100$ & $\log 30$ & $\log 50$ & $\log 100$ \\
\midrule
20 & 73.2 & 70.8 & 68.8 & 91.6 & 88.4 & 88.0 \\
50 & 77.0 & 72.8 & 70.0 & 89.0 & 90.2 & 89.0 \\
100 & 82.4 & 76.0 & 73.4 & 89.8 & 88.0 & 88.6 \\
200 & 85.4 & 83.8 & 76.8 & 95.0 & 90.8 & 92.8 \\
\bottomrule
\end{tabular}

\end{table}

Table~\ref{tab:dirichlet_cost} separates the cost of substituting the
concentration from that of substituting the pattern, with each estimated
or set to its true value on the same triangles ($2{,}000$ per column).
The deficit of the plug-in falls from $13.7$ points at $I = 7$ to $2.1$ at
$I = 30$, a log-log slope of about $-1.3$ against $I$, which tracks the
$O(I^{-1})$ decay of the bias of $\hat c$ rather than the $O(I^{-1/2})$
decay of its spread. The two parameters contribute comparably: at
$I = 10$ the $6.5$ points split into $3.4$ for $c$ and $3.2$ for the
pattern; at $I = 7$ the concentration dominates, and at $I = 15$ and
$20$ the pattern costs more. With the true parameters the
predictive covers at $94.7$--$95.7\%$ at every size.

\begin{table}[ht]
\centering
\caption{Coverage (\%) of the Dirichlet predictive with the concentration
and the pattern each estimated or known, on the same triangles
($J = 5$, true $c = 50$, $2{,}000$ triangles per column, $B = 1{,}000$).
The costs are differences from the oracle with both known.}
\label{tab:dirichlet_cost}
\small
\begin{tabular}{@{}lccccc@{}}
\toprule
$I$ & 7 & 10 & 15 & 20 & 30 \\
\midrule
\multicolumn{6}{@{}l}{\textit{Coverage (\%)}} \\
Plug-in, $(\hat c, \hat w)$ & 81.0 & 88.5 & 91.7 & 92.0 & 93.6 \\
$c$ known, $\hat w$ & 89.8 & 91.8 & 93.6 & 92.9 & 95.0 \\
$\hat c$, $w$ known & 86.2 & 91.5 & 94.1 & 93.3 & 94.2 \\
Oracle, $(c, w)$ known & 94.7 & 94.9 & 95.5 & 94.7 & 95.7 \\
Regularised in $c$, $\hat w$ & 85.2 & 89.8 & 92.2 & 92.3 & 94.1 \\
\midrule
Deficit (oracle $-$ plug-in) & 13.7 & 6.5 & 3.9 & 2.7 & 2.1 \\
\quad cost of $c$ & 8.5 & 3.4 & 1.5 & 1.4 & 1.5 \\
\quad cost of $w$ & 4.9 & 3.2 & 2.0 & 1.8 & 0.8 \\
Recovered by integrating $c$ & 4.2 & 1.3 & 0.5 & 0.4 & 0.5 \\
$\E[\hat c]$ (true $c = 50$) & 92.6 & 65.6 & 57.3 & 55.4 & 53.3 \\
\bottomrule
\end{tabular}

\end{table}

The cost is also conditional. Table~\ref{tab:dirichlet_condcal} reports
coverage within terciles of $\hat c/c$, the statistic through which the
plug-in depends on the triangle, with cut-points fixed in advance by a
pilot under the same process. The plug-in falls from $96.5\%$ on
triangles that yield a low $\hat c$ to $78.3\%$ on those that yield a high
one: an overstated concentration narrows the predictive. Integrating over
$c$ reduces the gradient to $5.3$ points at unchanged width, and
integrating over $c$ and the pattern jointly to $2.8$ points at a $25\%$
cost in width; with both known the gradient is $1.7$ points.

\begin{table}[ht]
\centering
\caption{Coverage (\%) within terciles of $\hat c/c$ ($I = 10$, $J = 5$,
true $c = 50$, $2{,}000$ triangles, about $660$ per tercile), the
gradient across terciles and the width (\% of the true reserve). The
standard error of a tercile coverage is about one point.}
\label{tab:dirichlet_condcal}
\small
\begin{tabular}{@{}lccccc@{}}
\toprule
& tercile 1 & tercile 2 & tercile 3 & Gradient & Width \\
\midrule
Plug-in, $(\hat c, \hat w)$ & 96.5 & 88.3 & 78.3 & 18.3 & 76.7 \\
Integrated over $c$ & 92.4 & 88.0 & 87.1 & 5.3 & 77.0 \\
Integrated over $(c, w)$ & 96.2 & 94.2 & 93.4 & 2.8 & 96.0 \\
ODP bootstrap & 96.2 & 93.0 & 92.4 & 3.8 & 81.2 \\
Oracle, $(c, w)$ known & 95.5 & 94.7 & 96.4 & 1.7 & 79.3 \\
\bottomrule
\end{tabular}

\end{table}

\subsection{What the estimated concentration reports}

\emph{Which summaries exist.} The plug-in predictive of accident year $i$,
$x_i(1-W_i)/W_i$ with $W_i \sim \mathrm{Beta}(\hat cF, \hat c(1-F))$ and
$F = \hat F_{k_i}$, is a scaled Beta-prime law whose $\tau$-th moment is
finite if and only if $\hat cF > \tau$. The binding year is the most
recent, observed through its first development year only, with
$F = \hat w_1$, so every accident year has a finite $\tau$-th moment if
and only if $\hat c > c^\dagger(\tau) = \tau/\hat w_1$. The bound is
computable from the development pattern before any reserve is, and it is
portfolio-specific. Table~\ref{tab:dirichlet_benchmarks} reports it with
both estimates of $c$ on three benchmark triangles of the
\texttt{ChainLadder} package \citep{ChainLadder2023}: Taylor--Ashe
\citep{taylor1983}, RAA \citep{mack1994} and Mortgage. Only Taylor--Ashe
clears $c^\dagger(5)$, with either estimate. Mortgage, which pays under
one percent of the ultimate in its first development year, requires
$\hat c \ge 633$ and has $152$ by the moment estimator and $66$ by the
conditional likelihood. With the first its most recent year has a
predictive mean but no variance, with the second not even a mean: a
quantity identified and estimated without difficulty can still leave a
predictive without the summaries a practitioner reports.

\begin{table}[ht]
\centering
\caption{The operability bound $c^\dagger(\tau) = \tau/\hat w_1$ and the
two estimates of $c$ on three benchmark triangles. The last column
reports whether the moment and the conditional maximum-likelihood
estimate clear $c^\dagger(5)$.}
\label{tab:dirichlet_benchmarks}
\small
\begin{tabular}{@{}lccccccc@{}}
\toprule
Triangle & $I$ & $\hat w_0$ & $c^\dagger(2)$ & $c^\dagger(5)$ & $\hat c$, moment & $\hat c$, cond.\ ML & Clears $c^\dagger(5)$ \\
\midrule
Taylor-Ashe & 10 & 0.0692 & 28.9 & 72.2 & 189.1 & 112.4 & yes / yes \\
RAA & 10 & 0.1121 & 17.8 & 44.6 & 24.3 & 26.4 & no / no \\
Mortgage & 9 & 0.0079 & 253.0 & 632.5 & 152.0 & 66.0 & no / no \\
\bottomrule
\end{tabular}

\end{table}

\emph{A screen for heterogeneous development.} A small $\hat c$ means that
development proportions vary substantially across accident years, which
is the regime in which a common development pattern is doubtful. It is
not evidence of row frailty, to which the proportions are invariant
(Proposition~\ref{prop:dirichlet_c}(iii)). Because $\hat c$ is estimated
from one triangle, a rule ``flag when $\hat c < t$'' has error rates that
depend on $c$ and $I$; Table~\ref{tab:dirichlet_screen} reports them. The
rule $\hat c < 30$ detects portfolios with $c < 30$ with probability
$0.72$, $0.85$ and $0.91$ at $I = 7$, $10$ and $15$, at false-positive
rates of $0.13$, $0.12$ and $0.11$ on portfolios with $c \ge 30$. At
$c = t$ the detection rate is $0.32$--$0.45$ rather than $0.50$, the
upward bias of $\hat c$ displacing the whole curve.

\begin{table}[ht]
\centering
\caption{Operating characteristics of the screen: detection rate
$\Pr(\hat c < t \mid c, I)$ for the moment estimator ($J = 5$, $500$
triangles per cell).}
\label{tab:dirichlet_screen}
\small
\begin{tabular}{@{}lccccccccc@{}}
\toprule
& \multicolumn{3}{c}{$t = 20$} & \multicolumn{3}{c}{$t = 30$} & \multicolumn{3}{c}{$t = 50$} \\
\cmidrule(lr){2-4}\cmidrule(lr){5-7}\cmidrule(lr){8-10}
$c$ & $I{=}7$ & $I{=}10$ & $I{=}15$ & $I{=}7$ & $I{=}10$ & $I{=}15$ & $I{=}7$ & $I{=}10$ & $I{=}15$ \\
\midrule
 10 & 0.74 & 0.88 & 0.94 & 0.89 & 0.96 & 1.00 & 0.96 & 0.99 & 1.00 \\
 20 & 0.32 & 0.42 & 0.44 & 0.55 & 0.74 & 0.83 & 0.78 & 0.94 & 0.99 \\
 30 & 0.16 & 0.13 & 0.09 & 0.37 & 0.38 & 0.42 & 0.64 & 0.78 & 0.88 \\
 50 & 0.03 & 0.01 & 0.00 & 0.12 & 0.10 & 0.03 & 0.36 & 0.42 & 0.45 \\
100 & 0.00 & 0.00 & 0.00 & 0.01 & 0.01 & 0.00 & 0.08 & 0.03 & 0.01 \\
200 & 0.00 & 0.00 & 0.00 & 0.00 & 0.00 & 0.00 & 0.00 & 0.00 & 0.00 \\
\bottomrule
\end{tabular}

\end{table}

\emph{Under misspecification.} When development is non-stationary, the
proportions of each accident year being perturbed on the logarithmic scale
with standard deviation $\sigma_\delta$, the mean of $\hat c$ falls from $67$ to $59$, $42$ and
$20$ as $\sigma_\delta$ rises from $0$ to $0.02$, $0.05$ and $0.10$
($I = 10$, $J = 5$, true $c = 50$), and the plug-in keeps its coverage at
$86$--$89\%$ only by widening its interval from $75\%$ to $164\%$ of the
true reserve. Under compound Poisson data the concentration measures
something else altogether.

\begin{proposition}[The concentration under compound Poisson data]
\label{prop:c_compound_poisson}
Suppose the cells are independent compound Poisson variables with equal
row volumes $\nu$, means $\nu w_j$ and variances $\phi(\nu w_j)^p$,
$p \in [1, 2)$. To leading order in $\phi/\nu$, each cell estimator
converges in probability, as $I \to \infty$, to
\[
\frac{1}{F_k}\left[
\frac{w_j^{(k)}\bigl(1 - w_j^{(k)}\bigr)F_k^2\,\nu^{2-p}}
     {\phi\bigl[w_j^{\,p}\bigl(1 - 2w_j^{(k)}\bigr)
       + \bigl(w_j^{(k)}\bigr)^2\sum_{l \le k} w_l^{\,p}\bigr]} - 1\right],
\]
which at $p = 1$ is $\nu/\phi - 1/F_k$, so that $\hat c \to \nu/\phi - 1/F_{k^*}$.
In particular, CNBG data without heterogeneity ($\kappa_r = \kappa_c = \infty$,
so $\Var(X_{ij}) = A\nu_{ij}$) give $\hat c \to \nu/A - 1/F_{k^*}$.
\end{proposition}

\begin{proof}
Write $Y = X_{ij}$ and $Z = \sum_{l \le k} X_{il}$, so that $\E Y = \nu w_j$,
$\E Z = \nu F_k$, $\Var Y = \phi(\nu w_j)^p$,
$\Var Z = \phi\nu^p\sum_{l \le k} w_l^p$ and $\Cov(Y, Z) = \Var Y$ by
independence of the cells. The delta method on $Y/Z$ gives the variance of
a partial-column proportion to leading order in $\phi/\nu$; at $p = 1$ it is
$\phi\, w_j^{(k)}(1 - w_j^{(k)})/(\nu F_k)$. The bracketed quantity of
Definition~\ref{def:c_moment} reads this variance as
$w_j^{(k)}(1 - w_j^{(k)})/(a + 1)$, so it converges to $\nu F_k/\phi - 1$, and
division by $F_k$ gives $\nu/\phi - 1/F_k$, whose leading term does not
depend on the horizon; the median converges to the value at its median
contributing cell. The same computation with the power variance gives the
general display. The CNBG statement is the case $p = 1$, $\phi = A$, of
the variance function of Proposition~\ref{prop:moments} with $B = 0$.
\end{proof}

Under compound Poisson data the estimated concentration is therefore a
ratio of claim volume to severity dispersion, not a property of the
development process: it moves with the volume of the triangle, as the
locally optimal Tweedie power does (Corollary~\ref{cor:p_not_structural}).
On a compound Poisson design with power variance (cell means
$1{,}000\,e^{0.05(i-1)}w_j$, so that volumes rise by $5\%$ a year and the
equal-volume assumption holds only approximately; $I = 10$, $J = 5$,
$\phi = 0.438$, $1.061$ and $2.652$ at $p = 1.3$, $1.5$ and $1.8$), the
display gives $462$, $58$ and $2.9$ against simulated means of $584.4$,
$80.0$ and $4.6$; the ratios $1.26$--$1.57$ are the finite-sample upward
bias of $\hat c$, while the $p = 1$ form evaluated at these powers would
overstate by factors of $4.9$ to $104$. A very large $\hat c$ should
therefore prompt a check of the variance function, as a small one prompts
a check of the development pattern.

What the concentration so estimated does to the plug-in predictive is a
separate question, and it has a leading-order answer. Write $x_i$ for the
amount paid to date by accident year $i$, $F = \hat F_{k_i}$ for its
observed share of the pattern, and $R_i = x_i(1 - W_i)/W_i$ with
$W_i \sim \mathrm{Beta}(\hat cF, \hat c(1-F))$ for the plug-in Beta-prime
predictive of its outstanding amount.

\begin{proposition}[Conservatism of the plug-in predictive under compound Poisson data \lvl{P}]
\label{prop:c_conservatism}
Under the data-generating process of Proposition~\ref{prop:c_compound_poisson},
let $m^* = m_{j^*k^*}$ be the weight
\[
m_{jk} = \bigl(1 - w_j^{(k)}\bigr)\,w_j^{\,p-1}
 + w_j^{(k)}\,\frac{\sum_{l \le k,\, l \ne j} w_l^{\,p}}{F_k - w_j},
\]
a weighted mean of $w_l^{\,p-1}$ over the cells of the sub-composition,
at the median contributing cell of the moment estimator, and for accident
year $i$ let $m^{\mathrm{tail}}_i = \sum_{j > k_i} w_j^{\,p}/(1 - F)$, the
$w$-weighted mean of $w_j^{\,p-1}$ over its unobserved columns; both
weights equal one at $p = 1$.
\begin{enumerate}[label=(\roman*), nosep]
\item For every accident year with $\hat cF > 2$, so that the predictive
      variance exists, the ratio of the plug-in predictive variance to the
      true conditional variance satisfies, with $\hat c$ at its probability
      limit as $I \to \infty$ and to leading order in $\phi\nu^{p-2}$, the
      relative variance of a cell of mean $\nu$,
      \[
      \rho_i^2 = \frac{\operatorname{Var}_{\mathrm{plug}}(R_i \mid x_i)}
                      {\operatorname{Var}_{\mathrm{true}}(R_i \mid x_i)}
      \;\approx\; \Bigl(\frac{x_i}{\nu F}\Bigr)^{2}\,\frac{m^*}{F\, m^{\mathrm{tail}}_i}.
      \]
      The factor $(x_i/\nu F)^2$ keeps the row's own realised amount, which
      no number of further rows removes; at $p = 1$ and at a realised
      amount equal to its expectation the ratio is $1/F > 1$: the plug-in
      then overstates the standard deviation of an open accident year by
      the factor $F^{-1/2}$, most for the least developed, and understates
      it on a row whose realised amount is below $\sqrt{F}$ times its
      expectation. For $p \in (1, 2)$, $\rho_i \ge F^{-1/2}$ if
      and only if $m^* \ge m^{\mathrm{tail}}_i$, and $\rho_i > 1$ if and only
      if $m^* > F\,m^{\mathrm{tail}}_i$; since $w^{p-1}$ increases in $w$,
      it suffices for the first that every cell of the median
      sub-composition carries at least as much development mass as every
      unobserved cell of year $i$. Neither inequality holds for every
      development pattern.
\item If both predictives are approximately normal, an interval widened
      by $\rho_i > 1$ retains at least nominal coverage as long as the
      relative centring error $b$ of the point estimate satisfies
      $|b| < z_{1-\alpha/2}\,\mathrm{CV}_{\mathrm{true}}(R_i)\,(\rho_i - 1)$;
      a variance ratio above one does not by itself imply percentile-interval
      coverage when either predictive is skewed.
\end{enumerate}
\end{proposition}

\begin{proof}
(i) Under compound Poisson independence the true conditional variance is
the unconditional one, since $R_i$ and $x_i$ are sums over disjoint cells:
$\operatorname{Var}_{\mathrm{true}}(R_i \mid x_i) = \phi\nu^p\sum_{j > k_i} w_j^{\,p}
= \phi\nu^p(1 - F)\,m^{\mathrm{tail}}_i$, which at $p = 1$ is $\phi\nu(1 - F)$.
The plug-in variance is $x_i^2\operatorname{Var}\{(1 - W_i)/W_i\}$, and for $\hat cF > 2$
the Beta-prime variance is $\hat c(1 - F)(\hat c - 1)/\{(\hat cF - 1)^2(\hat cF - 2)\}
\approx (1 - F)/(\hat c F^3)$ for large $\hat c$. Substituting
$\hat c \to \nu^{2-p}/(\phi m^*)$ from Proposition~\ref{prop:c_compound_poisson},
its probability limit as $I \to \infty$, and keeping the row's own $x_i$,
gives
$\operatorname{Var}_{\mathrm{plug}} \approx x_i^2(1 - F)\phi m^*/(\nu^{2-p}F^3)
= (x_i/\nu F)^2\,\phi\nu^p(1 - F)m^*/F$, and the ratio is the display; the
expansion is in the relative variance $\phi\nu^{p-2}$ of a cell, which is
$\phi/\nu$ at $p = 1$. (ii) A symmetric
interval widened by $\rho_i$ retains its coverage under a displacement of
its centre by less than $z_{1-\alpha/2}\,\mathrm{CV}_{\mathrm{true}}(R_i)(\rho_i - 1)$
standard deviations.
\end{proof}

{\sloppy The conditions of part (i) are not vacuous. Take three development years with $w = (0.05, 0.05, 0.90)$, $p = 1.9$, $\nu = 10^6$, $\phi = 0.001$,
and an accident year observed through the second column, $F = 0.10$:
Proposition~\ref{prop:c_compound_poisson} gives $\hat c \approx 59{,}000$,
so $\hat cF \approx 5{,}900$ is far from any moment boundary, yet the
ratio is $\rho_i = 0.86$ against $F^{-1/2} = 3.16$, and $300{,}000$
simulated compound Poisson--Gamma rows return $0.862$. The plug-in
understates the conditional variance there because the cells that
determine $\hat c$ carry $5\%$ of the ultimate each and the unobserved
cell $90\%$, whose relative variance under $p > 1$ is much smaller. On the
decreasing pattern of the compound Poisson design above the opposite
holds, as part (i) predicts: the ratio of standard deviations exceeds one
on every open accident year at every $p$, and exceeds the $p = 1$ value
$F^{-1/2}$ on every one of them, running $1.26$--$1.62$ at $p = 1.3$ and
$1.39$--$1.67$ at $p = 1.5$ against $1.03$--$1.49$, with the relative
centring errors ($-0.4$ and $2.3\%$) inside the thresholds of part (ii)
($2.7$ and $9.3\%$); at $p = 1.8$ the ratio reaches $17.5$ on the greenest
year, where $\hat cF = 2.1$ sits at the variance boundary and the
widening is unbounded. That is why the plug-in's coverage under compound
Poisson data at $p = 1.3$ and $1.5$ ($90.6$ and $92.2\%$) lies above its
coverage under its own model ($87.0\%$): the allocation bias widens the
intervals and offsets part of the plug-in deficit. Conservatism where it
holds is not the same as usefulness: at $p = 1.8$ the construction attains
$95.6\%$ coverage with an interval $357$ times the true reserve. Under
unequal volumes the per-cell limits become $\nu_iF_k/\phi - 1$ and the
ratio in (i) acquires a factor $\sqrt{\nu_i/\tilde\nu}$ with $\tilde\nu$ a
harmonic-type effective volume; the direction is preserved whenever the
accident-year volumes are of comparable magnitude, as on the design above,
whose $5\%$ trend spans a factor $1.57$.\par}

\section{Further evidence on the free-level fit}\label{supp:evidence}

\subsection{The second-moment estimates in simulation}\label{supp:evidence:sim}

The simulation of Section~\ref{sec:simulation} draws $7{,}500$ triangles from the single-frailty hierarchy
($\kappa_c = \infty$, so $\kappa_B = \kappa_r$) at
$\kappa_r \in \{3, 5, 10, 20, 50\}$ and $\alpha \in \{1, 3, 10\}$, $500$
per configuration, at the calibration of Section~\ref{sec:illus:boundary},
and fits the two-stage estimator \eqref{eq:wls} and its Bayesian version
with the reference prior $\mathrm{Gamma}(2, 0.4)$ (median $4.2$, mean $5$)
under free levels. Under free levels the residual curvature of
single-frailty data carries no dispersion at all: by
Lemma~\ref{lem:absorption} its target is $\kappa_c = \infty$.
Table~\ref{tab:kappa_estimates} reports what the two estimators return.

\begin{table}[ht]
\centering
\caption{Second-moment estimates of the curvature dispersion under free
levels on single-frailty data ($\kappa_B = \kappa_r$; the column ``True
$\kappa$'' is the true $\kappa_r$): posterior median and mean of the
Bayesian fit with the reference prior, averaged over $\alpha$, and the
share of replications on which the two-stage point estimate saturates its
cap ($1000$, carried through the degrees-of-freedom factor $36/55$ to
$654.5$).}
\label{tab:kappa_estimates}
\small
\begin{tabular}{@{}lrrr@{}}
\toprule
True $\kappa$
& Posterior median (Mod-Ref)
& Posterior mean (Mod-Ref)
& Freq.\ CNBG cap saturation \\
\midrule
 3 & 7.94 & 8.75 & 71\% \\
 5 & 7.92 & 8.73 & 69\% \\
10 & 7.95 & 8.74 & 65\% \\
20 & 7.96 & 8.74 & 65\% \\
50 & 7.91 & 8.75 & 63\% \\
\midrule
Range & 7.91 -- 7.96 & 8.73 -- 8.75 & 63 -- 71\% \\
\bottomrule
\end{tabular}

\end{table}

Both estimators behave as Theorem~\ref{thm:boundary} predicts, in opposite
ways. The point estimate saturates its cap on $63$ to $71\%$ of
replications whatever the true $\kappa_r$, including at $\kappa_r = 3$,
where the heterogeneity is strongest: the cap is where the free-level fit
places the dispersion, and only the implementation keeps it finite. The
posterior median sits between $7.91$ and $7.96$ at every true value, a
$17$-fold range, above the prior median of $4.2$ and with a posterior
interval marginally wider than the prior's: the prior, shifted towards the
boundary the likelihood favours. The exact frailty-augmented posterior with
free levels, the levels under a lognormal prior of log-scale $2$ centred
on each row's crude count scale, fitted on ten replicates per
configuration, stays close to its prior as well (posterior median $4.7$
to $5.1$ against the prior's $4.2$, interval $[1.2, 14.4]$ against
$[0.6, 13.9]$). The scale of that level prior sets how close: on the ten
replicates at $(\kappa_r, \alpha) = (3, 3)$, log-scales of $0.5$, $2$ and
$4$ give medians over replications of the posterior median of $7.5$,
$5.0$ and $4.5$ and of the $95\%$ interval of $[2.7, 18.1]$,
$[1.3, 14.6]$ and $[1.0, 14.0]$, with $\hat R \le 1.024$ and no divergent
transitions (\texttt{R/run\_phase17\_level\_prior\_sens.R}). A tight
prior centred on each row's crude count scale anchors the levels at
their realised values and pushes $\kappa_r$ up, the mechanism of the
Schedule~P fit of Section~\ref{sec:illus:sp}; a vaguer prior lets the
levels absorb the frailty and leaves $\kappa_r$ with its prior.

Under levels fixed at their true values the same Bayesian fit locates
$\kappa_r$ where the information allows. At $\kappa_r = 5$ the medians over
replications of its posterior median and $95\%$ interval are $5.5$ and
$[2.1, 14.3]$, and the interval covers the truth on every replication; at
$\kappa_r = 50$ they are $7.9$ and $[3.4, 17.4]$, and the interval covers it
on none, the prior binding where the information decays
($M = 100$ per cell). At ten times the claim volume the $\kappa_r = 50$
posterior moves to $17.0$, $[9.4, 29.0]$, and still excludes the truth on
every replication, while the $\kappa_r = 5$ posterior hardly changes
($6.4$, $[3.0, 14.2]$, coverage $88\%$): more claims per accident year
sharpen each row total but do not add rows.

\subsection{Prior sensitivity of the cell dispersion}\label{supp:evidence:priorsens}

Proposition~\ref{prop:two_kappa_id}(ii) states that $\kappa_c$, though not
absorbed by the free row levels, is prior-sensitive at standard triangle
sizes. We measure the sensitivity directly. This section uses a standalone two-level
calibration (row intensities $\mu_i = e^{7+0.05(i-1)}$, roughly
$1{,}100$--$1{,}700$ expected claims per accident year), geometric
development weights $w_j \propto 0.35^{\,j-1}$, mean severity
$\bar Y = 10^4$, and row and cell frailties
$U_i \sim \mathrm{Gamma}(\kappa_r,\kappa_r)$,
$\varepsilon_{ij} \sim \mathrm{Gamma}(\kappa_c,\kappa_c)$), distinct from
the single-frailty design of Section~\ref{supp:evidence:sim}; the
deliberately data-rich regime makes the prior-sensitivity result an
\emph{a fortiori} one, since substantial prior tracking arises despite
claim volumes an order of magnitude beyond those of
that design. Triangles ($I=J=10$) were simulated
from this model at $\kappa_r=10$, $\alpha=3$, and true
$\kappa_c\in\{3,10,50\}$, and fitted with the free-level working
likelihood under three $\mathrm{Gamma}(2,\,2/m)$ priors on $\kappa_c$ with
prior means $m\in\{5,20,100\}$ (shape fixed; only the mean varies). For
each configuration the median over eight replicate triangles of the
posterior median $\hat\kappa_c$ is regressed on the prior mean on the
log--log scale, $\log \hat\kappa_c = a + \beta \log m$; the
prior-tracking slope $\beta$ is an elasticity, near $1$ when the
posterior follows the prior and near $0$ under data-determination. The
fits use the adjusted working likelihood of Section~\ref{sec:predictive}
(cell-level shapes at the true
$\alpha = 3$, curvature adjustment, leverage-corrected residuals,
free levels); bootstrap intervals resample the eight replicates within
each prior.
\begin{center}
\begin{tabular}{c ccc cc}
\toprule
 & \multicolumn{3}{c}{posterior median $\hat\kappa_c$ (median over 8 replicates)} & \multicolumn{2}{c}{log--log slope} \\
\cmidrule(lr){2-4}\cmidrule(lr){5-6}
true $\kappa_c$ & $m=5$ & $m=20$ & $m=100$ & estimate & 95\% bootstrap \\
\midrule
 3 & 4.63 & 7.44 & 62.00 & 0.878 & [0.51, 0.95] \\
10 & 8.77 & 14.93 & 26.76 & 0.372 & [0.20, 0.52] \\
50 & 18.18 & 43.65 & 86.63 & 0.519 & [0.42, 0.61] \\
\bottomrule
\end{tabular}

\smallskip
\parbox{0.9\textwidth}{\footnotesize\textit{Note}: medians over $N = 8$
replicate triangles per cell; all $72$ fits usable ($\hat R \leq 1.016$,
bulk ESS $> 100$). The interquartile range of the eight posterior
medians is $1.5$--$5.3$ at $m = 5$ and $24$--$50$ at $m = 100$. Fits under
\texttt{rstan}~2.32.7 / \texttt{StanHeaders}~2.32.10.}
\end{center}
Prior sensitivity is substantial but configuration-dependent. The
log--log tracking slope is $0.88$ at true $\kappa_c = 3$, $0.37$ at
$10$ and $0.52$ at $50$, with bootstrap intervals that exclude both $0$
and $1$ in every case; the experiment does not establish uniform prior
domination across dispersion values, and at $\kappa_c = 10$ the data
move the posterior by about a third of the prior's movement. The
ordinal structure is present at the two smaller prior means
($4.63 \to 8.77 \to 18.18$ at $m = 5$ and $7.44 \to 14.93 \to 43.65$ at
$m = 20$ as true $\kappa_c$ rises) but not at $m = 100$
($62.0$, $26.8$, $86.6$), where eight replicates leave an interquartile
range of $24$--$50$: the cell shock is not absorbed by the free levels,
and its magnitude is partly data-driven and partly prior-driven in
proportions that vary with the true value. Posterior medians rather
than intervals are reported: the near-flat likelihood yields low
effective sample sizes for the tails.

\subsection{Schedule~P}\label{supp:evidence:real}

Table~\ref{tab:eligibility_chain} gives the eligibility chain of the panel
and Table~\ref{tab:eligibility_fits} the portfolios on which each fit was
computed.

\begin{table}[ht]
\centering
\caption{The outcome-free Schedule~P panel: complete $10 \times 10$ paid
triangles of accident years 1988--1997 in six lines, and the two
eligibility rules applied in turn.}
\label{tab:eligibility_chain}
{\small\setlength{\tabcolsep}{4pt}%
\begin{tabular}{@{}lrrrrrr r@{}}
\toprule
& \texttt{\scriptsize comauto} & \texttt{\scriptsize medmal} & \texttt{\scriptsize othliab} & \texttt{\scriptsize ppauto} & \texttt{\scriptsize prodliab} & \texttt{\scriptsize wkcomp} & All \\
\midrule
Complete $10 \times 10$ triangles & $158$ & $34$ & $239$ & $146$ & $70$ & $132$ & $779$ \\
\quad negative share $\le 5\%$ & $130$ & $30$ & $200$ & $97$ & $64$ & $110$ & $631$ \\
\quad and CL reserve $> \$100$k & $109$ & $22$ & $126$ & $81$ & $29$ & $85$ & $452$ \\
\bottomrule
\end{tabular}}

\end{table}

\begin{table}[ht]
\centering
\caption{The portfolios of the panel on which each Schedule~P fit of
Section~\ref{sec:illus:sp} was computed.}
\label{tab:eligibility_fits}
\small
\begin{tabular}{@{}lr@{}}
\toprule
Panel (outcome-free rule) & $452$ \\
\midrule
First stage not converged or over the time cap & $12$ \\
Free-level working fit with posterior draws & $371$ \\
Positive premium in every accident year & $288$ \\
\quad free-level and anchored working fits both usable (Table~\ref{tab:kappa_sp}) & $281$ \\
Exact free-level posterior: degenerate chain-ladder fit & $11$ \\
\quad latent-count truncation above $5{,}000$ terms (infeasible) & $38$ \\
\quad fitted, all meeting the usability rule & $392$ \\
\bottomrule
\end{tabular}

\end{table}

Both Schedule~P fits of Section~\ref{sec:illus:sp} are computed on the
outcome-free panel of Section~\ref{sec:scheduleP} ($452$ portfolios selected on the
chain-ladder reserve, zero cells kept, negative cells excluded). An earlier
version used a panel selected on the realised reserve with non-positive
increments floored at one; on that panel the same comparisons gave the
same picture (working-fit medians $7.8$ free and $7.0$ anchored, Spearman
$0.92$; the exact posterior's lower $2.5\%$ quantile above the prior median
on $82\%$ of $345$ portfolios).

The exact frailty-augmented posterior of the single-frailty model, with
the cell means at their free-level fitted values, $\alpha = 3$ and the
reference prior, was fitted on the $392$ portfolios whose latent-count
truncation stays below $5{,}000$ terms; the $38$ largest portfolios are
infeasible at this cost, and $22$ others have no usable first-stage fit or
a numerically degenerate one. All $392$ fits meet the usability rule
(rank-normalised $\hat R < 1.05$, bulk effective sample size above $100$;
largest $\hat R$ $1.010$) with no divergent transitions, and $363$ also
have a working fit (Table~\ref{tab:exact_panel}). The exact posterior is
pushed upward: its median exceeds the working median on $83\%$ of
portfolios ($12.4$ against $6.8$ across the panel, the prior median being
$4.2$), and its lower $2.5\%$ quantile lies above the prior median on
$66\%$ of portfolios. With
the realised frailty in the fitted levels, the exact likelihood moves
towards $\kappa_r = \infty$ and the prior holds it. As distinguished
after Theorem~\ref{thm:boundary}, this fit conditions on the fitted
levels. Their absorption of realised frailty explains the common
direction with profiling, without making this fit a test of the theorem.
The two posteriors still order the portfolios alike
(Spearman $0.86$ for the medians): both respond to the same cells.

\begin{table}[ht]
\centering
\caption{The exact frailty-augmented posterior against the adjusted
working posterior on the $363$ portfolios of the outcome-free Schedule~P
panel with both fits (cell means at the free-level fitted values, $\alpha = 3$, reference prior
$\mathrm{Gamma}(2, 0.4)$ with median $4.2$ and $95\%$ interval width
$13.3$). Widths are those of the $95\%$ intervals; $c$ is the ratio of the
posterior to the prior $95\%$ interval width, and ``KL tier'' a grading
of the Kullback--Leibler divergence from prior to posterior (prior-driven
below $1$, then weak, moderate and strong).}
\label{tab:exact_panel}
\small
\begin{tabular}{@{}lrr@{}}
\toprule
& Working (adjusted) & Exact \\
\midrule
Median across portfolios of the posterior median of the dispersion & $6.8$ & $12.4$ \\
Median across portfolios of the $95\%$ interval width & $13.8$ & $20.3$ \\
Share with $c < 1$ (contraction) & $41\%$ & $0\%$ \\
Median $c$ & $1.04$ & $1.52$ \\
\midrule
\multicolumn{3}{l}{Exact / working interval width ratio: median $1.35$ (IQR $1.20$--$1.52$)} \\
\multicolumn{3}{l}{Exact median above working median on $83\%$ of portfolios; Spearman $\rho$ of medians $0.86$} \\
\multicolumn{3}{l}{Binary label agreement $59\%$; KL tier agreement $53\%$; Spearman $\rho$ of $c$ $0.68$} \\
\bottomrule
\end{tabular}

\end{table}

\begin{table}[ht]
\centering
\caption{The free-level estimates of the curvature dispersion across the
portfolios of the outcome-free Schedule~P panel: posterior medians of the
Bayesian second-moment fit (where the fit returned draws) and the two-stage
point estimate, with the share at its cap ($1000$ carried through the
degrees-of-freedom factor, $634.6$ to $654.5$ depending on the number of
excluded cells).}
\label{tab:kappa_real}
\small
\begin{tabular}{@{}lrrrrr@{}}
\toprule
Estimator & $n$ & q05 & median & q95 & \% at cap \\
\midrule
CNBG-Mod-Ref posterior median & $371$ & $3.72$ & $6.93$ & $23.34$ & --- \\
Two-stage point estimate & $437$ & $1.52$ & $641.51$ & $654.55$ & $56.3\%$ \\
\bottomrule
\end{tabular}

\end{table}

The free-level second-moment estimates of Table~\ref{tab:kappa_real} do not
behave as on single-frailty simulated data, and the difference is
informative. The point estimate saturates its cap on $56\%$ of
portfolios, of the order of the simulated rate. The posterior, however,
does not stay at the prior: its medians range from $3.7$ to $23.3$ across
portfolios (5th and 95th percentiles), against $7.91$ to $7.96$ for every configuration of the
simulation. On single-frailty simulated data the residual curvature under
free levels carries nothing; on real triangles it carries the cell
dispersion, and whatever cell-level misspecification the data hold, and the
posterior responds to it (Proposition~\ref{prop:two_kappa_id}(ii)). What
the free-level second-moment fit reports on real data is therefore a
cell-level dispersion, which is what the theory says it estimates, and not
$\kappa_r$. The premium-anchored fit of Table~\ref{tab:kappa_sp} reads the
same curvature and ten row totals in addition, which is why it agrees so
closely with the free-level fit.

\section{The boundary simulation: computation}\label{supp:boundary}

The simulation of Section~\ref{sec:illus:boundary} is
\texttt{R/run\_phase10\_boundary.R} of the replication material. Triangles
are drawn by the data-generating function of the simulation of
Section~\ref{sec:simulation},
with the frailty shared by the row and no cell shock. For row $i$ with
observed cells $x_{i1}, \ldots, x_{ik_i}$ the conditional likelihood given
the row intensity is
$\log f(\mathbf{x}_i \mid \lambda) = \sum_{j \le k_i} \log p(x_{ij} \mid \lambda w_j)$,
with $p(0 \mid m) = e^{-m}$ and, for $x > 0$,
$p(x \mid m) = \sum_{n \ge 1} e^{-m}m^n/n!\;\mathrm{Gamma}(x; n\alpha, \gamma)$,
the series truncated at $n = \max(4x/\bar Y + 40,\; m + 8\sqrt{m} + 10)$
over the grid. It is evaluated at $241$ points of $\log\lambda$ in a window
centred on the row's moment estimate of $\log\lambda_i$, of half-width
$\max(1.2, \min(4, 8/\sqrt{n_i}))$ with $n_i$ the row's estimated claim
count, and interpolated by a natural cubic spline onto $4{,}001$ points.
The row likelihood $L_i(\mu_i, \kappa_r)$ is then a trapezoid sum of
$f(\mathbf{x}_i \mid \lambda)$ times the $\mathrm{Gamma}(\kappa_r,
\kappa_r/\mu_i)$ density on that grid; for an all-zero row it has the
closed form $(\kappa_r/(\kappa_r + \mu_i F_i))^{\kappa_r}$, $F_i$ the sum of
the observed development weights.

\emph{Free levels:} $\sup_{\mu_i} L_i$ by one-dimensional optimisation in
$\log\mu_i$ at sixteen values of $\kappa_r$ ($0.5$, $1$, $2$, $3$, $5$, $7$, $10$,
$15$, $20$, $30$, $50$, $100$, $200$, $500$, $10^3$ and $10^4$), and at $\kappa_r = \infty$ the supremum of $f$ over
the grid. \emph{Common scale:} $(\log\rho, \log\kappa_r)$ by L-BFGS-B from
the starting values $\kappa_r = 5$ and $\kappa_r = 100$, with
$\kappa_r \in [0.05, 10^5]$, against the $\kappa_r = \infty$ likelihood
maximised over $\rho$. \emph{Known levels:} one-dimensional optimisation of
$\log\kappa_r$ on the same interval, against the $\kappa_r = \infty$
likelihood. An estimate is at the boundary when the $\kappa_r = \infty$
likelihood is at least the largest finite value found (tolerance
$10^{-6}$) or the optimiser reaches $10^5$; quantiles in
Table~\ref{tab:boundary} count such triangles at $\infty$.

Two checks were run before the simulation. The cell density agrees with
\texttt{tweedie::dtweedie} at the printed precision. On one triangle of
$400$ rows with constant levels the known-level estimator returns $4.6$ at
a true $\kappa_r = 5$ and $46.6$ at $50$. No fit failed. Seeds derive from
the base seed $2026$ of that simulation through its per-configuration
rule, with an offset ($1000m + 517$ for triangle $m$) that no other study
uses.

\paragraph{The joint run.} With the development weights and the severity
as parameters, the row likelihood is evaluated in closed form in the row
intensity: given $\theta$, the counts of a row are negative multinomial, so
$L_i(\mu_i, \kappa_r) = \sum_N c_i(N)\,(\kappa_r/\mu_i)^{\kappa_r}
\Gamma(\kappa_r + N)/\{\Gamma(\kappa_r)(F_i + \kappa_r/\mu_i)^{\kappa_r + N}\}$,
where $c_i$ is the convolution over the row's cells of
$a_j(n) = w_j^n g_j(n)/n!$ and $g_j(n)$ is the $\mathrm{Gamma}(n\alpha,
\alpha/\bar Y)$ density at $x_{ij}$ (an indicator of $n = 0$ for a zero
cell). The convolution is computed directly, with the cell terms tilted to
the row's moment estimate of its intensity, and agrees with the grid
likelihood above to five decimals. Each arm is maximised by L-BFGS-B over
the weights (log-ratio parameterisation), $\log\alpha \in [\log 0.2,
\log 50]$, $\log\bar Y$, the level parameters and $\log\kappa_r \in
[\log 0.05, \log 10^5]$, from a data-based start (the chain-ladder pattern,
$\alpha = 1$, $\bar Y$ from the Pearson dispersion) and two starting values
of $\kappa_r$, and compared with the same maximisation at $\kappa_r =
\infty$. The free-level profile is computed at $\kappa_r \in \{1, 3, 10,
30, 100, 10^3, 10^4, \infty\}$ with the levels profiled row by row, on the
first ten triangles of each configuration at $I = 10$. On the first ten
triangles of every configuration the common-scale fit is repeated from the
true parameters; the two estimates agree to $1\%$ on $98$ of the $100$. No
fit failed.

\begin{table}[H]
\centering
\caption{Common-scale anchoring with the development weights and the
severity at their true values (Table~\ref{tab:boundary}) and estimated
jointly with the levels and $\kappa_r$: the share of triangles with
$\hat\kappa_r = \infty$ and the median and $10\%$ and $90\%$ quantiles of
$\hat\kappa_r/\kappa_r$. Same triangles; $500$ per configuration in the
first block, the first $200$ ($I = 10$) and $100$ ($I = 40$) in the
second.}
\label{tab:boundary_joint}
\footnotesize
\setlength{\tabcolsep}{4pt}
\begin{tabular}{@{}rrrrrrrr@{}}
\toprule
& & \multicolumn{3}{c}{$\theta$ at its true value} & \multicolumn{3}{c}{$\theta$ estimated jointly} \\
\cmidrule(lr){3-5}\cmidrule(lr){6-8}
$I$ & $\kappa_r$ & \% at $\infty$ & median & [q10, q90] & \% at $\infty$ & median & [q10, q90] \\
\midrule
$10$ & 3 & 0.4 & 1.21 & [0.62, 2.88] & 0.0 & 1.15 & [0.62, 3.00] \\
 & 5 & 1.0 & 1.23 & [0.60, 3.61] & 1.5 & 1.09 & [0.58, 2.80] \\
 & 10 & 3.8 & 1.33 & [0.57, 7.95] & 3.0 & 1.13 & [0.54, 4.65] \\
 & 20 & 13.0 & 1.64 & [0.55, $\infty$] & 11.0 & 1.22 & [0.43, $\infty$] \\
 & 50 & 36.8 & 2.75 & [0.37, $\infty$] & 30.0 & 1.48 & [0.32, $\infty$] \\
\midrule
$40$ & 3 & 0.0 & 1.03 & [0.79, 1.44] & 0.0 & 1.09 & [0.79, 1.43] \\
 & 5 & 0.0 & 1.04 & [0.77, 1.53] & 0.0 & 1.03 & [0.76, 1.45] \\
 & 10 & 0.0 & 1.06 & [0.74, 1.76] & 0.0 & 0.99 & [0.75, 1.66] \\
 & 20 & 0.2 & 1.14 & [0.70, 2.29] & 0.0 & 1.18 & [0.65, 2.62] \\
 & 50 & 5.4 & 1.14 & [0.55, 5.78] & 3.0 & 1.01 & [0.51, 2.86] \\
\bottomrule
\end{tabular}

\end{table}

\section{Estimation details}
\label{supp:estimation}

\subsection{The two-stage estimator: negative increments, within-row covariance, and the failure of frequentist corrections}
\label{supp:estimation:wls}

Negative incremental paid amounts need separate treatment. The CNBG
distribution has support $[0, \infty)$, yet paid loss triangles
occasionally record negative incremental amounts, arising from salvage
recoveries, subrogation receipts, or adjuster reserve revisions that
reduce previously recorded payments. The rule applied throughout the
paper is exclusion, stated in Section~\ref{sec:scheduleP}: portfolios
with more than $5\%$ negative observed increments are excluded from the
panel, every remaining negative cell is excluded from both estimation
stages of every procedure, the cumulative development of that accident
year is treated as running from the previous positive entry, and zero
cells are retained. Two sensitivity checks bear on the rule. On the
$255$ portfolios of the outcome-free panel without any negative cell,
every procedure's coverage is lower, since the subset is the sparser
part of the panel (ODP bootstrap $42.0\%$ on $255$, ODP-cond $23.5\%$,
Tweedie-cond $24.5\%$ on $253$, Frequentist CNBG $44.3\%$ on $253$,
WLS-cond $23.9\%$, CNBG-Mod $64.9\%$ on $191$), the paired lead of
CNBG-Mod over the ODP bootstrap is $12$ points on $191$ portfolios
against $10$ on the full $371$, and the misses stay one-sided ($54\%$
below and $4\%$ above the ODP bootstrap's interval, $26\%$ and $9\%$ for
CNBG-Mod). And the earlier implementation of Section~\ref{supp:evidence:real},
which floored non-positive increments at one instead of excluding them,
gave the same picture for the fits it reported. A distributional
extension that admits negative increments, a shifted CNBG or a mixture
with a point mass at the retention boundary, is not pursued.

The second-stage estimator~\eqref{eq:wls} treats cells as
independent, ignoring the within-accident-year covariance
$\Cov(X_{ij}, X_{ij'}) = \kappa_B^{-1}\nu_{ij}\nu_{ij'}$ for $j \neq
j'$ of Proposition~\ref{prop:moments}(iii). With the cell means known
it remains consistent, because each accident year's contribution to the
score is then unbiased, so within-year correlations affect efficiency but
not the estimating equation's unbiasedness; with fitted means its target
is the deflated one of Proposition~\ref{prop:two_stage_target}(iii). A GLS correction would treat each accident
year's vector of squared residuals as a correlated unit and weight by
the inverse fourth-moment covariance matrix; for standard triangles
($I = J = 10$) each row contributes at most $J - i + 1$ cells, so the
within-row correlation affects only a modest fraction of total
variation and the efficiency gain over WLS is small. The WLS
estimator is therefore adequate as a first stage and as a diagnostic.
The modular predictive distribution of
Section~\ref{sec:estimation:ppd} in any case propagates within-year
dependence correctly, by drawing a single frailty $U_i$ per accident
year and conditioning all future cells in that row on the same draw,
and is therefore unaffected by the WLS independence simplification.

Under the free-row first stage the estimator carries no row-frailty
signal (Lemma~\ref{lem:absorption}). In the single-frailty
regime simulated here there is no cell shock for the residual curvature
to detect either, so $\hat\kappa_B \to +\infty$ by construction; with a
finite cell shock it orders $\kappa_c$ without locating it
(Proposition~\ref{prop:two_kappa_id}). No frequentist correction changes this. The
REML correction divides each squared residual by $1 - h_{ij}$, with $h_{ij}$
the Stage-1 leverage; the corner cells $(I,1)$ and $(1,J)$ have
$h_{ij} = 1$ and are excluded. In a simulation with $M = 500$ replications
at $\alpha = 3$, the uncapped degrees-of-freedom-adjusted estimator
returns mean $\hat\kappa_B$ between $423{,}000$ and $461{,}000$ across the
five true values, and the leverage-corrected estimator excluding
$h_{ij} \ge 0.90$ between $240{,}000$ and $273{,}000$: a $1.1$-fold spread
against a $17$-fold variation in truth, with no ordering. The capped
estimator used by the Frequentist CNBG bootstrap ($\hat\kappa_B \le 1000$
before the degrees-of-freedom factor) saturates its cap on $63$ to $71\%$
of replications whatever the true value
(Section~\ref{supp:evidence:sim}). The Bayesian regularisation of
Section~\ref{sec:estimation:bayes}, and the prior-carried dispersion of
the predictive, follow from this.

\subsection{The working likelihood: cell-level shapes, curvature adjustment, and sandwich variance}
\label{supp:estimation:bayes}

This section continues Section~\ref{sec:estimation:bayes} of the main
text from the working density \eqref{eq:working_lik}.

\paragraph{The sensitivity prior.} Besides the reference prior, a second
prior, $\kappa_B \sim \mathrm{Gamma}(1.78, 0.22)$ with mean $8$ and standard
deviation about $6$, is used as a sensitivity check (CNBG-Mod-EmpB). Its
parameters were matched to the mean and variance of the cross-portfolio
distribution of negative binomial dispersion estimates in an unpublished
study by the author; the prior is fully specified by its two parameters,
and nothing in this paper depends on that study.

The choice of $s$ matters for the curvature of the working likelihood
and not for its location. For any fixed $s$ the working score with
respect to the parameters is proportional to $s\,(z_{ij} -
\mu_z^{ij})/(\mu_z^{ij})^2 \cdot \partial\mu_z^{ij}/\partial\theta$,
which has zero mean under the CNBG model because $\E[z_{ij}] =
\mu_z^{ij}$; the working density is therefore an unbiased estimating
equation whatever $s$ is, and the shape must be held fixed rather than
made a function of $(\kappa_B, A)$, since a parameter-dependent shape
introduces $\E[\log z_{ij}]$ terms into the score whose expectation the
CNBG model does not match. The shape sets the working squared
coefficient of variation $\mathrm{CV}^2_W(z_{ij}) = 1/s$ and hence the
curvature of the working log-likelihood, and that curvature should
match the true variability of the squared residuals. In the Gaussian
limit of a large cell, $(X_{ij} - \nu_{ij})^2$ is $\mathrm{Var}(X_{ij})$
times a $\chi^2_1$ variate, so $\mathrm{CV}^2(z_{ij}) = 2$ and the
matching shape is $s = 1/2$; Proposition~\ref{prop:fourth_moment} below
shows that under the CNBG model $\mathrm{CV}^2(z_{ij})
\ge 2$ for every cell, with the excess over $2$ largest for small cells
and small $\kappa_B$. A fixed shape of $s = 2$ would overstate the
per-cell precision by a factor of at least four and produce posteriors
that contract far more than the data warrant. The correction has two parts, both standard in the
composite-likelihood literature, and neither is a single scalar. First,
the shape is set cell by cell at a first-pass fit,
\begin{equation}\label{eq:magnitude_adj}
  s_{ij} \;=\; \frac{1}{\widehat{\mathrm{CV}}{}^2_{\mathrm{CNBG}}(z_{ij})}
  \;=\; \frac{(\hat\mu_z^{ij})^2}{\widehat{\E}[(X_{ij}-\nu_{ij})^4] - (\hat\mu_z^{ij})^2},
\end{equation}
with the fourth moment from \eqref{eq:fourth_assembled} at the
first-pass $(\hat A, \hat\kappa_B)$ and then held fixed, so that the
score is unbiased for the fixed shapes; the shapes are themselves
functions of the first pass, and the dependence this induces between them
and the residuals is part of what Remark~\ref{rem:twostep} leaves
underived. The fourth moment also involves the severity constants
$c_3$ and $c_4$ of Section~\ref{app:cnbg_moments}, which depend on
$\alpha$ and $\bar Y$ separately and are not identified by the
second-moment procedure used here, which identifies only their
composite $A$; in deployment they are evaluated
at $\bar Y = \hat A\alpha/(\alpha+1)$ for a stated $\alpha$, taken at
its true value in the simulation and at $\alpha = 3$, the convention of
the exact engine of Section~\ref{rem:exact_lik}, on real data. The
shapes depend on the convention (at $A = 1$, $\kappa_B = 10$, $\nu = 1$,
$s_{ij} = 0.19$ at $\alpha = 1$ and $0.23$ at $\alpha = 3$), but the
Schedule~P results hardly do: under the historical implementation, fitting
the panel with $\alpha = 1$ or $\alpha = 10$ in place of $\alpha = 3$
moved pooled CNBG-Mod coverage from $63.8\%$ to $62.2\%$ in both cases. The alternative
is a fixed working shape, $s = 1/2$, with the sandwich variance
estimated from the ten row-level scores; it needs no severity split but
rests its curvature estimate on ten independent rows. For a Gamma working density of shape $s_{ij}$ the
cell-level ratio $\mathcal{V}^{ij}/\mathcal{I}_W^{ij}$ equals
$s_{ij}\,\mathrm{CV}^2(z_{ij})$, so
\eqref{eq:magnitude_adj} makes it one for every cell: this is the
inverse-fourth-moment weighting of \citet{DavidianCarroll1987} written
as a working density, it corrects both parameter directions at once,
and it reduces to $s = 1/2$ in the Gaussian limit. Second, what
\eqref{eq:magnitude_adj} cannot absorb is the within-row score
covariance $\Delta$ of Proposition~\ref{prop:godambe}, which
enters $\mathcal{V}$ but not $\mathcal{I}_W$. For that remainder we
apply the curvature adjustment of \citet{Ribatet2012}: the working
log-likelihood is evaluated at $\theta^\ast = \hat\theta +
C(\theta - \hat\theta)$ with $C = M^{-1}M_A$, where $M^\top M =
\mathcal{I}_W$ and $M_A^\top M_A = \mathcal{I}_W\mathcal{V}^{-1}\mathcal{I}_W$
are Cholesky factors and $\mathcal{V}$ is the row-level score variance
of Proposition~\ref{prop:sandwich}, so that the adjusted posterior has
asymptotic covariance $\mathcal{I}_W^{-1}\mathcal{V}\mathcal{I}_W^{-1}$
in every direction. A scalar magnitude adjustment
\citep{Pauli2011}, dividing the log-likelihood by
$\operatorname{tr}(\mathcal{I}_W^{-1}\mathcal{V})/2$, matches only the
trace of the sandwich; it is retained as a diagnostic. For the
absorbed-mean fit that the predictive uses, whose working mean is
$A\hat\nu_{ij}$ with the quadratic term at zero, the parameter is the
scalar $\log A$ with $\kappa_B$ held at its prior, and the adjustment is
one-dimensional: $C = (\mathcal{I}_W/\mathcal{V})^{1/2}$ in that
coordinate. All estimates and diagnostics in
Sections~\ref{sec:simulation}--\ref{sec:scheduleP} are computed under
this specification, with two further elements of the free-level fit.
First, the squared residuals entering \eqref{eq:working_lik} are divided
by $1 - h_{ij}$, the leverage of the Stage-1 cell. This is the
linearised approximation $\E[(X_{ij} - \hat\nu_{ij})^2] \approx
\Var(X_{ij})(1 - h_{ij})$, the diagonal of $(I - H)\Sigma(I - H)^\top$
under working independence with the matched conditional variance; the
within-row covariance of the shared frailty enters the off-diagonal
terms and is neglected, and the correction is not a CNBG identity. Its
scope was checked on $500$ simulated triangles at $(\kappa_r, \alpha) =
(5, 3)$ and $(50, 3)$: by development position the raw squared
residuals average $0.40$--$0.89$ of the target variance under free
levels ($0.46$--$0.92$ anchored) and the corrected ones $0.91$--$1.12$
($0.89$--$1.07$ anchored), the target being the absorbed variance
$A\hat\nu_{ij}$ under free levels and $A\hat\nu_{ij} +
\hat\nu_{ij}^2/\kappa_B$ anchored; leverage-one cells remain excluded. Second, under free levels the row effects absorb
the frailty, so the working estimating equation for $A$ converges to
the absorbed target $A^\ast$ with the quadratic term at zero
($\kappa_B^\ast = \infty$ in every grid cell of
Section~\ref{sec:simulation}), a pseudo-true value whose distance from
$A$ is measured and is not zero: with the pattern known and the row
leverage removed exactly, the corrected response of a row with $N$
claims has conditional mean $A\,N\alpha/(N\alpha + 1)$, $0.99A$ at $30$
claims and $\alpha = 3$ but $0.58A$ at one claim per row on average, a
gap that replicating rows at fixed volume does not close; the $A$ posterior that enters the
predictive is therefore taken from a fit with $\mu_z^{ij} =
A\hat\nu_{ij}$, while the $\kappa_B$ diagnostic of
Section~\ref{supp:evidence:real} uses the full working mean. On
the simulation grid the adjusted $A$ interval has frequentist coverage
of $90\%$ pooled ($87$--$95\%$ by cell); with the full mean under free
levels it was $69$--$80\%$, the prior-allotted $\nu^2/\kappa_B$ term
displacing $A$ downward.

The working density is misspecified, since $z_{ij}$ is not Gamma under
the CNBG model; its score is nevertheless an unbiased estimating equation
for $(\kappa_B, A)$ when the cell means are known, because the working
mean then equals the true variance, which is what composite-likelihood
Bayes requires \citep{Lindsay1988,
VarinReidFirth2011, Pauli2011, Ribatet2012}. Its credible intervals are
those of an adjusted working likelihood, whose asymptotic covariance the
adjustments above match to the sandwich; the procedure is consistent under
anchored levels (Proposition~\ref{prop:sandwich}(i)) and converges to the
absorbed target under free levels, which is why the predictive takes $A$
from the absorbed fit and $\kappa_B$ from the prior. The construction is
used here because two conditions coincide: the dispersion is identified
only through the second moment of paid amounts, and the exact posterior
costs roughly $200$ times as much. Where either fails, as in the count
model of \citet{VanOirbeek2026nbcl}, the full likelihood is used instead. On
correctly specified data $14{,}937$ of the $15{,}000$ full-mean fits of
the grid return a usable posterior, the others failing at
initialisation; on Schedule~P the working fit returns no usable posterior
on $81$ of the $452$ portfolios of the outcome-free panel, mostly the
smallest triangles (median realised reserve \$$518$k against \$$1{,}760$k
elsewhere).

\begin{proposition}[Consistency and sandwich variance of the working
likelihood estimator, conditional on the first-stage means]\label{prop:sandwich}
Let $\hat\theta_n = (\hat\kappa_n, \hat{A}_n)$ denote the maximiser of
the working log-likelihood $\ell_W(\theta) = \sum_{(i,j)\in\mathcal{D}}
\log p_W(z_{ij};\theta)$ derived from~\eqref{eq:working_lik} at any fixed
shape $s$, and let $S_i(\theta) = \sum_j \nabla_\theta \log
p_W(z_{ij};\theta)$ be the row-level score. The asymptotic regime is
$I \to \infty$ with the development length $J$ fixed, so that the
number of observed cells $n = |\mathcal{D}|$ grows proportionally to
$I$; the normalisation is by the number of independent rows. Assume:
(i) the true data-generating process is the CNBG model of
Definition~\ref{def:cnbg} with $\kappa_0 > 0$ and $A_0 > 0$;
(ii) the cell means $\nu_{ij}$ are known, so that $z_{ij}$ is observed and
the row scores involve no common estimated quantity; with fitted means
the result is conditional on the first stage (Remark~\ref{rem:twostep});
under unrestricted row effects
the estimator of $\kappa_r$ is degenerate
(Theorem~\ref{thm:boundary}) and the result below applies
to the absorbed target of the equation, $\kappa_B^\ast = \infty$ and
the $A^\ast$ that solves the expected working score with the
quadratic term at zero, whose distance from the true $A$ is measured,
not assumed, in Section~\ref{supp:estimation:bayes};
(iii) accident years are mutually independent
(Proposition~\ref{prop:moments}(iv)), so that the row scores are
independent across $i$; cells within a row are positively correlated
through the shared frailty (Proposition~\ref{prop:moments}(iii)), and
the row is the independent unit; and
(iv) the row-level expected Hessian and score covariance,
$H_I = -I^{-1}\sum_i \E[\nabla_\theta S_i(\theta_0)]$ and
$V_I = I^{-1}\sum_i \mathrm{Var}[S_i(\theta_0)]$, converge to finite
limits $H$ and $V$ with $H$ invertible, and the row scores satisfy the
Lindeberg condition \citep[Section~2.8]{vanderVaart1998}; with $J$
fixed, $\mathrm{Var}[S_i(\theta_0)]$ is bounded uniformly in $i$ under
(i), so (iv) is a regularity rather than a growth condition; and
(v) the parameter space is the compact rectangle $[\kappa_{\min},
\kappa_{\max}] \times [A_{\min}, A_{\max}]$ with $\kappa_{\max}$ the
implementation's cap and $A_{\min} > 0$, and under anchored levels the
target $\theta_0$ lies in its interior.
Then, under anchored levels:
\begin{enumerate}[label=(\roman*), nosep]
\item $\hat\theta \xrightarrow{p} \theta_0 = (\kappa_0, A_0)$
      as $I \to \infty$.
\item $\sqrt{I}(\hat\theta - \theta_0)
      \xrightarrow{d} \mathcal{N}(0,\, \Sigma_{\mathrm{sandwich}})$,
      where the sandwich variance is
      \begin{equation}\label{eq:sandwich}
        \Sigma_{\mathrm{sandwich}}
        \;=\; H^{-1}\,V\,H^{-1},
      \end{equation}
      with $H$ the expected row-level working information and $V$ the
      true row-level score variance. In the cell-indexed notation of
      \eqref{eq:info_matrix}--\eqref{eq:score_var} used below,
      $\mathcal{I}_W$ and $\mathcal{V}$ are the row sums of the cell
      quantities (and, for $\mathcal{V}$, of the within-row cross
      terms), so $H = \mathcal{I}_W/I$ and $V = \mathcal{V}/I$.
\end{enumerate}
Under free levels, where the absorbed target has $\kappa_B^\ast = \infty$,
only a boundary statement is made: in the parameterisation $B = 1/\kappa_B
\in [0, B_{\max}]$ the maximiser satisfies $\hat B \xrightarrow{p} 0$
and $\hat A \xrightarrow{p} A^\ast$, and no Gaussian limit is claimed
for $\hat B$; the constrained asymptotics at the boundary are not
derived here. No claim is made for square triangles growing in both dimensions:
there $n \propto I^2$, and because the within-row score cross-terms of
the shared frailty grow with the row length, the row-score variance
need not be $O(J)$, so a $\sqrt{n}$ rate does not follow from a
Lindeberg condition and would require an explicit assumption on the
growth of $V_I$. The $10 \times 10$ triangles of this paper are far
from either regime; the proposition supplies the form of the
variance that the adjustment of Section~\ref{sec:estimation:bayes}
targets, and the finite-sample calibration of that adjustment is
measured by simulation.
\end{proposition}

\begin{remark}[Fitted means]\label{rem:twostep}
The fits of this paper use the first-stage estimates $\hat\nu_{ij}$ in
place of $\nu_{ij}$. They enter both the squared residuals and their
modelled variance, and they share the estimated development pattern
across rows, so the row scores are not independent as (iii) assumes and
the first-stage estimation error adds a term to $V$ that the division of
the squared residuals by $1 - h_{ij}$ removes only to first order and only
on the diagonal. The joint two-step sandwich is not derived here; the
proposition should be read as conditional on the first stage, and the
finite-sample calibration of the adjusted $A$ posterior, which covers the
true $A$ on $90\%$ of the simulation grid (Section~\ref{sec:predictive}),
is the measured substitute, and its five-point shortfall is reported
as such.
\end{remark}

\begin{proof}
\textit{(i)} The row-level score $S_i(\theta_0) =
\sum_j \partial \log p_W(z_{ij};\theta_0)/\partial\theta$
satisfies $\E_{\mathrm{CNBG}}[S_i(\theta_0)] = 0$ because the Gamma
working score is proportional to $z_{ij}/\mu_z^{ij} - 1$ and
$\E_{\mathrm{CNBG}}[z_{ij}] = \mu_z^{ij}$ by construction, so each
cell-level score has zero mean and the sum over cells within row $i$
inherits this property. Consistency follows from standard M-estimation
theory applied to the $I$ independent row contributions
\citep[Theorem~5.9]{vanderVaart1998}, provided the parameter space is
compact and the row-level working log-likelihood satisfies a uniform
law of large numbers, the first by (v) and the second under (i)--(iv)
with $J$ fixed.

\textit{(ii)} Asymptotic normality follows from the sandwich theorem
for misspecified likelihoods \citep[Section~5.3]{White1982}, applied
to the $I$ independent row scores $S_i(\theta)$ under (iv): by the
Lindeberg--Feller theorem and the usual Taylor expansion of the
estimating equation around $\theta_0$,
\[
  I^{-1/2}\sum_i S_i(\theta_0) \xrightarrow{d} \mathcal{N}(0, V),
  \qquad
  I^{-1}\sum_i \nabla_\theta S_i(\theta_0) \xrightarrow{p} -H,
  \qquad
  \sqrt{I}(\hat\theta - \theta_0) \xrightarrow{d}
  \mathcal{N}(0,\, H^{-1} V H^{-1}).
\]
The cell-indexed form $\mathcal{I}_W^{-1}\mathcal{V}
\mathcal{I}_W^{-1}$ used in Section~\ref{sec:estimation:bayes} equals
$H^{-1}VH^{-1}/I$, the asymptotic covariance of $\hat\theta$ itself
rather than of $\sqrt{I}(\hat\theta - \theta_0)$, since
$\mathcal{I}_W = I H$ and $\mathcal{V} = I V$ with $J$ fixed. The working information
$\mathcal{I}_W$ is invertible when the design matrix of the variance
regression~\eqref{eq:wls} has full rank, which requires at least two
distinct values of $\hat\nu_{ij}$, the condition under which the
variance function identifies $(A, B)$.
\end{proof}

Without the adjustments a posterior would not perform this correction:
under misspecification it is asymptotically normal about the working
maximiser with covariance $\mathcal{I}_W^{-1}$, the inverse working
curvature, not the sandwich \citep{KleijnvanderVaart2012}, which is why the
cell-level shapes \eqref{eq:magnitude_adj} and the curvature adjustment
are applied. At $I = J = 10$ the asymptotic regime is approached for $A$
but not for $\kappa_B$, whose posterior is the prior's statement more than
the data's; predictive coverage is therefore measured by simulation, not
inferred from this approximation.

For completeness, the explicit components of~\eqref{eq:sandwich} for a
Gamma working density of shape $s$ are
\begin{align}
  \mathcal{I}_W^{ij}
    &= \frac{s_{ij}}{(\mu_z^{ij})^2}
       \begin{pmatrix}
         (\hat\nu_{ij}^2/\kappa_0^2)^2 &
         -\hat\nu_{ij}^2\,\hat\nu_{ij}/\kappa_0^2 \\
         -\hat\nu_{ij}^2\,\hat\nu_{ij}/\kappa_0^2 &
         \hat\nu_{ij}^2
       \end{pmatrix},
       \label{eq:info_matrix}\\
  \mathcal{V}^{ij}
    &= \frac{s_{ij}^2\,\E_{\mathrm{CNBG}}[(z_{ij} - \mu_z^{ij})^2]}{(\mu_z^{ij})^4}
       \begin{pmatrix}
         (\hat\nu_{ij}^2/\kappa_0^2)^2 &
         -\hat\nu_{ij}^2\,\hat\nu_{ij}/\kappa_0^2 \\
         -\hat\nu_{ij}^2\,\hat\nu_{ij}/\kappa_0^2 &
         \hat\nu_{ij}^2
       \end{pmatrix},
       \label{eq:score_var}
\end{align}
the off-diagonal entries negative because $\partial\mu_z^{ij}/\partial\kappa_B
= -\hat\nu_{ij}^2/\kappa_B^2 < 0$ while $\partial\mu_z^{ij}/\partial A =
\hat\nu_{ij} > 0$; these are the diagonal (within-cell) contributions,
and $\mathcal{V}$ additionally carries the within-row cross terms.
The cell-level ratio $\mathcal{V}^{ij}/\mathcal{I}_W^{ij} =
s_{ij}\,\E[(z_{ij}-\mu_z^{ij})^2]/(\mu_z^{ij})^2 = s_{ij} \cdot
\mathrm{CV}^2(z_{ij})$ is the working shape times the
true squared coefficient of variation of $z_{ij}$ under the CNBG model.
Under a Gamma$(s,\cdot)$ working family the implied
$\mathrm{CV}^2_W(z_{ij}) = 1/s$, whereas under the CNBG model the true
value is at least $2$; the unadjusted working curvature is therefore
too sharp by the factor $s \cdot \mathrm{CV}^2(z_{ij})$,
which the cell-level shapes \eqref{eq:magnitude_adj} remove cell by
cell.

\begin{proposition}[Fourth central moment and the curvature of the
working likelihood]\label{prop:fourth_moment}
Under Definition~\ref{def:cnbg}, with $\nu_{ij} = \mu_{ij}\bar{Y}$,
$A = (1 + 1/\alpha)\bar Y$,
$B = \kappa_2(M) = \kappa_r^{-1}+\kappa_c^{-1}+(\kappa_r\kappa_c)^{-1}$, and
$\mu_z^{ij} = A\nu_{ij} + B\nu_{ij}^2$:
\begin{enumerate}[label=(\roman*), nosep]
\item The fourth central moment of $X_{ij}$ is
\begin{equation}\label{eq:fourth_assembled}
  \E[(X_{ij}-\nu_{ij})^4]
  \;=\; 3\bigl(A\nu_{ij} + B\nu_{ij}^2\bigr)^2
  \;+\; c_4\nu_{ij} \;+\; (4c_3 + 3A^2)\,B\,\nu_{ij}^2
  \;+\; 6A\,\kappa_3(M)\,\nu_{ij}^3 \;+\; \kappa_4(M)\,\nu_{ij}^4,
\end{equation}
with $c_3,c_4$ the severity constants of
Section~\ref{app:cnbg_moments} and $B=\kappa_2(M)$, $\kappa_3(M)$,
$\kappa_4(M)$ the cumulants of the cell multiplier $M=U_i\varepsilon_{ij}$
given in \eqref{eq:kappa2_M}--\eqref{eq:kappa4_M}; this is
equation~\eqref{eq:fourth_full_expanded}.
\item The true squared coefficient of variation satisfies
\begin{equation}\label{eq:cv_lower}
  \mathrm{CV}^2(z_{ij})
  \;=\; \frac{\E[(X_{ij}-\nu_{ij})^4]}{(\mu_z^{ij})^2} - 1
  \;\ge\; 2
\end{equation}
for all $\kappa_B > 0$, $\alpha > 0$, and $\nu_{ij} > 0$, with equality
only in the Gaussian limit. Consequently, for a Gamma working density of
shape $s$, writing $\mathcal{V} = \mathcal{V}_{\mathrm{diag}} +
\mathcal{V}_{\mathrm{cross}}$ with $\mathcal{V}_{\mathrm{diag}} =
\sum_{ij} \mathcal{V}^{ij}$ the within-cell (working-independence)
part and $\mathcal{V}_{\mathrm{cross}}$ the within-row
cross-covariances of the cell scores,
$\mathcal{V}_{\mathrm{diag}} \succeq 2s\,\mathcal{I}_W$ in the Loewner
order, and the working-independence sandwich
$\mathcal{I}_W^{-1}\mathcal{V}_{\mathrm{diag}}\mathcal{I}_W^{-1}
\succeq 2s\,\mathcal{I}_W^{-1}$: the unadjusted working posterior
understates that part of the sampling variability of the working
estimator by a factor of at least $2s$ in variance ($4$ for $s = 2$,
$1$ for $s = 1/2$). No Loewner bound is claimed for the full
$\mathcal{V}$: $\mathcal{V}_{\mathrm{cross}}$ is not positive
semi-definite in general, and the full sandwich is handled separately
by the curvature adjustment, which uses the estimated $\mathcal{V}$
(Section~\ref{sec:estimation:bayes}). At the simulation calibration
the cross terms are positive in the $\kappa_B$ direction
(Proposition~\ref{prop:godambe}) and the full ratio exceeds the
diagonal one (eigenvalues of $\mathcal{I}_W^{-1}\mathcal{V}$ of
$1.1$--$1.5$ against $1.04$--$1.08$ for the diagonal part with
cell-level shapes, Section~\ref{supp:evidence:sim}).
\end{enumerate}
\end{proposition}

\begin{proof}
\textit{(i)} This is equation~\eqref{eq:fourth_full_expanded}, derived in
Section~\ref{app:cnbg_moments} by the law of total cumulants applied to
the two-level multiplier $M=U_i\varepsilon_{ij}$; it is restated here as
\eqref{eq:fourth_assembled} for use in part~(ii).

\textit{(ii)} Every coefficient in \eqref{eq:fourth_assembled} is
strictly positive (the severity constants $c_3,c_4>0$, and
$B,\kappa_3(M),\kappa_4(M)>0$ for all $\kappa_r,\kappa_c>0$ by
\eqref{eq:kappa2_M}--\eqref{eq:kappa4_M}), so
$\E[(X_{ij}-\nu_{ij})^4] \ge 3(A\nu_{ij}+B\nu_{ij}^2)^2 = 3(\mu_z^{ij})^2$,
whence
\[
  \mathrm{CV}^2(z_{ij})
  = \frac{\E[(X_{ij}-\nu_{ij})^4]}{(\mu_z^{ij})^2} - 1
  \;\ge\; 2 .
\]
By \eqref{eq:info_matrix}--\eqref{eq:score_var}, $\mathcal{V}^{ij} =
s\,\mathrm{CV}^2(z_{ij})\,\mathcal{I}_W^{ij} \succeq
2s\,\mathcal{I}_W^{ij}$ for every cell, since each $\mathcal{I}_W^{ij}$
is a positive multiple of a rank-one positive semi-definite matrix;
summing over cells gives $\mathcal{V}_{\mathrm{diag}} \succeq
2s\,\mathcal{I}_W$, and conjugating by $\mathcal{I}_W^{-1}$ preserves
the Loewner order, hence
$\mathcal{I}_W^{-1}\mathcal{V}_{\mathrm{diag}}\mathcal{I}_W^{-1}
\succeq 2s\,\mathcal{I}_W^{-1}$. The cross terms are excluded: the
within-row score covariances form a symmetric matrix that need not be
positive semi-definite, so the inequality does not extend to
$\mathcal{V}$ by this argument.
\end{proof}

The proposition orders the unadjusted working posterior against the
sampling variability of its own estimator, not against the exact
posterior, which is compared empirically in
Section~\ref{rem:exact_lik}. The excess of $\mathrm{CV}^2$ over $2$
grows with the $\nu_{ij}^1$ term, which is proportional to $\bar{Y}^3$
and dominates for small cells, and with the $\kappa_4(M)$ coefficient of
$\nu_{ij}^4$, which grows as $\kappa_B \to 0$, so the unadjusted working
curvature is most over-sharp at high heterogeneity and for cells with
small expected amounts. With $s = 1/2$ the residual factor is
$\mathrm{CV}^2/2 \ge 1$, which the cell-level shapes
\eqref{eq:magnitude_adj} absorb; with a fixed $s = 2$ the factor is at
least $4$, and a posterior built on that shape contracts sharply around
a value the data do not support. Neither
statement is a statement about predictive coverage of the reserve,
which is measured directly in Section~\ref{sec:simulation}.

\subsection{The exact frailty-augmented posterior and the sampler settings}
\label{supp:estimation:exact}
\label{rem:exact_lik}

Throughout this subsection the generative model is the single-frailty
specialisation $\kappa_c\to\infty$ of Definition~\ref{def:cnbg}
($\varepsilon_{ij}\equiv1$), consistent with the simulation scope of
Section~\ref{sec:simulation}; the latent variables are then the row
frailties $\lambda_i$ alone and the conditional law given $\lambda_i$ is
compound Poisson--Gamma.

The Bayesian quasi-likelihood of Section~\ref{sec:estimation:bayes} is an
inference procedure based on the second-moment equation of the CNBG
model, not a Bayesian fit of the generative model itself. The latter
exists in principle and is theoretically the correct object: augment the
parameter space with the latent accident-year frailties
$(\lambda_i)_{i=1}^I$ and target the joint posterior
\begin{equation}\label{eq:exact_posterior}
  p(\kappa_B, A, \lambda_1,\ldots,\lambda_I \mid \mathcal{D})
  \;\propto\;
  \prod_{i=1}^I \left[
    p(\lambda_i \mid \kappa_B)\,
    \prod_{j:(i,j)\in\mathcal{D}} p(X_{ij} \mid \lambda_i, A)
  \right]
  \cdot p(\kappa_B)\,p(A),
\end{equation}
where $p(\lambda_i|\kappa_B) = \mathrm{Gamma}(\kappa_B, \kappa_B/\mu_i)$
is the frailty prior and $p(X_{ij}|\lambda_i, A)$ is the conditional
likelihood, which conditional on $\lambda_i$ is a Tweedie compound
Poisson-Gamma:
\begin{equation}\label{eq:tweedie_conditional}
  X_{ij} \mid \lambda_i \;\sim\;
  \mathrm{Tweedie}\!\left(\mu_{ij}^* = \lambda_i w_j \bar{Y},\;
  \phi^* = A\,(\mu_{ij}^*)^{1-p_c},\;
  p_c = \frac{\alpha+2}{\alpha+1}\right),
\end{equation}
so that $\phi^*(\mu_{ij}^*)^{p_c} = A\mu_{ij}^*$ reproduces the
conditional variance of \eqref{eq:k2_cond}; in terms of the
micro-level parameters, $\phi^* = (1+1/\alpha)\,\bar Y^{\alpha/(\alpha+1)}
(\lambda_i w_j)^{-1/(\alpha+1)}$. The conditional density is evaluated
by a truncated compound Poisson--Gamma summation over the latent claim
counts (numerically equivalent to the \citet{Dunn2005} series).
Reparametrising as $(A, p_c)$ gives a self-contained exact likelihood;
for zero cells, $p(X_{ij} = 0 \mid \lambda_i) = e^{-\lambda_i w_j}$
exactly. Note that $p_c$, and hence the zero mass, requires $\alpha$
separately from $\bar Y$, which the second-moment procedure does not
identify (Corollary~\ref{cor:cumulant}); the exact posterior is therefore a
reference object that needs either a count triangle or a prior on
$\alpha$ to be deployed. The exact frailty-augmented posterior is fit by Hamiltonian Monte Carlo
(\texttt{rstan}); a reference implementation is provided in the
\texttt{hgr} R package (\url{https://github.com/robin-vo/hgr}).

The posterior geometry of \eqref{eq:exact_posterior} is poorly
conditioned for HMC, the centred Gamma frailty creating a
$\kappa_B$--$\lambda$ funnel. The reparameterised reference implementation
resolves it on the benchmark triangles (Taylor--Ashe and RAA: all
$\hat R \le 1.001$, zero divergent transitions, effective sample sizes
above $4{,}000$) at $1.6$--$4.6$ minutes per triangle against
$0.6$--$1.1$ seconds for the working fit, a factor of $175$--$260$. That
cost, not convergence failure, is why the exact posterior is a reference
and not the production estimator. The working likelihood is validated by
its own properties, consistency and the sandwich covariance under anchored
levels (Proposition~\ref{prop:sandwich}) and the curvature relation of
Proposition~\ref{prop:fourth_moment}, and by the predictive coverage
measured in simulation, not by proximity to the exact posterior. On the
two benchmark triangles the exact $95\%$ interval for $\kappa_B$ is wider
than the adjusted working one by factors of $1.37$ (Taylor--Ashe) and
$1.51$ (RAA), and the medians differ, $14.7$ against $12.8$ and $12.6$
against $6.4$, with interval overlaps of $73\%$ and $47\%$ of the union:
on a single $10 \times 10$ triangle the split of variance between the
linear and quadratic terms is weakly determined, and two estimating
functions resolve it differently. Under free levels the exact posterior
is pushed upward in $\kappa_r$ by a mechanism related to
Theorem~\ref{thm:boundary}, since
its frailty prior is centred on fitted levels that already carry the
realised frailty.

The sampler settings differ by analysis. The exact-versus-working
comparison of Section~\ref{rem:exact_lik} runs four chains of 2{,}000
iterations (1{,}000 warmup) with \texttt{adapt\_delta = 0.95} and
\texttt{max\_treedepth = 12} for both engines; the benchmark-triangle
fits of Section~\ref{supp:evidence:real} use two chains of
1{,}200 iterations (600 warmup) with the same control settings. For
the large-scale simulation of Section~\ref{sec:simulation}, where
$M \times K = 7{,}500$ posterior fits are required, two chains of
1{,}000 iterations (500 warmup) with Stan default control suffice at a
manageable computational cost, and the Schedule~P backtest of
Section~\ref{sec:scheduleP} uses two chains of 800 iterations (400
warmup) with default control. A fit is scored as usable when the
rank-normalised $\hat R$ of $\kappa_B$ and $A$ is below $1.05$ and the
bulk effective sample size exceeds $100$; fits that fail the rule are
counted and reported, not dropped silently. In the simulation,
$22{,}436$ of the $22{,}500$ fits (two priors plus the absorbed-mean fit
for $A$, $7{,}500$ replications) returned draws and every one of them
meets the rule ($\hat R \leq 1.034$, bulk ESS $\geq 113$, tail ESS
$\geq 60$, two divergent transitions in total); the $64$ others failed
at initialisation: $34$ reference-prior and $29$ wide-prior full-mean
fits ($27$ replications failed under both priors, $36$ under at least
one) and one absorbed-mean fit, so the predictive, which uses the
absorbed-mean $A$ posterior, is available on $7{,}499$ replications
per prior. On a random sample of $300$ replications refitted
with their seeds, the worst rank-normalised $\hat R$ is $1.024$, the
smallest bulk and tail ESS are $191$ and $107$, and no transition hit
the maximum tree depth. On the outcome-free Schedule~P panel, $367$ of the
$371$ full fits that returned draws meet the rule, and the predictive,
which uses the absorbed-mean fit, returns an interval on the same $371$
portfolios; the four fits that fail the rule are retained in every Schedule~P table, so that the usable-fit denominator of $371$ counts them, a departure from the rule that is stated here and reported as a sensitivity: excluding the four moves pooled CNBG-Mod coverage from $68.5\%$ on $371$ to $68.4\%$ on $367$ (three of the four are covered), and under the historical implementation excluding its seven failing fits moved it by $0.4$ points. The per-fit
diagnostics, together with every per-replication and per-portfolio
result file behind the tables, are provided as replication materials
accompanying the paper. For the benchmark triangles, whose cell amounts are
several orders of magnitude larger than the simulated ones, the inputs
are rescaled by the mean fitted value $\bar\nu = \mathrm{mean}(\hat\nu_{ij})$
before passing to Stan ($\tilde z_{ij} = z_{ij}/\bar\nu^2$,
$\tilde\nu_{ij} = \hat\nu_{ij}/\bar\nu$), which keeps all inputs of
order 1 and prevents numerical overflow in the term
$\hat\nu_{ij}^2/\kappa_B$ during initialisation; the posterior draws for
$A$ are then back-transformed by $A^{(b)}_{\mathrm{raw}} =
A^{(b)}_{\mathrm{scaled}} \times \bar\nu$ before use in the predictive
simulation below. The simulation and Schedule~P fits pass the raw
inputs.

\section{The modular predictive and the conditional WLS bootstrap}
\label{supp:predictive}

\subsection{Justification of the modular predictive}\label{sec:estimation:ppd}

The modular approximation of Section~\ref{sec:predictive} combines a
prior draw for row dispersion, a sampling-distribution draw for Stage-1
coefficients, the working posterior for $A$, and a quasi-Poisson
conjugate update for row frailty. These components do not constitute the
posterior predictive of a single joint model. Free row coefficients
already absorb realised frailty (Lemma~\ref{lem:absorption}), so the use
of both draws needs empirical justification. In
Section~\ref{sec:simulation:oracle}, their contributions overlap: the
frailty draw moves coverage by about one point when the coefficient
draw is present and by twelve when it is absent (Table~\ref{tab:steps}).
The exact posterior of Section~\ref{supp:estimation:exact} provides a
reference for $(\kappa_r, A)$ with fitted means held fixed. A fully joint
treatment would instead place levels, frailties and dispersions in one
likelihood under anchored levels. That fit is not attempted here, so we
reserve the term ``posterior predictive'' for other constructions.

In the simulation we work with the \emph{normalised frailty}
$U_i = \lambda_i / \mu_i$ rather than the raw intensity $\lambda_i$ of
Definition~\ref{def:cnbg}. Under $\lambda_i
\sim \mathrm{Gamma}(\kappa_r,
\kappa_r/\mu_i)$, the rescaling gives $U_i \sim \mathrm{Gamma}(\kappa_r,
\kappa_r)$ with $\E[U_i] = 1$ and $\mathrm{Var}(U_i) = 1/\kappa_r$. The
accident-year mean $\mu_i$ is absorbed into the fitted $\hat\nu_{ij}$,
leaving the frailty draw dependent only on $\kappa_r$, which separates
frailty uncertainty from mean-structure uncertainty in the simulation
loop without changing the joint distribution.

The following CNBG simulation combines parameter uncertainty and process
variance while conditioning on the observed triangle. It propagates
mean-structure uncertainty through the sampling distribution of the
Stage-1 estimates; the oracle study in
Section~\ref{sec:simulation:oracle} measures how fully this captures it.
The severity composite $A$ is drawn from its
working-likelihood posterior $\{A^{(b)}\}_{b=1}^B$, since it is
identified (Proposition~\ref{prop:identification}). The predictive
$\kappa_r$ is drawn from its \emph{prior}, not its posterior, and the
procedure is labelled accordingly throughout: by
Lemma~\ref{lem:absorption} the non-diversifying reserve term
$\kappa_r^{-1}\sum_i S_i^2$ (equation~\eqref{eq:reserve_var}) dominates reserve
uncertainty, yet $\kappa_r$ is not recoverable by the free-level fit,
so its predictive contribution is supplied by the prior on the
frailty \emph{dispersion}. Whether that makes the interval too wide or
too narrow depends on where the true $\kappa_r$ sits relative to the
prior. Section~\ref{sec:simulation} measures the consequences at fixed
true values. The conjugate update~(16) uses the row-level signal $\sum_j X_{ij}$, so the prior supplies only the dispersion $\kappa_r$, not the realised row effects. The working-likelihood $\kappa_r$
posterior is computed for the diagnostic of
Section~\ref{supp:evidence:real} and nowhere enters the
predictive draw. CNBG-Mod interval width is therefore insensitive to
the location of the $\kappa_r$ posterior.
Both pipelines draw the predictive $\kappa_r$ from the prior.

\noindent Step~4 draws from a Gamma distribution matched to the first two
conditional moments of the CNBG distribution given $U_i$: mean
$\nu_{ij}^{(s)}U_i^{(s)}$ and variance $A^{(s)}\nu_{ij}^{(s)}U_i^{(s)}$. The
exact conditional distribution is compound Poisson--Gamma (Tweedie),
with a zero mass and a skewness that depend on $\alpha$ and $\bar Y$
separately; the second-moment procedure carries no information on that
split and a single triangle recovers it only through the third-moment
channel, whose precision is not established here
(Corollary~\ref{cor:cumulant}), so a two-moment
draw is what this procedure can do without a prior on $\alpha$ or a
count triangle. The
moment-matched Gamma is exact in the limit of large expected claim
counts per cell and is an approximation at the small expected counts
of the later development columns; the oracle study of
Section~\ref{sec:simulation:oracle} measures the cost of that
approximation directly. Step~3 is likewise an approximation, since the
conjugate update uses a Poisson working law for $X_{ij}/A$ in place of
the compound law. Step~3 uses the
conjugate posterior update for $U_i$ rather than the marginal prior draw
stated in some presentations of such constructions; this is the correct
conditional implementation, and matches the code in the
\texttt{hgr} package (\url{https://github.com/robin-vo/hgr}). The
observed triangle $\mathcal{D}$ is held fixed throughout; only future
cells are simulated. A lower bound $\kappa_r^{(s)} \ge 0.5$ is imposed on
the $\kappa_r$ draws entering this predictive simulation,
distinct from and stricter than the fit-stage truncation
$\kappa_r \ge 0.1$ of the working-likelihood fit
(Section~\ref{supp:estimation:bayes}); it bounds the
frailty variance $1/\kappa_r^{(s)} \le 2$ away from the degenerate
high-dispersion region. The bound truncates only the extreme left tail of the prior $\kappa_r$ draws and is reached by a small fraction of them, so it has no material effect on the reported coverage. 

We retain the modular construction as the principal amount-only CNBG
benchmark, subject to the approximations, prior-supplied dispersion and
convergence limitations described above.

\subsection{The conditional WLS bootstrap (WLS-cond)}\label{sec:boot}

WLS-cond conditions on the observed triangle and draws future cells from
the CNBG marginal variance at the point estimates $(\hat\kappa_B, \hat A)$,
with the mean structure held at $\hat\theta$ and the cells drawn
independently. It therefore omits parameter integration, mean-structure
uncertainty and the within-row covariance of
Proposition~\ref{prop:moments}(iii). This lower-edge ablation measures the
three omissions together: they cost $30.4$ coverage points against the modular predictive
($62.2\%$ against $92.6\%$, at $46.5\%$ against $99.0\%$ of the true
reserve). The design does not assign individual costs to them. Future cells are drawn
by matching the CNBG variance function locally to a Tweedie law at each
cell mean $\nu_0$:
\begin{equation}\label{eq:local_tw}
    p^*(\nu_0) = 1 + \frac{\nu_0/\kappa_B}{A + \nu_0/\kappa_B},
    \qquad
    \phi^*(\nu_0) = \frac{A\nu_0 + \nu_0^2/\kappa_B}{\nu_0^{p^*(\nu_0)}}
                 = A\,\nu_0^{1-p^*} + \frac{\nu_0^{2-p^*}}{\kappa_B}.
\end{equation}
This is exact: $\phi^*\nu_0^{p^*} = V(\nu_0)$ exactly. Note that
$\phi^* \neq A + \nu_0/\kappa_B$; the dispersion parameter equals the
variance divided by $\nu_0^{p^*}$.

\begin{algorithm}[ht]
\caption{WLS-cond: Fixed-Mean WLS Predictive Bootstrap for Reserve
Uncertainty}
\label{alg:bootstrap}
\begin{algorithmic}[1]
\REQUIRE Fitted $(\hat{\nu}_{ij}, \hat{\kappa_B}_{\mathrm{adj}}, \hat{A})$;
         replications $B$. \textbf{The point-estimate, no-frailty lower
         edge of the design space; used in Section~\ref{sec:simulation}
         as WLS-cond. The modular predictive of
         Section~\ref{sec:estimation:ppd} is the principal CNBG benchmark.}
\FOR{$b = 1, \ldots, B$}
    \STATE \textbf{// Condition on observed triangle as fixed; no refit,
           no frailty draw}
    \FOR{$(i,j) \notin \mathcal{D}$}
        \STATE $(p^*_{ij},\,\phi^*_{ij}) \gets$ local Tweedie match to
               $V(\hat\nu_{ij}) = \hat{A}\hat\nu_{ij} +
               \hat\nu_{ij}^2/\hat\kappa_{\mathrm{adj}}$
               via~\eqref{eq:local_tw}
        \STATE $X_{ij}^{*(b)} \sim
               \mathrm{Tweedie}(\mu = \hat\nu_{ij},\,
               p = p^*_{ij},\, \phi = \phi^*_{ij})$
               \quad\textbf{// drawn independently per cell}
    \ENDFOR
    \STATE $R^{*(b)} = \sum_{(i,j)\notin\mathcal{D}} X_{ij}^{*(b)}$
\ENDFOR
\RETURN $\{R^{*(b)}\}_{b=1}^B$
\end{algorithmic}
\end{algorithm}


\section{Simulation details}
\label{supp:simulation}

\subsection{The full comparison of eight procedures}\label{supp:simulation:full}

The main text reports means over the configurations
(Table~\ref{tab:two_axis}); this subsection gives the design space and the
per-configuration results.

We compare eight methods for reserve prediction interval
construction on simulated data drawn exactly from the CNBG
generative model of Definition~\ref{def:cnbg}. The methods populate
a methodological design space defined by two
organising axes, the treatment of mean-structure uncertainty (refit
bootstrap, fixed-parameter plug-in, or modular predictive draws) and
the variance function (linear or power, misspecified for the marginal
law, versus the CNBG law).

\begin{table}[ht]
\centering
\caption{Methodological design space for amount-only reserving
inference. Eight candidate procedures populate the combinations of
mean-structure-uncertainty treatment and variance specification. Each is
tested under correctly specified CNBG data in
Section~\ref{sec:simulation:headline}.}
\label{tab:design_space}
\small
\begin{tabular}{p{3.4cm}p{4.3cm}p{6cm}}
\toprule
& \textbf{Misspecified variance} & \textbf{CNBG-form variance} \\
\midrule
\textbf{Mean-structure uncertainty propagated by refitting}
& Standard ODP \citep{EnglandVerrall1999};\newline
  Standard Tweedie \citep{SmythJorgensen2002}
& Frequentist CNBG (Section~\ref{sec:free:fits};
  weighted squared-residual regression, refit on each
  bootstrap pseudo-triangle)
\\
\addlinespace[3pt]
\textbf{Plug-in (mean structure fixed at $\hat\theta$)}
& ODP-cond; Tweedie-cond
  (Section~\ref{sec:estimation:ppd})
& WLS-cond (Algorithm~\ref{alg:bootstrap}, point estimate, correct
  marginal cell variance but no shared frailty)
\\
\addlinespace[3pt]
\textbf{Propagated by posterior and sampling-distribution draws}
& ---
& CNBG-Mod-Ref; CNBG-Mod-EmpB
  (Sections~\ref{sec:estimation:bayes}--\ref{sec:estimation:ppd})
\\
\bottomrule
\end{tabular}
\end{table}

The refit procedures resample the observed cells and refit the mean
structure on each pseudo-triangle; the plug-in procedures fit once and
simulate with the mean structure at $\hat\theta$; CNBG-Mod holds the
triangle fixed and draws the mean-structure coefficients and $A$, and is
listed with the plug-in procedures in Table~\ref{tab:simulation} because
it never refits.

\begin{table}[ht]
\centering
\caption{Simulation results: coverage of $95\%$ predictive intervals
(\%) and median relative width (\% of $R_{\mathrm{true}}$). $M =
500$ replications, $I = J = 10$, $B = 500$ predictive draws.
Methods grouped by conditioning regime and ordered within each
group. The conditioning axis produces a $\sim 24$ percentage point
coverage gap (ODP vs ODP-cond; Tweedie vs Tweedie-cond) that
dominates the $\sim 1$ percentage point variance-function axis
within the misspecified pair. Among the conditional
methods with the correct CNBG variance, the two Bayesian procedures
reach $92$--$93\%$ coverage with the modular predictive draw of Section~\ref{sec:predictive}, an approximate construction that draws the mean structure, a working posterior of the dispersion and a conjugate frailty update, whereas WLS-cond, which plugs in a point estimate, is the lowest-coverage method in the entire comparison at $62.2\%$.}
\label{tab:simulation}
\scriptsize
\renewcommand{\arraystretch}{1.12}
\setlength{\tabcolsep}{2.6pt}
\begin{tabular}{@{}llrrrrrrrrrrrrrrrr@{}}
\toprule
& & \multicolumn{6}{c}{\textit{Refit bootstrap}}
  & \multicolumn{10}{c}{\textit{Plug-in and hybrid draws}} \\
\cmidrule(lr){3-8}\cmidrule(lr){9-18}
& & \multicolumn{2}{c}{ODP} & \multicolumn{2}{c}{Tweedie}
  & \multicolumn{2}{c}{F-CNBG}
  & \multicolumn{2}{c}{ODP-c} & \multicolumn{2}{c}{Twe-c}
  & \multicolumn{2}{c}{WLS-c} & \multicolumn{2}{c}{CNBG-Mod-R}
  & \multicolumn{2}{c}{CNBG-Mod-E} \\
\cmidrule(lr){3-4}\cmidrule(lr){5-6}\cmidrule(lr){7-8}
\cmidrule(lr){9-10}\cmidrule(lr){11-12}\cmidrule(lr){13-14}
\cmidrule(lr){15-16}\cmidrule(lr){17-18}
$\kappa_r$ & $\alpha$
  & C & W & C & W & C & W & C & W & C & W & C & W & C & W & C & W \\
\midrule
 3 &  1 & 89 & 109.5 & 94 & 125.1 & 84 & 92.1 & 66 & 58.6 & 69 & 62.7 & 62 & 52.9 & 93 & 113.8 & 91 & 114.2 \\
 3 &  3 & 91 & 90.9 & 92 & 97.2 & 86 & 75.7 & 68 & 50.2 & 69 & 51.5 & 62 & 43.9 & 92 & 95.1 & 91 & 95.3 \\
 3 & 10 & 93 & 85.8 & 93 & 88.7 & 88 & 70.2 & 71 & 46.5 & 73 & 47.0 & 66 & 40.1 & 93 & 89.0 & 93 & 89.3 \\
\midrule
 5 &  1 & 92 & 110.6 & 95 & 119.0 & 86 & 92.5 & 66 & 60.3 & 67 & 60.8 & 61 & 54.9 & 94 & 115.6 & 94 & 115.3 \\
 5 &  3 & 93 & 92.3 & 93 & 96.0 & 87 & 76.9 & 69 & 50.2 & 69 & 50.6 & 64 & 44.2 & 94 & 98.4 & 93 & 96.7 \\
 5 & 10 & 92 & 86.1 & 93 & 90.8 & 85 & 70.7 & 69 & 46.3 & 69 & 47.5 & 62 & 40.3 & 93 & 89.7 & 93 & 88.4 \\
\midrule
10 &  1 & 92 & 108.8 & 94 & 122.9 & 86 & 91.4 & 67 & 60.1 & 71 & 63.1 & 65 & 56.0 & 94 & 113.5 & 93 & 112.7 \\
10 &  3 & 92 & 95.5 & 94 & 98.3 & 84 & 77.3 & 66 & 50.7 & 66 & 51.7 & 61 & 45.5 & 92 & 97.7 & 93 & 96.8 \\
10 & 10 & 92 & 84.4 & 92 & 86.4 & 84 & 69.6 & 68 & 46.3 & 69 & 46.6 & 62 & 40.6 & 92 & 87.1 & 91 & 87.4 \\
\midrule
20 &  1 & 91 & 106.6 & 95 & 120.8 & 84 & 91.8 & 66 & 59.6 & 68 & 61.9 & 63 & 55.4 & 92 & 112.3 & 92 & 111.0 \\
20 &  3 & 90 & 91.9 & 93 & 96.2 & 82 & 75.4 & 64 & 50.0 & 64 & 50.7 & 58 & 44.2 & 92 & 94.3 & 92 & 92.5 \\
20 & 10 & 92 & 84.1 & 91 & 85.6 & 84 & 68.1 & 65 & 45.8 & 64 & 46.6 & 60 & 40.0 & 92 & 86.7 & 91 & 85.9 \\
\midrule
50 &  1 & 90 & 107.0 & 94 & 123.6 & 85 & 90.4 & 67 & 59.8 & 69 & 62.8 & 65 & 55.0 & 93 & 112.3 & 92 & 112.3 \\
50 &  3 & 91 & 91.9 & 92 & 96.8 & 83 & 75.5 & 67 & 49.7 & 67 & 50.9 & 62 & 45.3 & 91 & 94.4 & 91 & 94.9 \\
50 & 10 & 92 & 83.7 & 94 & 86.4 & 86 & 68.2 & 69 & 45.6 & 69 & 46.3 & 63 & 39.9 & 92 & 85.9 & 92 & 85.1 \\
\midrule
\multicolumn{2}{l}{Mean}
   & 91.5 & 95.3 & 93.2 & 102.2 & 84.9 & 79.0 & 67.1 & 52.0 & 68.2 & 53.4 & 62.2 & 46.5 & 92.6 & 99.0 & 92.2 & 98.5 \\
\bottomrule
\end{tabular}

\smallskip
\footnotesize\textit{Note}: C = coverage (\%), W = median relative
width (\% of $R_{\mathrm{true}}$). F-CNBG = Frequentist CNBG;
ODP-c / Twe-c / WLS-c = conditional-target variants;
CNBG-Mod-R / CNBG-Mod-E = CNBG-Mod under reference and empirical-Bayes
priors. Per-cell MC SE at $M = 500$ is $\sim 1.2$pp; pooled means
(final row) have MC SE $\sim 0.3$pp.
\end{table}

\subsection{The two draws of the modular predictive switched off}
\label{supp:simulation:steps}

Table~\ref{tab:steps} completes the attribution of
Section~\ref{sec:simulation:oracle} by switching the two draws of the
modular predictive off one at a time on the same triangles.

\begin{table}[ht]
\centering
\caption{The two draws of the modular predictive switched off one at
a time ($M = 500$ at each configuration; same triangles and absorbed
fits as Table~\ref{tab:oracle}). C = coverage (\%), W = median relative
width (\% of $R_{\mathrm{true}}$), E = median signed centre error (\%).
``No step 2'' holds the mean structure at $\hat\theta$; ``no step 3''
holds the per-row frailty at $U_i = 1$; ``neither'' does both.}
\label{tab:steps}
{\scriptsize\setlength{\tabcolsep}{2.5pt}%
\begin{tabular}{@{}llccccccccccccccc@{}}
\toprule
& & \multicolumn{3}{c}{Full (steps 2 and 3)} & \multicolumn{3}{c}{No step 2} & \multicolumn{3}{c}{No step 3} & \multicolumn{3}{c}{Neither} & \multicolumn{3}{c}{Oracle-moment} \\
\cmidrule(lr){3-5}\cmidrule(lr){6-8}\cmidrule(lr){9-11}\cmidrule(lr){12-14}\cmidrule(lr){15-17}
$\kappa_r$ & $\alpha$ & C & W & E & C & W & E & C & W & E & C & W & E & C & W & E \\
\midrule
 3 & 3 & 92 & 95.1 & $-2.1$ & 83 & 69.6 & $-0.7$ & 93 & 98.1 & $-0.6$ & 71 & 53.7 & $-0.4$ & 94 & 71.1 & $-0.0$ \\
10 & 1 & 94 & 113.5 & $-3.2$ & 83 & 87.2 & $-0.4$ & 95 & 117.7 & $-1.2$ & 72 & 67.5 & $-0.7$ & 97 & 81.2 & $+2.5$ \\
50 & 3 & 91 & 94.4 & $-1.6$ & 82 & 71.0 & $-0.7$ & 92 & 96.8 & $-0.7$ & 70 & 54.0 & $-0.4$ & 95 & 59.7 & $+1.3$ \\
\bottomrule
\end{tabular}}

\end{table}

\subsection{A larger claim volume}\label{sec:simulation:volume}

The grid of Table~\ref{tab:simulation} has $30$--$47$ expected claims
per accident year. Table~\ref{tab:less_sparse} repeats two
configurations at ten times that volume ($M = 200$). Every width falls
by a factor of about three, CNBG-Mod reaches $94\%$ at both values of
$\kappa_r$, its width gap to the oracles narrows from $24$--$35$ to
$8$--$9$ points, and ODP-cond, which carries no estimation-error term,
stays at $64$--$73\%$. The curvature channel improves with volume as
Proposition~\ref{prop:godambe} predicts, but slowly. Under
anchored levels, where $\kappa_B$ is estimable, the adjusted $\kappa_B$
posterior at $\kappa_r = 5$ contracts from $0.92$ to $0.85$ of the prior
width and its median moves from $5.5$ to $6.4$ (interval
$[3.0, 14.2]$), while at $\kappa_r = 50$ the posterior median moves from
$7.9$ to $17.0$ (interval $[9.4, 29.0]$) and still excludes the truth on
every replication ($M = 100$ per cell). That interval is not wide; it
is narrow in the wrong place. Where the likelihood is weak the
reference prior binds, and the curvature adjustment, which matches the
posterior's width to the sampling variability of the working
estimator, does not repair a misplaced prior. At $\kappa_r = 50$ the
$O(\kappa_B^{-4})$ decay of the information leaves the channel weak even
at this volume, which is also the regime in which the quadratic term is
inconsequential for the reserve (equation~\eqref{eq:reserve_var}).

\begin{table}[ht]
\centering
\caption{Coverage (C, \%) and median relative width (W, \% of
$R_{\mathrm{true}}$) at the base claim volume ($\times 1$:
$\mu_i = 30e^{0.05(i-1)}$, $30$--$47$ expected claims per accident year,
$M = 500$) and at ten times that volume ($\times 10$:
$\mu_i = 300e^{0.05(i-1)}$, $M = 200$), $\alpha = 3$. Oracles at the base calibration are available
for $\kappa_r = 50$ only (Table~\ref{tab:oracle}).}
\label{tab:less_sparse}
{\scriptsize\setlength{\tabcolsep}{3pt}%
\begin{tabular}{@{}llcccccccccccc@{}}
\toprule
& & Volume & \multicolumn{2}{c}{CNBG-Mod} & \multicolumn{2}{c}{ODP} & \multicolumn{2}{c}{ODP-c} & \multicolumn{2}{c}{Oracle-moment} & \multicolumn{2}{c}{Oracle-exact} \\
\cmidrule(lr){4-5}\cmidrule(lr){6-7}\cmidrule(lr){8-9}\cmidrule(lr){10-11}\cmidrule(lr){12-13}
$\kappa_r$ & $\alpha$ & & C & W & C & W & C & W & C & W & C & W \\
\midrule
 5 & 3 & $\times 1$ & 94 & 98.4 & 93 & 92.3 & 69 & 50.2 & --- & --- & --- & --- \\
 5 & 3 & $\times 10$ & 94 & 32.4 & 95 & 31.4 & 64 & 17.1 & 96 & 23.7 & 96 & 23.8 \\
\midrule
50 & 3 & $\times 1$ & 91 & 94.4 & 91 & 91.9 & 67 & 49.7 & 95 & 59.7 & 95 & 60.0 \\
50 & 3 & $\times 10$ & 94 & 31.6 & 94 & 30.9 & 73 & 16.6 & 94 & 22.4 & 94 & 22.5 \\
\bottomrule
\end{tabular}}

\end{table}

\subsection{Prior sensitivity}\label{sec:simulation:prior_sensitivity}

The Bayesian under-coverage in Table~\ref{tab:simulation} could in
principle reflect the prior. The pooled equivalence of CNBG-Mod-Ref and
CNBG-Mod-EmpB (Section~\ref{sec:simulation:headline}), two priors with
different locations producing pooled coverage agreeing to $0.4$ percentage
point, argues against this directly, and holds configuration by
configuration: across all $15$ $(\kappa_r, \alpha)$ cells the two
priors' coverage differs by at most $1.8$ percentage points (mean
$0.6$), and their median widths by at most $1.8$ percentage points,
despite a prior-mean change from $5.0$ to $8.1$. Coverage is
insensitive to the prior on $\kappa_r$ at this calibration, and the oracle
study of Section~\ref{sec:simulation:oracle} is what says why.

\subsection{The linear coefficient of the refit bootstrap corrected}
\label{supp:simulation:leverage}

Proposition~\ref{prop:two_stage_target}(iii) gives the Frequentist CNBG
bootstrap a second defect besides the capped dispersion: the linear
coefficient it refits is the constrained intercept of the second-moment
regression, $0.54$ to $0.56$ of $A$ at $I = 10$
(Table~\ref{tab:fcnbg_leverage}), whereas the Pearson dispersion of the
ODP bootstrap divides by $n - q$ and is not deflated. To separate the two
defects, the variant F-CNBG-lev divides the squared residuals by
$1 - \min(h_{ij}, 0.95)$ before the second-moment regression, the
leverage correction of the working fit, in the original fit and in every
refit, and changes nothing else. Table~\ref{tab:fcnbg_leverage} runs the
two variants and the ODP bootstrap on the same $200$ triangles at
$(\kappa_r, \alpha) = (3, 3)$, $(10, 3)$ and $(50, 3)$ (the seeds of
Section~\ref{sec:simulation}, $B = 300$ replications; Monte Carlo
standard error of a coverage about $2$ points;
\texttt{R/run\_phase18\_fcnbg\_leverage.R}). The correction raises the
linear coefficient to $0.71$ to $0.75$ of $A$, the remaining shortfall
being the leverages capped at $0.95$ and the factor $N\alpha/(N\alpha+1)$
of Section~\ref{supp:estimation:bayes}, and raises coverage by $9.0$,
$9.0$ and $10.5$ points (paired standard errors $2.0$ to $2.2$) to
$88.0$, $94.5$ and $91.5\%$, against the ODP bootstrap's $87.0$, $94.0$
and $89.0\%$ on the same triangles (paired differences $+1.0$, $+0.5$ and
$+2.5$ points, standard errors $0.9$ to $1.7$), at a median width $6$ to
$7\%$ greater. The deficit of the Frequentist CNBG bootstrap in
Table~\ref{tab:two_axis} is therefore the deflated linear coefficient and
not the variance law: with the dispersion at its cap on a third to a half
of the corrected fits, the refitted CNBG law and the linear law cover
alike. What the correct form adds when its dispersion comes from the
prior, about a point over the ODP bootstrap for CNBG-Mod, is the part the
free-level fit does not supply.

\begin{table}[ht]
\centering
\caption{The Frequentist CNBG bootstrap with its linear coefficient
corrected by the leverage factor (F-CNBG-lev) against the uncorrected
version and the ODP bootstrap on the same triangles ($\alpha = 3$,
$M = 200$ per row, $B = 300$): coverage of nominal $95\%$ intervals,
median relative width, the mean refitted linear coefficient relative to
$A$, and the share of fits with the dispersion at the cap.}
\label{tab:fcnbg_leverage}
\footnotesize
\setlength{\tabcolsep}{3pt}
\begin{tabular}{@{}rccccccccc@{}}
\toprule
 & \multicolumn{3}{c}{Coverage (\%)} & \multicolumn{3}{c}{Median width (\% of $R_{\mathrm{true}}$)} & \multicolumn{2}{c}{mean $\hat A/A$} & at cap \\
\cmidrule(lr){2-4}\cmidrule(lr){5-7}\cmidrule(lr){8-9}
$\kappa_r$ & ODP & F-CNBG & F-CNBG-lev & ODP & F-CNBG & F-CNBG-lev & F-CNBG & F-CNBG-lev & F-CNBG / -lev \\
\midrule
3 & 87.0 & 79.0 & 88.0 & 89.1 & 71.2 & 96.9 & 0.54 & 0.74 & 0.68 / 0.45 \\
10 & 94.0 & 85.5 & 94.5 & 92.2 & 75.5 & 101.3 & 0.56 & 0.75 & 0.63 / 0.41 \\
50 & 89.0 & 81.0 & 91.5 & 93.6 & 76.1 & 102.3 & 0.54 & 0.71 & 0.54 / 0.36 \\
\bottomrule
\end{tabular}

\end{table}

\subsection{A finite cell shock}\label{supp:simulation:cellshock}

The simulations above are run in the single-frailty case, in which the
cell shock is absent and the free levels absorb all heterogeneity. The
two-level hierarchy behaves differently, since the cell shock survives
free levels (Proposition~\ref{prop:two_kappa_id}(ii)) but no procedure
estimates it well: the free-level second stage reads it only below the
threshold of Remark~\ref{rem:cellshock_target}, which $\kappa_c = 20$
exceeds. Table~\ref{tab:cellshock} repeats the comparison on
data with a finite cell shock, at $(\kappa_r, \kappa_c) = (5, 5)$, $(5, 20)$
and $(50, 5)$ with $\alpha = 3$ and $200$ triangles each. Every procedure
covers less than on single-frailty data: the propagating procedures
$79$ to $90\%$, the plug-in procedures $54$ to $69\%$. The ordering by
treatment of mean-structure uncertainty is unchanged, with gaps of $22$
to $33$ points, while the lead of CNBG-Mod over the ODP bootstrap is
$2.0$ to $2.5$ points with paired standard errors of $1.1$ to $1.6$, and
over the Frequentist CNBG bootstrap $5.5$ to $7.5$. A cell shock that none of the procedures models as such costs them all.
Coverage is lowest not where the shock is strongest but at $(5, 20)$,
where it is weakest and the row frailty strong: a within-row shock
inflates the dispersion that every free-level fit estimates from its
residuals and so widens every interval, whereas the row frailty is
absorbed by the levels and widens none, which is
Proposition~\ref{prop:two_kappa_id} read off the coverages. Unequal
development patterns, exposure misspecification and non-Gamma severities
are not tested here.

\begin{table}[H]
\centering
\caption{The procedures on data with a finite cell shock
($\varepsilon_{ij} \sim \mathrm{Gamma}(\kappa_c, \kappa_c)$ multiplying the
cell rate), $\alpha = 3$, $M = 200$ triangles per configuration, $B = 500$:
coverage of nominal $95\%$ intervals (C, \%; Monte Carlo standard error
$2$ to $3.5$ points), median relative width (W, \% of $R_{\mathrm{true}}$)
and the paired difference of CNBG-Mod minus the procedure in coverage
points ($\Delta$, standard error in parentheses).}
\label{tab:cellshock}
{\small\setlength{\tabcolsep}{4pt}%
\begin{tabular}{@{}lrrrrrrrrr@{}}
\toprule
& \multicolumn{3}{c}{$(\kappa_r, \kappa_c) = (5, 5)$} & \multicolumn{3}{c}{$(\kappa_r, \kappa_c) = (5, 20)$} & \multicolumn{3}{c}{$(\kappa_r, \kappa_c) = (50, 5)$} \\
\cmidrule(lr){2-4}\cmidrule(lr){5-7}\cmidrule(lr){8-10}
Procedure & C & W & $\Delta$ & C & W & $\Delta$ & C & W & $\Delta$ \\
\midrule
ODP-cond & 63.0 & 64.5 & $+27.0$ (3.1) & 55.5 & 55.5 & $+31.0$ (3.3) & 65.0 & 64.7 & $+25.5$ (3.1) \\
Tweedie-cond & 62.5 & 64.8 & $+27.5$ (3.2) & 56.5 & 56.7 & $+30.0$ (3.2) & 63.5 & 64.1 & $+27.0$ (3.1) \\
WLS-cond & 67.0 & 70.8 & $+23.0$ (3.0) & 54.0 & 50.3 & $+32.5$ (3.3) & 69.0 & 73.8 & $+21.5$ (2.9) \\
ODP bootstrap & 88.0 & 116.6 & $+2.0$ (1.2) & 84.5 & 102.9 & $+2.0$ (1.6) & 88.0 & 115.7 & $+2.5$ (1.1) \\
Frequentist CNBG & 84.0 & 112.8 & $+6.0$ (1.8) & 79.0 & 85.9 & $+7.5$ (1.9) & 85.0 & 106.1 & $+5.5$ (1.6) \\
CNBG-Mod & 90.0 & 130.5 & --- & 86.5 & 107.7 & --- & 90.5 & 125.7 & --- \\
\bottomrule
\end{tabular}}

\end{table}

\subsection{A downward calendar-year change in the claim rate}\label{supp:simulation:drift}

The experiment of Section~\ref{sec:simulation} on a misspecified mean
structure draws triangles from the single-frailty hierarchy at
$\kappa_r = 5$ and $\alpha = 3$, the calibration of
Section~\ref{sec:illus:boundary}, with the cell rate
$\lambda_i w_j$ multiplied by $\exp\{-\delta\max(0, t - 7)\}$,
$t = i + j - 1$ and $\delta \ge 0$ the magnitude of the decline per
calendar year: a log-linear downward trend that starts in the eighth
calendar year, so that the last three observed diagonals and every
future cell carry it. A trend present in every calendar year would be
absorbed exactly by the free levels and the development effects, which
is why the change is placed inside the triangle. Four values
$\delta \in \{0, 0.05, 0.10, 0.20\}$ per calendar year, $M = 200$
triangles each on common seeds, $B = 500$ draws; four procedures, one
per treatment of the estimation error of the mean structure and the
Frequentist CNBG bootstrap beside the ODP bootstrap. Table~\ref{tab:drift}
reports coverage, the share of realised reserves below the interval,
the median relative width, the median ratio of the interval centre to
the realised reserve and the paired coverage difference against CNBG-Mod.
Every fit converged. The script is \texttt{R/run\_phase16\_calendar\_drift.R}.

\begin{table}[H]
\centering
\caption{A downward calendar-year change in the claim rate from the
eighth calendar year: the rate multiplied by $\exp\{-\delta\max(0, t - 7)\}$
with $\delta \ge 0$ the decline per year,
$\kappa_r = 5$, $\alpha = 3$, $M = 200$ triangles per $\delta$,
$B = 500$. Coverage of nominal $95\%$ intervals (C; Monte Carlo standard
error $1.5$ to $3.5$ points), share of realised reserves below the
interval, median width (W, \% of $R_{\mathrm{true}}$), median ratio of
the interval centre to $R_{\mathrm{true}}$, and the paired difference of
CNBG-Mod minus the procedure in coverage points ($\Delta$, standard
error in parentheses).}
\label{tab:drift}
\small
\begin{tabular}{@{}llcccccc@{}}
\toprule
$\delta$ & Procedure & $n$ & C (\%) & Below (\%) & W (\%) & Centre$/R_{\mathrm{true}}$ & $\Delta$ (s.e.) \\
\midrule
$0.00$ & ODP-cond (plug-in) & 200 & 67.0 & 18.5 & 51.6 & 1.04 & 28.0 (3.2) \\
 & ODP bootstrap (refit) & 200 & 93.5 & 3.0 & 95.3 & 1.09 & 1.5 (1.1) \\
 & Frequentist CNBG (refit) & 200 & 87.0 & 6.0 & 79.5 & 1.07 & 8.0 (1.9) \\
 & CNBG-Mod (draws) & 200 & 95.0 & 2.0 & 97.0 & 1.07 & --- \\
\addlinespace
$0.05$ & ODP-cond (plug-in) & 200 & 66.0 & 26.5 & 61.2 & 1.13 & 25.5 (3.1) \\
 & ODP bootstrap (refit) & 200 & 90.0 & 8.5 & 110.3 & 1.21 & 1.5 (0.9) \\
 & Frequentist CNBG (refit) & 200 & 82.0 & 14.0 & 90.4 & 1.17 & 9.5 (2.1) \\
 & CNBG-Mod (draws) & 200 & 91.5 & 7.0 & 114.4 & 1.20 & --- \\
\addlinespace
$0.10$ & ODP-cond (plug-in) & 200 & 60.5 & 34.5 & 74.5 & 1.27 & 30.5 (3.3) \\
 & ODP bootstrap (refit) & 200 & 85.0 & 14.0 & 126.9 & 1.35 & 6.0 (2.0) \\
 & Frequentist CNBG (refit) & 200 & 76.5 & 22.0 & 101.8 & 1.31 & 14.5 (2.5) \\
 & CNBG-Mod (draws) & 200 & 91.0 & 7.5 & 134.8 & 1.34 & --- \\
\addlinespace
$0.20$ & ODP-cond (plug-in) & 200 & 45.5 & 52.5 & 108.0 & 1.57 & 32.5 (3.3) \\
 & ODP bootstrap (refit) & 200 & 68.5 & 31.0 & 173.9 & 1.69 & 9.5 (2.1) \\
 & Frequentist CNBG (refit) & 200 & 54.5 & 44.0 & 139.2 & 1.64 & 23.5 (3.0) \\
 & CNBG-Mod (draws) & 200 & 78.0 & 21.5 & 179.2 & 1.65 & --- \\
\bottomrule
\end{tabular}

\end{table}

\section{Schedule~P details}
\label{supp:scheduleP}

\subsection{The outcome-free panel and its preprocessing}
\label{supp:scheduleP:six}

Of the $779$ complete $10 \times 10$ paid triangles of the six lines, $148$
have a negative incremental share above $5\%$; of the remaining $631$, the
$452$ with a chain-ladder reserve above \$$100$k form the panel: $126$ in
\texttt{othliab}, $109$ in \texttt{comauto}, $85$ in \texttt{wkcomp}, $81$
in \texttt{ppauto}, $29$ in \texttt{prodliab} and $22$ in \texttt{medmal}.
Zero cells, $8{,}449$ of the observed cells ($34\%$, in $352$ portfolios),
are kept, since the CNBG support includes zero; the $333$ negative cells (in
$196$ portfolios) are excluded for every procedure. The first stage fails
to converge on $11$ portfolios and one portfolio exceeds the five-minute
cap, so ODP, ODP-cond and WLS-cond return intervals on $440$. The
zeros-kept refit of the mean structure is numerically
degenerate (a reserve above $50$ times the observed paid total) on $11$
portfolios, on ten of which every procedure misses. The main run records
Tweedie-cond and Frequentist CNBG as not computed on three of them ($437$
intervals); the Tweedie bootstrap, the premium-anchored fit and the exact
posterior skip all eleven, so the Tweedie bootstrap returns $429$
intervals. CNBG-Mod returns an interval on $371$
portfolios; on the other $81$ ($18\%$), mostly the smallest triangles of the
panel, the working likelihood yields no usable posterior. The
reconciliation of every denominator in the paper is therefore: $440$
(ODP, ODP-cond, WLS-cond), $437$ (Tweedie-cond, Frequentist CNBG), $429$
(Tweedie bootstrap), $392$ (exact frailty posterior), $371$ (CNBG-Mod;
$370$ in the scores, which leave out its one degenerate refit, and $361$
when paired with the ODP bootstrap after all ten degenerate refits are
left out), $363$ (both the exact and the working fit), $281$ (positive
premium in every accident year and both working fits) and $286$
(positive premium and both predictive intervals). The outcome-free rule admits near-empty triangles, up to $53$ zero cells of $55$, which is why the chain-ladder error of Section~\ref{sec:scheduleP:bias} is measured against the chain-ladder reserve that defines the panel rather than against the realised reserve; the same denominator is used for the scores of Table~\ref{tab:scoring}, since $17$ realised reserves are zero or negative and a score divided by them is undefined or reversed. The degenerate refits of the previous paragraph are ten by the rule used there (a chain-ladder reserve above $50$ times the paid total); the eleventh of the main text is a borderline case that the rule does not catch. Tweedie-cond's profiled power ranges from
$1.1$ to $1.9$ with median $1.34$, inside the compound Poisson--Gamma range.

Table~\ref{tab:defensible} gives, for the six procedures of the main
run, coverage on each procedure's own denominator and on the full panel,
median width, the relative Winkler interval score, and the paired
difference between CNBG-Mod and each other procedure with an
insurer-clustered interval; the Tweedie bootstrap, computed in a separate
run with the same panel, preprocessing and seeds, is in
Table~\ref{tab:two_axis}. Table~\ref{tab:cl_ratio_by_line} gives the
chain-ladder error by line.

\begin{table}[H]
\centering
\caption{The six procedures of the main run on the outcome-free panel
($452$ portfolios, zero cells kept, negative cells excluded for every
procedure). Cov.\ = coverage of nominal $95\%$ intervals on the procedure's
own denominator $n$; Cov.\ all = on the full panel with portfolios lacking an interval scored as misses; Width = median $95\%$ interval width (\% of $R_{\mathrm{actual}}$); relative Winkler score ($\alpha = 0.05$) = median relative interval score (lower is better), both over the portfolios with $R_{\mathrm{actual}} > 0$. The last two columns give CNBG-Mod minus the procedure on the $361$ portfolios with both intervals finite, with an insurer-clustered bootstrap $95\%$ interval; a negative Winkler difference favours CNBG-Mod. Table~\ref{tab:scoring} restates these scores with every denominator explicit.}
\label{tab:defensible}
{\scriptsize\setlength{\tabcolsep}{2.5pt}%
\begin{tabular}{@{}lrrrrrll@{}}
\toprule
Procedure & $n$ & Cov.\ (\%) & Cov.\ all (\%) & Width (\%) & Winkler (\%) & \multicolumn{2}{c}{CNBG-Mod minus procedure} \\
& & & & & & coverage (pp) & Winkler (\%) \\
\midrule
ODP & $440$ & $50.7$ & $49.3$ & $122.1$ & $317.0$ & $+9.4$ [$+4.9$, $+13.8$] & $-3.1$ [$-8.1$, $-0.8$] \\
F-CNBG & $437$ & $51.7$ & $50.0$ & $102.9$ & $393.0$ & $+10.2$ [$+5.8$, $+14.8$] & $-9.6$ [$-40.5$, $+1.1$] \\
ODP-cond & $440$ & $28.0$ & $27.2$ & $40.7$ & $827.5$ & $+36.3$ [$+31.2$, $+42.0$] & $-250.2$ [$-394.7$, $-132.9$] \\
Tweedie-cond & $437$ & $32.3$ & $31.2$ & $43.9$ & $790.2$ & $+31.6$ [$+26.2$, $+36.5$] & $-133.6$ [$-234.8$, $-29.5$] \\
WLS-cond & $440$ & $30.9$ & $30.1$ & $59.2$ & $762.3$ & $+33.0$ [$+27.9$, $+38.4$] & $-171.1$ [$-322.9$, $-70.6$] \\
CNBG-Mod & $371$ & $68.5$ & $56.2$ & $108.2$ & $200.0$ & --- & --- \\
\bottomrule
\end{tabular}}

\end{table}

\begin{table}[H]
\centering
\caption{The six procedures of the main run scored with every denominator
stated. $n$ = portfolios with an interval, after the ten numerically
degenerate refits (chain-ladder reserve above $50$ times the paid total)
are left out of every score, so $370$ of the $371$ usable CNBG-Mod fits; Cov.\ = coverage of nominal $95\%$ intervals.
Fourth column: median Winkler score ($\alpha = 0.05$) divided by the realised reserve, on the portfolios where it is positive. Next three columns: median width and median Winkler score divided by the chain-ladder reserve that
defines the outcome-free panel, which is positive on every portfolio and
does not involve the outcome; the number of explosive intervals, whose
Winkler score exceeds $100$ times the chain-ladder reserve; and the
untrimmed mean Winkler score in thousands of dollars, a proper score,
which the explosive intervals dominate. Bottom: CNBG-Mod minus the procedure on the
portfolios where both have an interval, with insurer-clustered bootstrap
$95\%$ intervals ($2{,}000$ resamples): coverage, the mean log ratio of
the two Winkler scores, which needs no denominator, and the untrimmed
mean dollar difference.}
\label{tab:scoring}
\tiny
\setlength{\tabcolsep}{2pt}
\begin{tabular}{@{}lrr r rrr r@{}}
\toprule
Procedure & $n$ & Cov.\ & $/R_{\mathrm{actual}}$ & \multicolumn{3}{c}{$/R_{\mathrm{CL}}$, all $n$} & Winkler \\
\cmidrule(lr){4-4}\cmidrule(lr){5-7} & & (\%) & Winkler & Width & Winkler & $> 100 R_{\mathrm{CL}}$ & (\$000) \\
\midrule
CNBG-Mod & 370 & 68.4 & 188.6 & 99.5 & 148.3 & 4 & $+114$ \\
ODP bootstrap & 430 & 51.9 & 272.9 & 103.5 & 223.7 & 9 & $1.9 \times 10^{9}$ \\
Frequentist CNBG & 430 & 52.3 & 363.9 & 90.0 & 298.9 & 22 & $1.4 \times 10^{19}$ \\
ODP-cond & 430 & 28.6 & 765.5 & 31.0 & 697.8 & 5 & $+288$ \\
Tweedie-cond & 430 & 32.8 & 668.1 & 35.7 & 537.1 & 5 & $+281$ \\
WLS-cond & 430 & 31.6 & 669.4 & 36.7 & 600.2 & 4 & $+234$ \\
\midrule
\multicolumn{8}{@{}l}{\emph{CNBG-Mod minus procedure, common portfolios}} \\
 & $n$ & \multicolumn{2}{l}{Coverage (pp)} & \multicolumn{2}{l}{Mean log ratio of Winkler scores} & \multicolumn{2}{l}{Mean difference (\$000)} \\
ODP bootstrap & 366 & \multicolumn{2}{l}{$+9.3$ [$+5.1$, $+13.6$]} & \multicolumn{2}{l}{$-0.31$ [$-0.46$, $-0.17$]} & \multicolumn{2}{l}{$-76$ [$-167$, $-12$]} \\
Frequentist CNBG & 366 & \multicolumn{2}{l}{$+9.6$ [$+5.1$, $+13.9$]} & \multicolumn{2}{l}{$-1.07$ [$-1.60$, $-0.62$]} & \multicolumn{2}{l}{$-1.6 \times 10^{19}$ [$-5.0 \times 10^{19}$, $-2.0 \times 10^{7}$]} \\
ODP-cond & 366 & \multicolumn{2}{l}{$+35.8$ [$+30.5$, $+40.9$]} & \multicolumn{2}{l}{$-0.70$ [$-0.86$, $-0.54$]} & \multicolumn{2}{l}{$-172$ [$-347$, $-62$]} \\
Tweedie-cond & 366 & \multicolumn{2}{l}{$+30.9$ [$+25.7$, $+35.7$]} & \multicolumn{2}{l}{$-0.55$ [$-0.71$, $-0.39$]} & \multicolumn{2}{l}{$-162$ [$-323$, $-53$]} \\
WLS-cond & 366 & \multicolumn{2}{l}{$+32.5$ [$+27.3$, $+37.5$]} & \multicolumn{2}{l}{$-0.62$ [$-0.79$, $-0.47$]} & \multicolumn{2}{l}{$-113$ [$-219$, $-45$]} \\
\multicolumn{8}{@{}l}{Winkler by $R_{\mathrm{actual}}$ on the 358 to 415 portfolios with $R_{\mathrm{actual}} > 0$; dollar means and differences untrimmed on all $n$,} \\
\multicolumn{8}{@{}l}{explosive intervals included; the paired differences without them, a descriptive summary, are in the replication report.} \\
\bottomrule
\end{tabular}

\end{table}

\begin{table}[H]
\centering
\caption{The realised reserve as a fraction of the chain-ladder reserve by
line of business, outcome-free panel: median and share of portfolios below
one.}
\label{tab:cl_ratio_by_line}
\small
\begin{tabular}{@{}lrrr@{}}
\toprule
Line & $n$ & median $R_{\mathrm{actual}}/R_{\mathrm{CL}}$ & below one (\%) \\
\midrule
\texttt{comauto} & $109$ & 0.67 & 80.7 \\
\texttt{medmal} & $22$ & 1.02 & 50.0 \\
\texttt{othliab} & $126$ & 0.69 & 73.8 \\
\texttt{ppauto} & $81$ & 0.73 & 90.1 \\
\texttt{prodliab} & $29$ & 0.70 & 69.0 \\
\texttt{wkcomp} & $85$ & 0.59 & 78.8 \\
\bottomrule
\end{tabular}

\end{table}

\subsection{Premium-anchored levels}

Premium anchoring (Section~\ref{sec:scheduleP:anchored}) uses net earned
premium by accident year (\texttt{EarnedPremNet}, direct premium as
fallback on $6$ portfolios); the $152$ portfolios without a positive
premium in every accident year are excluded. The mean structure
$\mu_i = \pi_i \cdot \mathrm{ELR}$ has one common expected loss ratio and
free development effects; the working likelihood is that of
Section~\ref{sec:estimation:bayes} with the dispersion now estimable, and
the predictive draws it from its posterior. Preprocessing, seeds and the
reference prior are those of the main run.

\begin{table}[H]
\centering
\caption{Premium-anchored CNBG-Mod against the free-level fit on the
$286$ portfolios of the outcome-free panel with a positive net earned
premium in every accident year and both intervals: pooled coverage, median
$95\%$ interval width (\% of $R_{\mathrm{actual}}$), and the median across
portfolios of the posterior median of $\kappa_B$ under anchoring.}
\label{tab:anchored}
\begin{tabular}{@{}rrrrrr@{}}
\toprule
& \multicolumn{2}{c}{Premium-anchored} & \multicolumn{2}{c}{Free levels} & \\
\cmidrule(lr){2-3}\cmidrule(lr){4-5}
$n$ & Cov.\ (\%) & Width (\%) & Cov.\ (\%) & Width (\%) & median $\kappa_B$ (anchored) \\
\midrule
$286$ & 68.2 & 91.4 & 72.0 & 97.0 & 6.6 \\
\bottomrule
\end{tabular}

\end{table}

\subsection{The historical implementation}

An earlier implementation selected portfolios on the realised reserve
(above \$$100$k, $398$ portfolios after one timeout) and set every
non-positive observed increment to $1$ (thousand USD), zeros included.
Table~\ref{tab:eligibility} records the pooled coverage of the six
procedures under that selection rule and under two outcome-free rules: the
realised-reserve screen flatters every procedure by three to five points,
uniformly across procedures. Apart from a few sensitivity statements that
name it, no result of this paper uses the historical implementation.

\begin{table}[H]
\centering
\caption{Pooled empirical $95\%$ coverage (\%) of the six procedures under
the realised-reserve selection rule of the historical implementation and
under two outcome-free rules (chain-ladder reserve, latest-diagonal
cumulative paid), all with the historical preprocessing. CNBG-Mod is
reported on its own denominator (number of usable fits in parentheses) and
on the full panel with the portfolios that have no interval scored as
misses.}
\label{tab:eligibility}
{\scriptsize\setlength{\tabcolsep}{2.5pt}%
\begin{tabular}{@{}lrccccccc@{}}
\toprule
Eligibility rule & $n$ & ODP & F-CNBG & ODP-c & Twe-c & WLS-c & CNBG-Mod ($n$) & CNBG-Mod/all \\
\midrule
Realised reserve $> \$100$k (reported) & $398$ & $58.3$ & $54.0$ & $28.4$ & $36.7$ & $32.9$ & $63.8$ ($370$) & $59.3$ \\
Chain Ladder reserve $> \$100$k & $452$ & $53.8$ & $51.8$ & $25.4$ & $33.0$ & $29.9$ & $60.9$ ($419$) & $56.4$ \\
Latest-diagonal paid $> \$100$k & $474$ & $55.1$ & $53.8$ & $26.8$ & $32.7$ & $30.4$ & $60.0$ ($455$) & $57.6$ \\
\bottomrule
\end{tabular}}

\end{table}

\subsection{Pairs of procedures}

Frequentist CNBG and CNBG-Mod share the CNBG variance law and differ in
how parameter uncertainty is propagated; they also differ in the cell
law (Tweedie-matched against moment-matched Gamma), in the dispersion
input (the point estimate at its cap against the prior) and in the
treatment of the levels (refit against the conjugate update), so the
pair ranks two implementations within one variance family and does not
isolate the axis. Frequentist CNBG and WLS-cond share the point
estimator, and their two independent implementations return identical
$(\hat\kappa_B, \hat A)$ on every portfolio of the historical panel.


\begin{table}[H]
\centering
\caption{The procedures on the outcome-free panel and on its $303$
portfolios with at least $28$ positive observed cells: $n$ intervals,
coverage of nominal $95\%$ intervals on the procedure's own denominator (C)
and on the panel with missing intervals scored as misses (C all), and
median width (W, \% of $R_{\mathrm{actual}}$). The $10$ portfolios of the
restricted panel without any interval are first-stage failures.}
\label{tab:sp_positive}
\small
\begin{tabular}{@{}lrrrrrrrr@{}}
\toprule
& \multicolumn{4}{c}{Outcome-free panel} & \multicolumn{4}{c}{At least $28$ positive cells} \\
\cmidrule(lr){2-5}\cmidrule(lr){6-9}
Procedure & $n$ & C & C all & W & $n$ & C & C all & W \\
\midrule
ODP-cond & $440$ & 28.0 & 27.2 & 40.7 & $293$ & 38.6 & 37.3 & 43.0 \\
Tweedie-cond & $437$ & 32.3 & 31.2 & 43.9 & $293$ & 41.6 & 40.3 & 45.5 \\
WLS-cond & $440$ & 30.9 & 30.1 & 59.2 & $293$ & 39.6 & 38.3 & 45.1 \\
ODP bootstrap & $440$ & 50.7 & 49.3 & 122.1 & $293$ & 65.9 & 63.7 & 99.6 \\
Tweedie bootstrap & $429$ & 55.2 & 52.4 & 118.8 & $293$ & 70.0 & 67.7 & 103.2 \\
Frequentist CNBG & $437$ & 51.7 & 50.0 & 102.9 & $293$ & 61.1 & 59.1 & 87.3 \\
CNBG-Mod & $371$ & 68.5 & 56.2 & 108.2 & $293$ & 71.0 & 68.6 & 94.9 \\
\bottomrule
\end{tabular}

\end{table}

\section{Further limitations}\label{supp:limitations}

The working likelihood differs in location from the exact posterior
(Section~\ref{rem:exact_lik}). For the central reserve estimate this is
benign, but it matters in stress scenarios and for tail-loss inference,
where the two-moment working predictive cannot reproduce the tail of
the compound law and the exact fit, given $\alpha$, can. The exact
posterior now fits reliably on individual triangles at a roughly
$200\times$ runtime premium, so the natural refinement is a hybrid
deployment: the working likelihood for population-scale screening,
with the exact fit substituted on the individual portfolios where tail
accuracy is required and a count triangle or a prior on $\alpha$ is
available. Bringing the exact fit's runtime down, through
marginalisation of the latent $\lambda_i$ or further
reparameterisation, would widen the range of that substitution. A
related limitation is that the oracle study of
Section~\ref{sec:simulation:oracle} is run at selected configurations
only; extending it across the grid is cheap for the oracle procedures
and expensive only for the exact posterior.

The claim process also assumes calendar-year stationarity. The chain-ladder errors of the Schedule~P backtest
period (1988--1997 observed, 1998--2007 actual), under which the median
portfolio realises $70\%$ of its chain-ladder reserve and the median
absolute deviation is $44\%$ of it (Section~\ref{sec:scheduleP:bias}), are consistent with late-1990s
calendar-year changes violating that assumption, though they also contain
process variation and the backtest does not separate the two. The CNBG
framework as developed here does not address calendar-year effects
directly, and an extension incorporating calendar-year heterogeneity
would be necessary for application to long backtest windows or to
portfolios undergoing structural change. The CAS extension of the
Schedule~P compilation \citep{cas2025}, covering accident years
1998--2007 squared through 2016, provides a fully independent backtest
window on which the calendar-year attribution advanced here is directly
testable, and a replication on that window is in progress. The
horizon at which a predictive distribution is scored matters for how
these results compose with the wider literature. \citet{aslan2026} scores
\emph{one-calendar-year-ahead} development-factor forecasts on the same
database by probability integral transform and finds them
over\-dispersed, that is, too wide. The present paper's claim concerns the
\emph{ultimate-horizon} reserve distribution and runs in the opposite
direction. The two are consistent: a variance law not derived from a
generative process has no reason to be calibrated at both horizons, and
being wrong in opposite directions at different horizons is what one
should expect from it. Related to this is the
treatment of the exposure triangle, since the framework assumes that
the exposure-adjusted mean structure $\mu_i w_j$ is correctly
specified whereas in practice exposure is often imperfectly observed
(premium volume, policy-year counts, earned premium). The interaction
between exposure misspecification and the $\kappa_B$ identification
problem is not addressed here; joint frequency--severity modelling
provides a partial route, but the amount-only path remains exposed to
this source of misspecification.

The empirical analysis is also restricted to $10 \times 10$ triangles,
the Schedule~P standard. Larger triangle sizes, typical for long-tail
reinsurance where 20--30 development years are common, carry more
information on the curvature dispersion, though less than in proportion,
since the information is additive over independent rows and more
development years add correlated cells to existing rows
(Section~\ref{sec:precision:curvature}), but encounter
compounding finite-sample issues with seasonal effects,
late-development extrapolation, and exposure non-stationarity. The
behaviour of the diagnostic and predictive procedures on these larger
triangles remains to be studied in the long-tail reinsurance setting.

A more conceptual limitation concerns the plug-in conditional target
under mean-structure misspecification. The plug-in procedures target
the distribution of the reserve given the observed triangle under the
fitted mean structure, which would be the decision-relevant object if
that structure were known; the refit and posterior-draw procedures add the uncertainty
of that structure. When the chain-ladder mean structure carries large
per-portfolio errors, as the $44\%$ median absolute deviation of
Section~\ref{sec:scheduleP:bias} suggests for the post-1997
Schedule~P window, the plug-in intervals faithfully report the wrong
quantity, correctly characterising process uncertainty around a biased
centre, and the refit-based procedures' coverage advantage on
Schedule~P comes from width that carries the estimation-error
term and, on real data, absorbs mean-structure variability beyond what
a correct model generates. Every procedure, including CNBG-Mod, requires a correctly specified mean
structure. Reporting how its uncertainty is handled allows readers to
interpret the resulting width.

Finally, a zero-inflated extension is left open. The structural zero probability $\Pr(X_{ij} = 0) = (\kappa_r/(\kappa_r + \mu_{ij}))^{\kappa_r}$
(the zero mass of the negative binomial count in the single-frailty
regime) is
a testable prediction of the CNBG frailty model, and sparse triangles,
particularly excess-of-loss reinsurance data where cells below
retention are structurally empty, may exhibit zero rates exceeding
this structural prediction. A zero-inflated extension (ZI-CNBG) with a
parsimonious logistic specification for the excess zero probability,
together with a formal $Z_0$ goodness-of-fit test distinguishing
structural from excess zeros, is left to future work on the
reinsurance setting, where sparsity is the
operative case rather than the exception.

\end{document}